\documentclass[12pt]{article}

\usepackage{algorithm, algorithmic} 
\usepackage{amsmath,amssymb, dsfont}
\usepackage{graphicx}
\usepackage{caption}
\usepackage{dcolumn}
\usepackage{bm}
\usepackage{float}
\usepackage{url}
\usepackage{subfig}
 \usepackage[normalem]{ulem} 
\usepackage{array}
\usepackage[table]{xcolor}
\usepackage{multicol}
\usepackage{multirow}
\usepackage{tabularx,ragged2e} 
\newcolumntype{L}[1]{>{\raggedright\let\newline\\\arraybackslash\hspace{0pt}}m{#1}}
\newcolumntype{C}[1]{>{\centering\let\newline\\\arraybackslash\hspace{0pt}}m{#1}}
\newcolumntype{R}[1]{>{\raggedleft\let\newline\\\arraybackslash\hspace{0pt}}m{#1}}
\usepackage{booktabs}
\usepackage[authoryear, round]{natbib}
 \usepackage[
                  breaklinks = true,
                 colorlinks = true,
                 linkcolor = blue,
                 urlcolor  = black, 
                 citecolor = blue,
                 anchorcolor = green,
                 ]{hyperref}

\newcommand{\Bs}[1]{\boldsymbol{#1}}
\newcommand{\nbb}[3][0]{\underset{\quad\hspace{#1 mm}(#2)}{#3}}
\newcommand{\nb}[3][0]{\underset{\hspace{#1 mm}(#2)}{#3}}
\newcommand{\nBB}[3][0]{\underset{\quad\hspace*{#1 mm}(\Bs{#2})}{\Bs{#3}}}
\newcommand{\nB}[3][0]{\underset{\hspace*{#1 mm}(\Bs{#2})}{\Bs{#3}}}
\DeclareMathSymbol{\sminus}{\mathbin}{AMSa}{"39}
\newcommand{\splus}{\raisebox{.4\height}{\scalebox{.6}{+}}}

\newcommand{\bc}{\mathbf{c}}

\newcommand{\bA}{\mathbf{A}}

\newcommand{\br}{\mathbf{r}}

\newcommand{\bs}{\mathbf{s}}

\newcommand{\bS}{\mathbf{S}}

\newcommand{\bv}{\mathbf{v}}
\newcommand{\bV}{\mathbf{V}}

\newcommand{\bw}{\mathbf{w}}

\newcommand{\bx}{\mathbf{x}}
\newcommand{\bX}{\mathbf{X}}
\newcommand{\by}{\mathbf{y}}

\newcommand{\bone}{\mathbf{1}}
\newcommand{\rr}{\mathrm{r}}
\newcommand{\rx}{\mathrm{x}}

\newcommand{\bsb}{\boldsymbol{b}}

\newcommand{\bsu}{\boldsymbol{u}}

\newcommand{\bsx}{\boldsymbol{x}}
\newcommand{\bsX}{\boldsymbol{X}}
\newcommand{\bsy}{\boldsymbol{y}}

\newcommand{\cN}{\mathcal{N}}

\newcommand{\cL}{\mathcal{L}}

\newcommand{\cX}{\mathcal{X}}

\newcommand{\cT}{\mathcal{T}}

\newcommand{\cD}{\mathcal{D}}

\newcommand{\bsalpha}{\boldsymbol{\alpha}}
\newcommand{\bsbeta}{\boldsymbol{\beta}}
\newcommand{\bsgamma}{\boldsymbol{\gamma}}

\newcommand{\bsomega}{\boldsymbol{\omega}}

\newcommand{\bssigma}{\boldsymbol{\sigma}}

\newcommand{\bstheta}{\boldsymbol{\theta}}

\newcommand{\bseta}{\boldsymbol{\eta}}

\newcommand{\bspsi}{\boldsymbol{\psi}}
\newcommand{\bsPsi}{\boldsymbol{\Psi}}
\newcommand{\bsvPsi}{\boldsymbol{\varPsi}}

\newcommand{\bstau}{\boldsymbol{\tau}}
\newcommand{\bszeta}{\boldsymbol{\zeta}}
\newcommand{\bsxi}{\boldsymbol{\xi}}

\newcommand{\E}{\mathbb{E}}

\newcommand{\Pro}{\mathbb{P}}
\newcommand{\R}{\mathbb{R}}

\newcommand{\sign}{\mathrm{sign}}

\newcommand{\WT}[1]{\widetilde{#1}}

\title{
{Functional Mixtures-of-Experts}} 

\author{
\hspace{-.75cm} Fa\"icel Chamroukhi$^{1,2}$,\, Nhat Thien Pham$^2$\\ Van H\`a Hoang$^{3}$,\, Geoffrey J. McLachlan$^4$}
\date{
{\small {$^1$IRT SystemX, 2 boulevard Thomas Gobert, 91120 Palaiseau, France}\\$^2$Normandie Univ, UNICAEN, CNRS, LMNO, 14000 Caen, France\\
$^3$University of Science, Vietnam National University, Ho Chi Minh City, Vietnam\\ 
$^4$School of Mathematics and Physics, University of Queensland, Brisbane, Australia.
\\[2ex]}
\today
}
\usepackage{lipsum}
\usepackage{tikz}
\usepackage{eso-pic}
\begin{document}

\maketitle

\begin{center}

\hfill \break
\thanks

\vspace{-1cm}
\subsubsection*{Abstract}
\end{center} 
We consider the statistical analysis of heterogeneous  data for prediction in situations where the observations include functions, typically time series. We extend the modeling with Mixtures-of-Experts (ME), as a framework of choice in modeling heterogeneity in data for prediction  with vectorial observations, to this functional data analysis context. We first present a new family of ME models, named functional ME (FME) in which the predictors are potentially noisy observations, from entire functions. Furthermore, the data generating process of the predictor and the real response, is governed by a  hidden discrete variable representing an unknown partition. Second, by imposing sparsity on derivatives of the underlying functional parameters via Lasso-like regularizations, we provide sparse and interpretable functional representations of the FME models called iFME. We develop dedicated expectation--maximization  algorithms for Lasso-like (EM-Lasso) regularized maximum-likelihood parameter estimation strategies to fit the models. The proposed models and  algorithms are studied in simulated scenarios and in applications to two real data sets, and the obtained results demonstrate their performance in accurately capturing complex nonlinear relationships and in clustering the heterogeneous regression data. \\

\noindent {\bf Keywords:} 
Mixtures-of-Experts; 
Functional Data Analysis;
EM algorithms;
Maximum likelihood Estimation;
Lasso regularization;
Clustering

\pagebreak
{
\footnotesize
\tableofcontents
}

\renewcommand{\baselinestretch}{1}
\normalsize

\newpage

\section{Introduction}


Mixture-of-experts (ME), introduced by \citep{jacobsME}, is a successful and flexible supervised learning architecture that  allows one to  efficiently represent complex non-linear relationships in observed pairs of heterogeneous data $(\bsX,Y)$. The ME model relies on the divide and conquer principle, so that the response $Y$ is gathered from soft-association of several expert responses, each targeted to a homogeneous sub-population of the heterogeneous population, given the input covariates (predictors or features) $\bsX$.
From the statistical modeling point of view, a ME model is an extension of the finite mixture model \citep{McLachlan2000FMM} which explores the unconditional (mixture) distribution of a given set of the features $\bsX$. It is thus more tailored to unsupervised learning than to supervised learning and it has the fully conditional mixture model of the form
\begin{eqnarray}
\text{ME}(y|\bsx) = \sum_{k=1}^K \text{G}_k(\bsx) \text{E}_k(y|\bsx)\cdot
\label{eq: general ME model}
\end{eqnarray}In model  \eqref{eq: general ME model} the ME distribution
of the response $y$ given the predictors $\bsx$ is a conditional mixture distribution with predictor-dependent mixing weights, referred to as gating functions, $\text{G}_k(\bsx)$, and conditional mixture components, referred to as experts $\text{E}_k(y|\bsx)$, $K$ being the number of experts.  

Mixture of experts (ME) models  thus allow one to better capture more complex relationships between $y$ and $\bsx$ in heterogeneous situations in non-linear regression $y\in \R$, classification $y\in \{1,\ldots,G\}$, and in clustering the data by associating each expert component to a cluster. The richness of the class of ME models in terms of conditional density approximation capabilities has been recently demonstrated 
by proving denseness results \citep{nguyen2020approximationMoE,nguyen2019multMoGGE}. 

They have been investigated in their simple form, as well as in their hierarchical form \citep{Jordan-HME-1994},  for non-linear regression and model-based cluster and discriminant analyses and in different application domains.  The inference in this case can be performed by maximum likelihood estimation (MLE) via the expectation--maximization (EM) algorithm \citep{Jordan-HME-1994,McLachlanEM2008,dlr} or,  when $p$ is possibly larger than the sample size $n$, by regularized MLE via dedicated EM-Lasso algorithms as in \cite{khalili2010MoE,montuelle2014pen-MoE,chamroukhi2019rMoGGE,chamroukhi2019regularized-MoE,Chamroukhi-prEMME-2019,nguyen2020Oracle-l1-ME} which include Lasso-like penalties \citep{Tibshirani96LASSO}. 
For a more complete review of ME models, the reader is referred to \cite{YukselWG12} and \cite{NguyenChamroukhi-MoE}.
More recent theoretical results about the ME estimation and model selection for different families of ME models, can be found in \cite{nguyen2021nonBLoMPE,nguyen2021nonGLoME,nguyen2020Oracle-l1-ME}.

To the best of our knowledge, ME models have been exclusively studied in multivariate analysis when the inputs are vectors, i.e. $\bsX \in \cX = \R^p$. 
 However, in many problems, the predictors and/or the responses are observed from smooth functions.  
Indeed, in many situations, unlike in predictive and cluster analyses of multivariate and potentially high-dimensional heterogeneous data, which have been studied with the ME modeling in \eqref{eq: general ME model},
the observed data may arise from continuously observed processes, e.g. time series. Thus, a  multivariate (vectorial) analysis does not allow one to enough capture the inherent functional  structure of the data. 
In such situations, classical multivariate models are not adapted as they ignore the underlying intrinsic nature and structure of the data. Functional Data Analysis (FDA) \citep{RamsayAndSilvermanFDA2005,FerratyANDVieuBook} in which the individual data units are assumed to be functions, rather than vectors, offers an adapted framework to deal with continuously observed data, including in regression, classification and clustering. FDA considers the observed data as (discretized) values of smooth functions,  rather than multivariate observations represented in the form of ``simple'' vectors. 

The study of functional data has been 
considered in most of the statistical modeling and inference problems including regression, classification, clustering, functional graphical models \citep{qiao2019FGM}, among others. 
In regression, functional linear models have been introduced
including  
penalized functional regression
\citep{BRUNEL2016208, 
Goldsmith2011} 
and in particular the FLiRTI approach, a functional linear regression constructed upon interpretable regularization  \citep{FLIRTI}, and more generally  generalized linear models with functional predictors \citep{muller2005GFLM,James2002}, which cover functional logistic regression for classification.  
In classification, we can also cite functional linear discriminant analysis \citep{James2001},
and, as a penalized model, Lasso-regularized  functional logistic regression  \citep{MousaviFMLR}. 
To deal with heterogeneous functional data, the construction of mixture models with functional data analytic aspects have been introduced for model-based clustering
(\cite{liuANDyangFunctionalDataClustering}, \textcolor{black}{\cite{Jacques2014survey}}) 
including Lasso-regularized  mixtures for functional data \citep{Devijver2015-MBC-FDA,James2003,Jacques2014,Chamroukhi-FDA-2019}. 
The resulting functional mixture models are better able to handle functional data structures compared to standard multivariate mixtures. 

The problem of clustering and prediction in  presence of functional observations from heterogeneous populations, leading to complex distributions, is still however less investigated. 
In this paper, we investigate the framework of Mixtures-of-Experts (ME) models, as models of choice in modeling heterogeneity in data for prediction and clustering with vectorial observations, and extend it to the functional data framework. \textcolor{black}{The main novelty of our paper is to get interpretable results for functional ME (FME). First, the ME framework is extended to a functional data setting to learn from functional predictors, and the statistical inference in the resulting setting (ie. of the FME model) looks for estimating a sparse and interpretable FME model parameters. The key technical challenges brought by this setting are addressed by considering sparsity on the derivatives of the underlying functional parameters of the FME model, thanks to Lasso-like regularizations, solved by a dedicated EM-Lasso algorithm. 
To the best of our knowledge, this is we first time the ME model is constructed upon functional predictors, and provides interpretable sparse estimates. 
}

Firstly, we introduce in Section \ref{sec:FME} a new family of ME models, referred to as FME, to relate a functional predictor to a scalar response, and develop a dedicated EM algorithm for the maximum-likelihood parameter estimation.
Secondly, to deal with potential high-dimensional setting of the introduced FME model, we develop in Section \ref{sec:Lasso-MLE-FME}
a Lasso-regularized approach, which consists of a penalized MLE estimation via an hybrid EM-Lasso algorithm, which integrates an optimized coordinate ascent procedure to efficiently implement the M-Step. 
Thirdly, we in particular present in Section \ref{ssec:iFME} and extended FME model, which is constructed upon a sparse and highly-interpretable regularization of the functional expert and gating parameters. The resulting model, abbreviated as iFME, is fitted by regularized MLE via an dedicated EM algorithm.
The developed algorithms for the two introduced ME models are  applied and evaluated in Section \ref{sec:Experiments} on several simulated scenarios and on real data sets, in both clustering and non-linear regression.

 

\section{Functional Mixtures-of-Experts (FME)}
\label{sec:FME}
We wish to derive and fit new mixture-of-experts (ME) models in presence of functional predictors, \textcolor{black}{and potentially, functional responses}. 
In this paper, we first consider ME models with a  functional predictor $X(\cdot)$ and a real response $Y$ where the pair arises from heterogeneous population composed of unknown $K$ homogeneous sub-populations. 
%
To the best of our knowledge, this is the first time ME models are considered for functional data. 
\subsection{ME with functional predictor and scalar response}

Let $\{X_i(t),Y_i\}_{i=1}^n$, be a  sample of $n$  independently and identically distributed (i.i.d.) data pairs where $Y_i$ is a real-valued response and $X_{i}(t)$ is a functional predictor with $t \in \cT \subset \R$, for example the time in time series. First, to model the conditional relationships between the continuous response $Y$ and the functional predictor $X(\cdot)$, given an expert $z$, we formulate each expert component $\text{E}_z(y|\bsx)$ in \eqref{eq: general ME model} as a functional regression model (cf. \cite{muller2005GFLM}, \cite{FLIRTI}). The resulting functional expert regression model for the $i$th observation 
takes the following stochastic representation 
\begin{eqnarray}
Y_i = \beta_{z_i,0} + \int_{\cT} X_i(t) \beta_{z_i}(t) dt + \varepsilon_i, \quad i\in[n],
\label{eq:stochastic FME}
\end{eqnarray}where $\beta_{z_i,0}$ is an unknown constant intercept, $\beta_{z_i}(t),\, t\in \cT$ is the function of unknown coefficients of functional expert $z_i$, and $ \varepsilon_i \sim \cN(0,\sigma^2_{z_i})$ are independent Gaussian errors, $z_i\in [K]$ being the unknown label of the expert generating the $i$th observation. 
The notation $z_i\in[K]$ means $z_i=1,\ldots,K$ which is used throughout this paper.
In this context, the response $Y$ is related to the entire trajectory of $X(\cdot)$. 
Let \textcolor{black}{$\bsbeta = \{\beta_{z,0},\beta_{z}(t), t\in\cT\}_{z=1}^K$} represents the set of unknown functional parameters for the experts network.

Now consider the modeling of the gating network in the proposed  FME model. 
As in the context of ME for vectorial data, different choices are possible to model the gating network function, typically softmax-gated or Gaussian-gated ME (e.g. see \cite{NguyenChamroukhi-MoE}, \cite{xu1994alternative}, \cite{chamroukhi2019rMoGGE}). 
A standard choice as in \cite{jacobsME} to model the gating network $\text{G}_z(\bsx)$ in \eqref{eq: general ME model} is to use the multinomial logistic (softmax) function as a distribution of the latent variable $Z$.  
In this functional data modeling context with $K\geq 2$ experts, we use a multinomial logistic function as an extension of the functional logistic regression presented in \cite{mousavi2018functional} for linear classification. 
The resulting functional softmax gating network then takes the following form
\begin{eqnarray}\label{eq:gating Net FME}
\pi_z\left(X(t),t\in \cT; \bsalpha\right) &=&\Pro(Z=z|X(t),t\in\cT;\bsalpha)  \nonumber \\
&=& \frac{\exp\{\alpha_{z,0} + \int_{\cT} X(t) \alpha_{z}(t) dt\}}{1+\sum_{z^\prime=1}^{K-1}\exp\{\alpha_{z\prime,0} + \int_{\cT} X(t) \alpha_{z\prime}(t) dt\}},
\end{eqnarray}where 
$\bsalpha = \{\alpha_{z,0},\alpha_{z}(t),\, t\in \cT\}_{z=1}^K$ is the set of unknown constant intercept coefficients $\alpha_{z,0}$ and functional parameters $\alpha_{z}(t), t\in \cT$ for each expert $z\in[K]$. 
Note that model \eqref{eq:gating Net FME} is equivalent to assuming that each expert $z$ is related to the entire trajectory $X(\cdot)$ via the following functional linear predictor for the gating network 
\begin{eqnarray}\label{eq:linear-predictor-gating-network}
h_z(X(t),t\in\cT;\bsalpha)  
&=& 
 \log\left\{ 
 \frac{\pi_z\left(X(t),t\in \cT;\bsalpha\right)}{\pi_K\left(X(t),t\in \cT;\bsalpha\right)} 
 \right\}\nonumber \\
&=&
  \alpha_{z,0} + \int_{\cT} X(t) \alpha_{z}(t) dt.
\end{eqnarray}

The objective is to estimate the functional parameters  $\bsalpha$ and $\bsbeta$ of the FME model defined by \eqref{eq:stochastic FME}-\eqref{eq:gating Net FME}, 
from an observed sample. 
In this setting with functional predictors, this requires estimating a possibly infinite number of coefficients (as many as the number of temporal observations for the predictor). In order to reduce the complexity of the problem, the observed functional predictor can be projected onto a fixed number of basis functions so that we sufficiently capture enough the functional structure of the data, and sufficiently reduce enough the number of coefficients to estimate. 

\subsection{Smoothing representation of the functional experts}
 
Here we consider the case of fixed design, that is, the covariates $X_i(t)$ are non-random functions.
We suppose that the $X_i(\cdot)$'s are measured with error at any given time $t$.  
Hence, instead of observing directly $X_i(t)$, one has a noisy version of it $U_i(t)$, defined as
\begin{equation}
    U_i(t) = X_i(t) + \delta_i(t), \quad i \in[n],
    \label{eq:Ui(t)}
    \nonumber
\end{equation}
where $\delta_i(\cdot) \sim \mathcal{N}(0, \sigma^2_\delta)$ are measurement errors assumed to be independent of the $X_i(\cdot)$'s and the $Y_i$'s. 
%
 Since the functional predictors $X_i(t)$ are not directly observed, we first construct an approximation of $X_i(t)$ from the noisy predictors $U_i(t)$ by projecting the latter onto a set of continuous basis functions.
Let $\bsb_r(t) = \left[b_1(t), \ldots, b_r(t)\right]^\top$ be a $r$-dimensional (B-spline, Fourier, Wavelet) basis, then with $r$ sufficiently large $X_i(t)$ can be represented as
\begin{equation}
X_i(t) = \sum_{j=1}^r x_{ij} b_j(t) = \bsx_i^\top \bsb_r(t),
\label{eq: X projection}
\end{equation}
where 
$x_{ij} = \int_{\cT} X_i(t)b_j(t)dt$ for $j \in[r]$ and $\bsx_i = ( x_{i1}, \ldots,  x_{ir})^\top$. 
Since $X_i(t)$ is not observed,  the representation coefficients $x_{ij}$'s are unknown. Hence we propose an unbiased estimator of $x_{ij}$ defined as
\begin{equation}
\widehat x_{ij} := \int_\cT U_i(t)b_j(t)dt. 
\nonumber
\end{equation}
Thus, an estimate $\widehat{X}_i(t)$ of $X_i(t)$ is given by
\begin{equation}\label{eq:estimate-X(t)}
\widehat{X}_i(t) = \widehat\bsx_i^\top \bsb_r(t),\quad i\in[n],
\end{equation}
with $\widehat\bsx_i = (\widehat x_{i1}, \ldots, \widehat x_{ir})^\top$.

Similarly, to represent the regression coefficient functions $\beta_z(\cdot)$, consider a $p$-dimensional basis $\bsb_p(t) = \left[b_1(t),b_2(t),\ldots,b_p(t)\right]^\top$. Then the function $\beta_z(t)$ can be represented as
\begin{equation}
\beta_z(t) = \bseta^{\top}_z \bsb_p(t)
\label{eq: experts projection}
\end{equation}where 
$\bseta_z = (\eta_{z,1},\eta_{z,2},\ldots,\eta_{z,p})^\top$ is the vector of unknown  coefficients and the choice of $p$ should ensure the tradeoff between smoothness of the functional predictor and complexity of the estimation problem. 
We select $r \ge  p$ to satisfy the identifiability constraint (see for instance \cite{Goldsmith2011}, \cite{ramsayandsilvermanAppliedFDA2002}).
Furthermore, rather than assuming a perfect fit of $\beta_z(t)$ by $\bsb_p(t)$ as in (\ref{eq: experts projection}), we use for each Gaussian expert regressor $z$, the following error model as proposed by  \cite{FLIRTI} for functional linear regression
\begin{equation*}
\beta_z(t) = \bseta^{\top}_z \bsb_p(t) + e(t)
\label{eq: flirti experts}
\end{equation*}
where $e(t)$ represents the approximation error of $\beta_z(t)$ by the linear projection (\ref{eq: experts projection}). 
As we choose $p\gg n$, $|e(t)|$ can be assumed to be small.

\subsection{Smoothing  representation of the functional gating network}
 
Since here we are examining functional predictors,   an appropriate representation has also to be given for the gating network \eqref{eq:gating Net FME} with functional parameters $\{\alpha_{z}(t),\, t\in \cT\}_{z=1}^K$. Due to the infinite number of these parameters, we also represent the gating network by a finite set of basis functions similarly as for the experts network.  
For the representation of the functional predictors $X_i(t)$, $i\in[n]$, we use $\widehat{X}_i(t)$ established in \eqref{eq:estimate-X(t)}. The coefficients function $\alpha_{z}(t)$ is represented similarly as for the $\beta$ coefficients function of the experts network, 
by using a $q$-dimensional basis $\bsb_q(t) = \left[b_1(t),b_2(t),\ldots,b_q(t)\right]^\top$, $q \le r$, via the projection
\begin{equation}
\alpha_z(t) = \bszeta^\top_z \bsb_q(t),
\label{eq: gates projection}
\end{equation}
where $\bszeta_z = (\zeta_{z1},\zeta_{z2},\ldots,\zeta_{zq})^\top$ is the vector of softmax coefficients function. 
Note that here we use the same type of basis functions for both representations of $\beta_z$ and $\alpha_z$, but one can use different types of bases if needed. 
Then, by using the representations \eqref{eq:estimate-X(t)} and \eqref{eq: gates projection} of $X(t)$ and $\alpha_z(t)$, respectively, in the linear predictor $h_z(\cdot)$ defined in \eqref{eq:linear-predictor-gating-network} for $i\in[n]$, the latter is thus approximated as 
\begin{eqnarray}
h_z\left(U_i(t),t\in\cT;\bsalpha\right) 
&=& \alpha_{z_i,0} + \bszeta_{z_i}^\top \br_i, \label{eq: linear predictor gating}
\end{eqnarray}
where $\br_i = \left[\int_{\cT} \bsb_r(t) \bsb_q(t)^\top dt\right]^\top \widehat{\bsx}_i$. 
Thus, following its definition in (\ref{eq:gating Net FME}), the functional softmax gating network is  approximated as
\begin{equation}
\pi_{k}(\br_i;\bsxi) = \frac{\exp{\{\alpha_{k,0} + \bszeta^\top_k\br_i\}}}{1+\sum_{k^\prime=1}^{K-1}\exp{\{\alpha_{k^\prime,0} + \bszeta^\top_{k^\prime}\br_i \}}},
\label{eq:vectorial softmax gating net}
\end{equation}
where $\bsxi = \left((\alpha_{1,0},\bszeta^\top_{1}),\ldots, (\alpha_{K-1,0},\bszeta^\top_{K-1})\right)^\top\in \R^{(q+1)(K-1)}$ is the unknown parameter vector of the functional gating network to be estimated.

\subsection{The FME model conditional density}
We now have appropriate representations for the functional predictors, as well as for both the functional gating network and the functional experts network, involved in the construction of the functional ME (FME) model \eqref{eq:stochastic FME}-\eqref{eq:gating Net FME}.  
Gathering \eqref{eq:estimate-X(t)} and \eqref{eq: experts projection}, the stochastic representation (\ref{eq:stochastic FME}) of the FME model can thus be defined as follows,
\begin{eqnarray}
Y_i|u_i(\cdot) &=&  \beta_{z_i,0} + \bseta_{z_i}^\top\bx_i + \varepsilon^\star_i, \quad i\in[n],
\label{eq: stochastic FME-experts}
\end{eqnarray}where 
$\bx_i = \left[\int_{\cT} \bsb_r(t) \bsb_p(t)^\top dt\right]^\top \widehat{\bsx}_i$ and $\varepsilon^\star_i =\varepsilon_i + \widehat\bsx_i^\top \int_{\cT} \bsb_r(t)e(t) dt$. 
\textcolor{black}{We can see that the problem is now
reduced at the standard version of the ME model.}
 From this stochastic representation under the Gaussian assumption for the error variable $\varepsilon_i$, the  conditional density of each  approximated functional expert $z_i =k$ is thus given by
\begin{equation}
f\big(y_i|u_i(\cdot),z_i=k;\bstheta_k\big) = \phi(y_{i};\beta_{k,0} + \bseta_{k}^{\top} \bx_{i},\sigma^2_{k}),
\label{eq:conditional FME regression}
\end{equation}where
$\phi(\,\cdot\, ;\mu, v)$ 
 is the Gaussian probability density function with mean $\mu$ and variance $v$, 
$\beta_{k,0} + \bseta_{k}^{\top} \bx_{i}$ is the mean of the approximated functional regression expert, $\sigma^2_{k}$ its variance,  
and $\bstheta_k = (\beta_{k,0},\bseta^\top_{k}, \sigma^2_k)^\top\in \R^{p+2}$ the unknown parameter vector of expert density $k$, $k\in[K]$  to be estimated. 
Finally, combining (\ref{eq:conditional FME regression}) and (\ref{eq:vectorial softmax gating net}) in the ME model
\eqref{eq: general ME model}, leads to the  the following conditional density defining the FME  model,
\begin{equation}
f(y_i|u_i(\cdot);\bsvPsi) = \sum_{k=1}^{K} 
\pi_{k}(\br_i;\bsxi) 
\phi(y_{i};\beta_{k,0} + \bseta_{k}^{\top} \bx_{i},\sigma^2_k),
\label{eq:FME density}
\end{equation}where $\bsvPsi  = (\bsxi^{\top},\bstheta_1^{\top},\ldots,\bstheta_K^{\top})^{\top}$ is the parameter vector of the model to be estimated. 
%

\subsection{Maximum likelihood estimation via the EM algorithm} \label{ssec: EM-FME}

The FME model \eqref{eq:FME density} is now defined upon an adapted finite representation of the functional predictors, and its parameter estimation can then be performed given an observed data sample. 
We first consider the maximum likelihood estimation framework via the EM algorithm \citep{dlr,jacobsME} which has many desirable properties including stability and   convergence guarantees (e.g. see  \cite{McLachlanEM2008} for more details). Note that here we use the term maximum likelihood estimation to not unduly clutter the clarity of the text, while as it will be specified later, we refer to the conditional maximum likelihood estimator.

In practice, the data are  available in the form of discretized values of functions. The noisy functional predictors $U_i(t)$  are usually observed at discrete sampling points $t_{i1}<\ldots<t_{im_i}$ with $t_{ij}\in \cT$ for $j\in[m_i]$. 
We suppose that $U_i(t)$ is scaled such that $0\le t \le 1$ and divide the time period $[0,1]$ up into a fine grid of $m_i$ points $t_{i1}, \ldots, t_{im_i}$. 
 Thus, in (\ref{eq: stochastic FME-experts}) we have  
$\bx_{i} = \left[\sum_{j=1}^{m_i} \bsb_r(t_{ij}) \bsb_p(t_{ij})^\top\right]^\top\hspace*{-4pt}\widehat\bsx_i$,  
$\br_{i} = \left[\sum_{j=1}^{m_i} \bsb_r(t_{ij}) \bsb_q(t_{ij})^\top\right]^\top\hspace*{-4pt}\widehat\bsx_i$,  
where $\widehat  x_{ij} = \sum_{j=1}^{m_i} U_i(t_{ij})b_j(t_{ij})$. Note that if we choose $p = q = r$, then $\bx_{i} = \br_{i} = \widehat\bsx_i$. 
Let $\cD = \{(\bsu_1,y_1),\ldots,(\bsu_n, y_n)\}$ be an i.i.d. sample of $n$ observed data pairs
where $\bsu_i = (u_{i,1},\ldots,u_{i,m_i})$ is the observed functional predictor for the $i$th response $y_i$. 

We use $\cD$ to estimate the parameter vector $\bsvPsi$ by iteratively maximizing the observed data log-likelihood, 
\begin{equation}
\log L(\bsvPsi) = \sum_{i=1}^{n}\log \sum_{k=1}^{K} 
\pi_{k}(\br_i;\bsxi) 
\phi(y_{i};\beta_{k,0} + \bseta_{k}^{\top} \bx_{i},\sigma^2_k),
 \label{eq:loglik FME}
\end{equation}via the EM algorithm. 
As detailed in Appendix, the EM algorithm for the FME model is implemented as follows. After starting with an initial solution $\bsvPsi^{(0)}$, it alternates, at each iteration $s$, between the two following steps, until convergence (when there is no longer a significant change in the values of the log-likelihood (\ref{eq:loglik FME})).

\paragraph{E-step.}
Calculate the following conditional probability memberships  $\tau^{(s)}_{ik}$ (for all $i\in[n]$),  that the observed pair $(u_i, y_i)$ originates from  the $k$th expert, given the observed data and the current parameter estimate $\bsvPsi^{(s)}$,
\begin{equation}
\tau_{ik}^{(s)}= \Pro(Z_i=k|y_i,u_i(\cdot);\bsvPsi^{(s)}) = \frac{
\pi_{k}(\br_i;\bsxi^{(s)}) 
\phi(y_{i};\beta^{(s)}_{k,0} + \bx^{\top}_{i}\bseta^{(s)}_{k},{\sigma^2_k}^{(s)})
}{f(y_{i}|u_i(\cdot);\bsvPsi^{(s)})}\cdot
\label{eq:FME post prob}
\end{equation}

\paragraph{M-step.}
\label{ssec: M-step EM-FME} 
Update the value of the parameter vector $\bsvPsi$ by maximizing the $Q$-function (\ref{eq:Q-function FME}) with respect to $\bsvPsi$. 
The maximization is performed by separate maximizations with respect to (w.r.t.) the gating network parameters $\bsxi$ and, for each expert $k$, w.r.t. the expert network parameters $\bstheta_k$,  for each of the $K$ experts.\\

\noindent {\it Updating the gating network's parameters} $\bsxi$ consists of maximizing w.r.t. $\bsxi$ the part of \eqref{eq:Q-function FME} that is a function of $\bsxi$. 
 Since we use a softmax-gated expert network in \eqref{eq:vectorial softmax gating net}, this maximization problem consists of a weighted  multinomial logistic problem for which there is no a closed-form solution.  
We then use a Newton-Raphson (NR) procedure, which iteratively maximizes (\ref{eq:Q-function gating-network}) 
after starting from an initial parameter vector $\bsxi^{(0)}$, by updating, at each NR iteration $t$, the values of the parameter vector $\bsxi$ according to the following updating formula: 
\begin{eqnarray}
\label{eq. NR for Q gating}
    \bsxi^{(t+1)} = \bsxi^{(t)} - 
    \Big[H(\bsxi;\bsvPsi^{(s)})\Big]_{\bsxi=\bsxi^{(t)}}^{-1}
    g(\bsxi; \bsvPsi^{(s)})\Big| _{\bsxi=\bsxi^{(t)}}
\end{eqnarray}where $H(\bsxi;\bsvPsi^{(s)})$ and $g(\bsxi; \bsvPsi^{(s)})$ are, respectively, the Hessian matrix and the gradient vector of $Q(\bsxi; \bsvPsi^{(s)})$, and are provided in Appendix \ref{Appendix EM}. 
At each NR iteration, the Hessian matrix and gradient vector are evaluated at the current value of $\bsxi$. 
We keep updating the gating network parameter $\bsxi$ according to \eqref{eq. NR for Q gating} until there is no significant change in $Q(\bsxi;\bsvPsi)$. The maximization then provides $\bsxi^{(s+1)}$ for the next EM iteration.
\textcolor{black}{Alternatively, instead of using the NR procedure, one can employ the Minorize-Maximization (MM) algorithm to update the parameter $\bsxi$ of the gating network. In this approach, the Hessian matrix in \eqref{eq. NR for Q gating} is approximated by a square matrix that is independent of $\bsxi$. This can offer an improved computational efficiency, although it comes at the cost of requiring a greater number of iterations. For further details, refer to \cite{GormleyMurphy2008}.}

\noindent {\it Updating the experts network parameters}
$\bstheta_k$ consists of 
solving $K$ independent weighted regression problems where the weights are the conditional expert memberships $\tau^{(s)}_{ik}$ given by \eqref{eq:FME post prob}. 
The updating formulas for the regression parameters $(\beta_{k,0},\bseta_k)$ and the noise variances $\sigma_{k}^{2}$ for each expert $k$ are straightforward and correspond to weighted versions of those of standard Gaussian linear regression, i.e., weighted ordinary least squares.
The updating rules for the experts network parameters are given by the following formulas:
\begin{equation}
\begin{aligned}
\beta^{(s+1)}_{k,0} &=  \frac{1}{n^{(s)}_k}\sum_{i=1}^{n}  \tau_{ik}^{(s)}(y_i - \bx_{i}^{\top}\bseta^{(s)}_{k}),\quad 
\bseta^{(s+1)}_{k} = \frac{1}{\sum_{i=1}^{n}\tau^{(s)}_{ik} \bx_i\bx_i^\top} \sum_{i=1}^{n}\tau^{(s)}_{ik} (y_{i} - \beta_{k,0}^{(s+1)}) \bx_i, \\
{\sigma_{k}^2}^{(s+1)} &= \frac{1}{n^{(s)}_k} \sum_{i=1}^{n}  \tau_{ik}^{(s)}\left[y_i -  (\beta^{(s+1)}_{k,0} + \bx_{i}^{\top}\bseta^{(s+1)}_{k})\right]^2, 
 \end{aligned}
 \label{eq:MLE FME beta and sigma2}
\end{equation}
where $n^{(s)}_k = \sum_{i=1}^n\tau_{ik}^{(s)}$ 
represents the expected cardinal number of component $k$. 

This EM algorithm, designed here for the FME that is constructed upon smoothing of the functional data, can be seen as a direct extension of the vectorized version the ME model. 
While it  can hence be expected to provide accurate estimations as in the vector predictors setting, the number of parameters to estimate here in the case of the  FME can still be high, for example when a big number of basis functions is used to have more accurate approximation of the functional predictors. In that case, it is better to regularize the 
maximum likelihood estimator
in order to establish a compromise between the quality of fit and complexity.

\section{Regularized maximum likelihood estimation} 
\label{sec:Lasso-MLE-FME}
We  rely on the LASSO \citep{Tibshirani96LASSO} as a successful procedure to encourage sparse models in high-dimensional linear regression based on an $\ell_1$-penalty, and include it in this mixture of experts modeling framework for functional data. The $\ell_1$-regularized ME models have demonstrated their performance from a computational point of view \citep{chamroukhi2019regularized-MoE,Chamroukhi-prEMME-2019}
and enjoy good theoretical properties \textcolor{black}{of universal approximation and consistent model selection in the high-dimensional setting.}
\citep{nguyen2020Oracle-l1-ME,nguyen2021nonasympME}. 
 
\subsection{\texorpdfstring{$\ell_1$}{}-regularization and the EM-Lasso algorithm} 
\label{ssec: rEM-FME}

We propose an $\ell_1$-regularization of the observed-data log-likelihood \eqref{eq:pen-loglik FME} to be maximized, along with coordinate ascent algorithms to deal with the high-dimensional setting when updating the parameters within the resulting EM-Lasso algorithm.  
The objective function in this case is given by the following $\ell_1$-regularized observed-data log-likelihood, 
\begin{equation}
\cL(\bsvPsi) = \log L(\bsvPsi) - \text{Pen}_{\lambda,\chi}(\bsvPsi),
 \label{eq:pen-loglik FME}
\end{equation}where $\log L(\bsvPsi)$ is the observed-data log-likelihood of  $\bsvPsi$ defined by 
\eqref{eq:loglik FME}, 
and $\text{Pen}_{\lambda,\chi}(\bsvPsi)$ is a LASSO 
regularization term encouraging sparsity for the expert  and the gating network parameters, defined by
\begin{equation}
\text{Pen}_{\lambda,\chi}(\bsvPsi) = \lambda \sum_{k=1}^{K} \sum_{j = 1}^{p} |\eta_{k,j}| + \chi \sum_{k=1}^{K-1}\sum_{j = 1}^{q} |\zeta_{k,j}|,
\label{eq:Pen term}
\end{equation}
where $\lambda$ and $\chi$ are positive real values representing tuning parameters. 
The maximization of (\ref{eq:pen-loglik FME}) cannot be performed in a closed form but again  
the  EM algorithm  can be adapted to iteratively solve the maximization problem.  
The resulting algorithm for the FME model, called EM-Lasso, takes the following form, as detailed in Appendix \ref{Appendix EM-Lasso}. 
After starting with an initial solution $\bsvPsi^{(0)}$, it alternates between the two following steps, until convergence, i.e., when there is no longer a significant change in the values of the $\ell_1$-penalized log-likelihood (\ref{eq:pen-loglik FME}).
 
\paragraph{E-step.}
\label{ssec: E-step EM-Lasso FME} 
The E-Step in this EM-Lasso algorithm is unchanged compared to the previously presented EM algorithm, and only requires  the computation of the conditional expert memberships $\tau^{(s)}_{ik}$  according to \eqref{eq:FME post prob}. 

\paragraph{M-step.}
\label{ssec: M-step EM-FME-Lasso} 
In this regularized MLE context, the parameter vector $\bsvPsi$ is now updated by maximizing the regularized $Q$-function (\ref{eq:Q-function FMElasso}), i.e.,  
$\bsvPsi^{(s+1)} = \arg \max_{\bsvPsi}\left\{ Q(\bsvPsi;\bsvPsi^{(s)}) - \text{Pen}_{\lambda,\chi}(\bsvPsi)\right\}.$ 
This is performed by separate maximizations w.r.t. the gating network parameters $\bsxi$ and, for each expert $k$, w.r.t. the expert network parameters $\bstheta_k$, $k\in[K]$.

\noindent {\it Updating the gating network parameters} at iteration $s$ of the  EM-Lasso algorithm  consists of maximizing the following regularized $Q$-function w.r.t. $\bsxi$, 
\begin{eqnarray}
\label{eq:Q-function gating-network lasso}
\mathcal{Q}_{\chi}(\bsxi;\bsvPsi^{(s)}) =  Q(\bsxi;\bsvPsi^{(s)}) - \chi \sum_{k=1}^{K-1} \Vert\bszeta_{k}\Vert_1,
\end{eqnarray}
where
$
Q(\bsxi;\bsvPsi^{(s)}) =   \sum_{i=1}^n\left[\sum_{k=1}^{K-1} \tau^{(s)}_{ik} \left(\alpha_{k,0} + \bszeta^\top_{k}\br_i\right)  - \log\left(1 +   \sum_{k^\prime =1}^{K-1}  \exp\{\alpha_{k^\prime,0} + \bszeta^\top_{k^\prime}\br_i\} \right) \right]$. One can see this is equivalent to solving a weighted regularized multinomial logistic regression problem for which $\mathcal{Q}_{\chi}(\bsxi;\bsvPsi^{(s)})$ is its penalized log-likelihood, the weights being the conditional probabilities $\tau^{(s)}_{ik}$. 
There is no closed-form solution for this kind of problem. We then use an iterative optimization algorithm  to seek for a maximizer of $\mathcal{Q}_{\chi}(\bsxi;\bsvPsi^{(s)})$, i.e., an update for the parameters of the gating network. 
To be effective when the number of parameters to estimate is high, we propose a coordinate ascent algorithm to update the softmax gating network coefficients in this regularized context. 

\paragraph{\it Coordinate ascent for updating the gating network.}
The idea  of the coordinate ascent  algorithm (e.g. see \cite{hastie2015statistical}, \cite{Chamroukhi-prEMME-2019}), implemented in our context 
at the $s$th EM-Lasso iteration  to maximize $\mathcal{Q}_{\chi}(\bsxi;\bsvPsi^{(s)})$ at the M-Step, is as follows. 
The gating function parameter vectors $\bsxi_k=(\alpha_{k,0},\bszeta^\top_{k})^\top$  as components of the whole gating network parameters $\bsxi=(\bsxi^\top_1,\ldots,\bsxi^\top_{K-1})^\top$, are updated one at a time, while fixing the other gate's parameters to their previous estimates. 
Furthermore, to update a single gating parameter vector $\bsxi_{k}$, we only update its coefficients $\xi_{kj}$ one at a time, while fixing the other coefficients to their previous values.   
More precisely, for each single gating function $k$,  we partially approximate the smooth part of $\mathcal{Q}_{\chi}(\bsxi;\bsvPsi^{(s)})$ with respect to $\bsxi_k$ at the current EM-Lasso estimate, say $\bsxi^{(t)}$, 
then optimize the resulting objective function 
(with respect to $\bsxi_k$). This corresponds to solving penalized weighted least squares problems using coordinate ascent. 
Thus, this results into an inner loop, indexed by $m$, that cycles over the components of $\bsxi_k$ and updates them one by one, 
while the others are kept fixed to their previous values, i.e.,  $\xi_{kh}^{(m+1)} = \xi_{kh}^{(m)}$ for all $h\neq j$, until the objective function \eqref{gating optimization problem} is not significantly increased.

The obtained closed form updates
for each coefficient $\zeta_{kj}$, $j\in[q]$,
and for the intercept $\alpha_{k,0}$, are as follows 
\begin{eqnarray*}
    \zeta_{kj}^{(m+1)} = \frac{\mathcal S_{\chi}
    \Big(
        \sum_{i=1}^n w_{ik}\rr_{ij}
        (c_{ik} - \tilde{c}_{ikj}^{(m)})
    \Big)}
    { \sum_{i=1}^n w_{ik} \rr_{ij}^2 } \text{ for } j\in[q],
    \quad 
    \alpha_{k,0}^{(m+1)} = \frac{ \sum_{i=1}^n w_{ik} 
    (c_{ik} - \br_i^{\top} \bszeta_k^{(m+1)}) 
    }{\sum_{i=1}^n w_{ik}},
\end{eqnarray*}where $\tilde{c}_{ikj}^{(m)} = \alpha_{k0}^{(m)} + \br_i^{\top}\bszeta_k^{(m)} - \zeta_{kj}^{(m)}\rr_{ij}$ is the fitted value excluding the contribution from the $j$th component of the $i$th vector $\rr_{ij}$ in the design matrix of the gating network and 
$\mathcal S_{\chi}(\cdot)$ is a soft-thresholding operator defined by $\mathcal S_{\chi}(u) = \sign(u)(\vert u \vert - \chi)_+$ with $(v)_+$ is a shorthand for $\max\{v,0\}$.  
The values $(\alpha^{(m+1)}_{k,0},\bszeta^{(m+1)}_{k})$ obtained at convergence of the coordinate ascent inner loop for the $k$th gating function
are taken as the fixed values of that gating function, in the procedure of updating the next  parameter vector $\bsxi_{k+1}$. 
Finally, when all the gating functions have their values updated, i.e., after $K-1$ inner coordinate ascent loops, to avoid numerical instability, we perform a backtracking line search, before actually updating the gating network's parameters for the next EM-Lasso iteration. 
More precisely, the update is $\bsxi^{(t+1)} = (1-\nu)\bsxi^{(t)} + \nu\bar{\bsxi}^{(t)}$, where $\bar{\bsxi}^{(t)}$ is the output after $K-1$ inner loops and $\nu$ is backtrackingly determined to ensure $\mathcal{Q}_{\chi}(\bsxi^{(t+1)};\bsvPsi^{(s)}) \geq \mathcal{Q}_{\chi}(\bsxi^{(t)};\bsvPsi^{(s)})$.

We keep cycling these updated iterates for the parameter vectors $\bsxi_k$,  until convergence of the whole coordinate ascent (CA) procedure inside the M-Step, i.e., 
\textcolor{black}{when the relative increase in the Lasso-regularized objective $\mathcal{Q}_{\chi}(\bsxi;\bsvPsi^{(s)})$ related to the softmax gating network is not significant, e.g., less than a small tolerance.}
Then, the next EM-Lasso iteration is calculated with the updated gating network's parameters $\bsxi^{(s+1)} = (\widetilde\alpha_{1,0},\widetilde\bszeta^{\top}_{1},\ldots, \widetilde\alpha_{K-1,0},\widetilde\bszeta^{\top}_{K-1})^\top$ where the values $\widetilde\alpha_{k,0}$ and $\widetilde\zeta_{kj}$ 
for all $k\in [K-1], j \in [q]$ are those obtained for 
the $\alpha_{k,0}$'s and the $\zeta_{kj}$'s 
at convergence of the CA algorithm.

\paragraph{\it Updating the experts network parameters}
The maximization step for updating the expert parameters $\bstheta_k$ consists of maximizing the function 
$\mathcal{Q}_{\lambda}(\bstheta_k;\bsvPsi^{(s)})$ given by
\begin{eqnarray}
\label{eq:Q-function expert-network lasso}
 \mathcal{Q}_{\lambda}(\bstheta_k;\bsvPsi^{(s)}) &=& Q(\bstheta_k;\bsvPsi^{(s)}) - \lambda\Vert \bseta_k\Vert_1,
\end{eqnarray}where
$Q(\bstheta_k;\bsvPsi^{(s)}) = - \frac{1}{2\sigma_k^2}\sum_{i=1}^{n} \tau_{ik}^{(s)} \left(y_{i} - (\beta_{k,0} + \bseta_{k}^{\top} \bx_{i})\right)^2 - \frac{n}{2}\log(2\pi \sigma_k^2)\cdot$
This corresponds to solving a weighted LASSO problem  where the weights are the conditional experts memberships $\tau_{ik}^{(s)}$. 
We then solve it by an iterative optimization algorithm similarly to the previous case of updating the gating network parameters. 
As it can be seen in Appendix \ref{Appendix: expert updates for EM-Lasso}, updating $(\beta_{k,0}, \bseta_k)$ according to \eqref{eq:Q-function expert-network lasso} is obtained by coordinate ascent as follows. For each $j\in[p]$, the closed-form update for $\eta_{kj}$ 
at the $m$th iteration of the coordinate ascent algorithm within the M-Step of EM-Lasso, 
is given by
\begin{eqnarray*}
\eta_{kj}^{(m+1)} = \frac{\mathcal S_{\lambda{\sigma_k^2}^{(s)}}
    \Big(
        \sum_{i=1}^n \tau_{ik}^{(s)}\rx_{ij}
        (y_{i} - \tilde{y}_{ij}^{(m)})
    \Big)}
    { \sum_{i=1}^n \tau_{ik}^{(s)} \rx_{ij}^2 },
    \quad
    \beta_{k,0}^{(m+1)} = \frac{\sum_{i=1}^n\tau_{ik}^{(s)}(y_i - \bx_i^{\top}\bseta_{k}^{(m+1)})}{\sum_{i=1}^n\tau_{ik}^{(s)}},
\end{eqnarray*}
in which $\tilde{y}_{ij}^{(m)} = \beta_{k,0}^{(m)} + \bx_i^{\top}\bseta_k^{(m)} - \eta_{kj}^{(m)}\rx_{ij}$ is the fitted value excluding the contribution from $\rx_{ij}$ and $\mathcal S_{\lambda{\sigma_k^2}^{(s)}}(\cdot)$ is the soft-thresholding operator. 
 We keep updating the components of $(\beta_{k,0},\bseta_k)$ cyclically until no enough increase in objective function 
 \eqref{eq:Q-function expert-network lasso}. 
 Then, once $(\beta_{k,0}, \bseta_{k})$ are updated while fixing the variance $\sigma^2_k$, the latter is then updated straightforwardly as in the case of standard weighted Gaussian regression, and it is update is given by
\begin{eqnarray*}
{\sigma^2_k}^{(s+1)} = \frac{\sum_{i=1}^n \tau_{ik}^{(s)} \left(y_i - \beta_{k,0}^{(s+1)} - \bx_i^{\top}\bseta_k^{(s+1)}\right)^2}{\sum_{i=1}^n \tau_{ik}^{(s)}},
\end{eqnarray*}where $(\beta_{k,0}^{(s+1)}, \bseta_{k}^{(s+1)}) = (\widetilde\beta_{k,0}, \widetilde\bseta_{k})$ is the solution obtained at convergence of the CA algorithm, which is taken as the update in the next EM-Lasso iteration.

This completes the parameter vector update  $\bsvPsi^{(s+1)} = \big(\bsxi^{(s+1)}, \bstheta_1^{(s+1)}, \ldots, \bstheta_K^{(s+1)}\big)$ of the regularized FME model, where $\bsxi^{(s+1)}$ and $\bstheta_k^{(s+1)}, k\in[K]$, are, respectively, the updates of the gating network parameters and the experts network parameters, calculated 
by the coordinate ascent algorithms.  

The EM-Lasso algorithm provides an estimate of the FME parameters with sparsity constraints on the values of the parameter vectors $\bsxi$ and $\bstheta_k$, $k\in[K]$. 
Actually, since here these parameter vectors do not relate directly the original functional inputs, to the  output,  assuming some of their values is zero is not \textcolor{black}{necessarily justified, as there is no indeed any reason that these values are zero, nor} easily interpretable, compared to the sparsity in  vectorial  generalized linear models, mixture of regressions and ME models.

From now on, we refer to FME and FME-Lasso, respectively, the FME model fitted by EM algorithm in Section \ref{ssec: EM-FME} and the regularized FME model fitted by EM-Lasso algorithm, in Section \ref{ssec: rEM-FME}. In the following Section, we introduce a regularization that is interpretable and encourages sparsity in the FME model.
  

\subsection{Interpretable Functional Mixture of Experts (iFME)} 
\label{ssec:iFME}

In this section, we provide a sparse and, especially highly-interpretable fit, for the coefficient functions $\{\beta_k(t), t\in\cT\}$ and $\{\alpha_k(t), t\in\cT\}$ representing each of the $K$ functional experts and gating network. We call our approach Interpretable Functional Mixture of Experts (iFME). The presented iFME allows us to control the expert and gating parameter functions while still providing performance as with the standard FME model presented previously.

\subsubsection{Interpretable sparse regularization}

We rely on the methodology of Functional Linear Regression That's Interpretable (FLiRTI) presented in \cite{FLIRTI}. The idea of the FLiRTI methodology is as follows. We use variable selection with sparsity assumption on appropriate chosen derivatives of the coefficient function, say $\beta_{z_i}(t)$ here, in the case of the functional expert network, to produce a highly interpretable estimate for the coefficient functions $\beta_{z_i}(t)$. For instance, $\beta_{z_i}^{(0)}(t) = 0$ implies that the predictor $X_i(t)$ has no effect on the response $Y_i$ at $t$, $\beta_{z_i}^{(1)}(t) = 0$ means that $\beta_{z_i}(t)$ is constant in $t$, $\beta_{z_i}^{(2)}(t) = 0$ shows that $\beta_{z_i}(t)$ is linear in $t$, etc. Assuming sparsity in higher-order derivatives of $\beta_{z_i}(t)$, for instance when $d=3$ or $d=4$, will however give us a less easily interpretable fit. Hence, for example, if one believes that the expert parameter function $\beta_{z_i}(t)$ is exactly zero over a certain region and exactly linear over other region of $t$, then it is necessary to estimate $\beta_{z_i}(t)$ such that $\beta_{z_i}^{(0)}(t)=0$ and $\beta_{z_i}^{(2)}(t)=0$ over those regions, respectively. In this situation, we need to model $\beta_{z_i}(t)$ assuming that its zeroth and second derivatives are sparse.
However, with the EM-Lasso method derived above via the Lasso regularization, there is no actually any reason that we could obtain those desired properties, which may result in an estimate for $\beta_{z_i}(t)$ that is rarely exactly zeros (and/or linear), and making the sparsity and coefficient curves hard to interpret. The same situation may occur with the gating parameter functions. 
To handle this, we describe in what follows the construction of our iFME model that produces flexible-shape and highly-interpretable estimates for the expert and gating coefficient functions, by simultaneously constraining any two of their derivatives to be sparse.

We start by selecting a $p$-dimensional basis $\bsb_p(t)$ and a $q$-dimensional basis $\bsb_q(t)$ for approximating the experts and gating networks, respectively.
For the expert network, if we divide the time domain into a grid of $p$ evenly spaced points, and let $D^d$ be the $d$th finite difference operator defined recursively as
\begin{eqnarray*}
D \bsb_p(t_j) &=& p\left[\bsb_p(t_j) - \bsb_p(t_{j-1})\right], \\
D^2 \bsb_p(t_j) &=& D\left[D \bsb_p(t_j)\right] = p^2\left[\bsb_p(t_j) - 2\bsb_p(t_{j-1}) + \bsb_p(t_{j-2})\right], \\
&\vdots \\
D^d \bsb_p(t_j) &=& D\left[D^{d-1} \bsb_p(t_j)\right],
\end{eqnarray*}
then $D^d \bsb_p(t_j)$ provides an approximation for $\bsb_p^{(d)}(t_j) = [b_1^{(d)}(t_j), \ldots, b_p^{(d)}(t_j)]^\top$, $j\in[p]$. Let
\begin{eqnarray*}\label{eq:approx-derivatives-B(t)}
\bA_p &=& \Big[
\underbrace{D^{d_1} \bsb_p(t_1), D^{d_1} \bsb_p(t_2), \ldots, D^{d_1} \bsb_p(t_p)}_{\textcolor{black}{\bA_p^{[d_1]}}},\  
\underbrace{D^{d_2}\bsb_p(t_1), D^{d_2} \bsb_p(t_2), \ldots, D^{d_2} \bsb_p(t_p)}_{\textcolor{black}{\bA_p^{[d_2]}}}
\Big]^{\top}
\end{eqnarray*}
be the matrix of approximate $d_1$th and $d_2$th derivative of the basis $\bsb_p(t)$. We denote $\bA_p^{[d_1]}$ the first $p$ rows of $\bA_p$ and $\bA_p^{[d_2]}$ the remainder, i.e., $\bA_p = \big[\bA_p^{[d_1]}, \bA_p^{[d_2]}\big]^{\top}$. One can see such matrix $\bA_p$ is in $\R^{2p\times p}$ and $\bA_p^{[d_1]}$ is a square invertible matrix. 

Now, if we consider the following representation for the expert network coefficient function
\begin{equation}\label{eq:approx-derivatives-gamma}
\boldsymbol{\gamma}_{z_i} = \bA_p \Bs{\eta}_{z_i}
\end{equation}
with  $\bsgamma_{z_i} = \big({\bsgamma^{[d_1]}_{z_i}}^\top\hspace{-2pt},{\bsgamma^{[d_2]}_{z_i}}^\top\big)^\top$, where $\bsgamma^{[d_1]}_{z_i} = \big(\gamma^{[d_1]}_{1,{z_i}}, \ldots, \gamma^{[d_1]}_{p,{z_i}}\big)^\top$, $\bsgamma^{[d_2]}_{z_i} = \big(\gamma^{[d_2]}_{1,{z_i}}, \ldots, \gamma^{[d_2]}_{p,{z_i}}\big)^\top$, 
and $\Bs{\eta}_{z_i}$ defined as in relation to $\beta_{z_i}(t)$ as in \eqref{eq: experts projection},
then one can see that $\bsgamma^{[d_1]}_{z_i}$ provides an approximation to $\beta_{z_i}^{(d_1)}(t)$, i.e. the $d_1$th derivative of $\beta_{z_i}(t)$. Respectively, $\bsgamma^{[d_2]}_{z_i}$ provides an approximation to $\beta_{z_i}^{(d_2)}(t)$, the $d_2$th derivative of $\beta_{z_i}(t)$.
Therefore, enforcing sparsity in $\bsgamma_{z_i}$ will constrain $\beta_{z_i}^{(d_1)}(t)$ and $\beta_{z_i}^{(d_2)}(t)$ to be zero at most time points.

Similarly, \textcolor{black}{let $\bA_q = \big[\bA_q^{[d_1]}, \bA_q^{[d_2]}\big]^{\top} \in \R^{2q\times q}$ be the corresponding matrix defined for the gating network.} If we consider the following  representation for the gating network coefficient function 
\begin{equation}\label{eq:approx-derivatives-omega}
\bsomega_{z_i} = \bA_q \Bs{\zeta}_{z_i}
\end{equation}
with
$\bsomega_{z_i} = \big({\bsomega^{[d_1]}_{z_i}}^\top\hspace{-2pt}, {\bsomega^{[d_2]}_{z_i}}^\top\big)^\top$\hspace{-2pt},
where
$\bsomega^{[d_1]}_{z_i} = \big(\omega^{[d_1]}_{1,{z_i}}, \ldots, \omega^{[d_1]}_{q,{z_i}}\big)^\top$\hspace{-2pt},
$\bsomega^{[d_2]}_{z_i} = \big(\omega^{[d_2]}_{1,{z_i}},\ldots,\omega^{[d_2]}_{q,{z_i}}\big)^\top$\hspace{-2pt},
and $\Bs{\zeta}_{z_i}$ defined as in relation to $\alpha_{z_i}(t)$ as in \eqref{eq: gates projection},
then we can derive the same interpretation as above for the coefficient vector $\bsomega_{z_i}$ and the gating parameter function $\alpha_{z_i}(t)$.

 Using simple calculations, the representations \eqref{eq:approx-derivatives-gamma} and \eqref{eq:approx-derivatives-omega} imply the following relations that subsequently used in the iFME model:
 \begin{equation}\label{eq:flirti constrains expert}
\bseta_{z_i} = {\big(\bA_p^{[d_1]}\big)}^{-1} \bsgamma_{z_i}^{[d_1]}, \qquad
\bsgamma_{z_i}^{[d_2]} = \bA_p^{[d_2]}{\big(\bA_p^{[d_1]}\big)}^{-1} \bsgamma_{z_i}^{[d_1]},
\end{equation}
and
\begin{equation}\label{eq:flirti constrains gating}
\bszeta_{z_i} = {\big(\bA_q^{[d_1]}\big)}^{-1} \bsomega_{z_i}^{[d_1]}, \qquad
\bsomega_{z_i}^{[d_2]} = \bA_q^{[d_2]}{\big(\bA_q^{[d_1]}\big)}^{-1} \bsomega_{z_i}^{[d_1]},
\end{equation}
respectively. 

In fact, one can construct $\bA_p$ and $\bA_q$ with only one derivative. Then the constraints involved to the $d_2$th derivative will be eliminated making the estimation easier, but also limiting the flexibility in the shapes of the functions. That is why in this construction and in our experimental studies, $\bA_p$ and $\bA_q$ are constructed with multiple derivatives in order to produce curves of $\beta_{z_i}(\cdot)$ and $\alpha_{z_i}(\cdot)$ with such many desired properties.

\subsubsection{The iFME model}

Combining the stochastic representation of the FME model
in \eqref{eq: stochastic FME-experts} for the experts model, the linear predictor definition in \eqref{eq: linear predictor gating}, and the relations \eqref{eq:flirti constrains expert}-\eqref{eq:flirti constrains gating}, we obtain the following iFME model construction
\begin{eqnarray}
Y_i|u_i(\cdot) 
&=& \beta_{z_i,0} + {\Bs\gamma_{z_i}^{[d_1]}}^\top  \bv_i + \varepsilon_i^\star, \label{eq:FLiRTI expert} \\
h_{z_i}(u_i(\cdot);\bsalpha) 
&=& \alpha_{z_i,0} +  {\Bs\omega_{z_i}^{[d_1]}}^\top \bs_i, \label{eq:FLiRTI linear predictor gating}
\end{eqnarray}
subject to
\begin{eqnarray}\label{eq: FLiRTI constrains}
\bsgamma_{z_i}^{[d_2]} = \bA_p^{[d_2]}{\bA_p^{[d_1]}}^{-1} \bsgamma_{z_i}^{[d_1]}
\quad\text{and}\quad 
\bsomega_{z_i}^{[d_2]} = \bA_q^{[d_2]}{\bA_q^{[d_1]}}^{-1} \bsomega_{z_i}^{[d_1]},  
\end{eqnarray}
where $\bv_i =  \big({\bA^{[d_1]}_p}^{-1}\big)^{\hspace{-2pt}\top}\bx_i$ is the new design vector for the experts 
and $\bs_i = \big({\bA^{[d_1]}_q}^{-1}\big)^{\hspace{-2pt}\top}\br_i$ the new one for the gating network. 
 Hence, from \eqref{eq:FLiRTI expert} and \eqref{eq:FLiRTI linear predictor gating}, the conditional density of each expert and the gating network are now written as
\begin{equation}
f(y_i|u_i(\cdot);\bspsi_k) = \phi(y_{i};\beta_{k,0} + {\bsgamma_{k}^{[d_1]}}^{\top} \bv_{i},\sigma^2_{k})
\label{eq:vectorial expert sparse}
\end{equation}
and
\begin{equation}
\pi_{k}(\bs_i;\bw) = \frac{\exp{\{\alpha_{k,0}+ {\bsomega_k^{[d_1]}}^\top \bs_i \}}}{1+\sum_{k^\prime=1}^{K-1}\exp{\{\alpha_{k^\prime,0}+{\bsomega^{[d_1]}_{k^\prime}}^\top\bs_i \}}},
\label{eq:vectorial softmax gating sparse}
\end{equation}
where $\bspsi_k=\big(\beta_{k,0}, {\bsgamma^{[d_1]}_k}^\top, \sigma^2_k\big)^\top$ is the unknown parameter vector of expert component density $k$ and 
$\bw=\left(\alpha_{1,0},{\bsomega^{[d_1]}_1}^\top,\ldots,\alpha_{K-1,0},{\bsomega^{[d_1]}_{K-1}}^{\hspace{-4pt}\top}\right)^{\hspace{-2pt}\top}$, with $\big(\alpha_{K,0}, {\bsomega^{[d_1]}_{K}}^{\top}\big)^\top$ a null vector, is the unknown parameter vector of the gating network. Finally, gathering \eqref{eq:vectorial expert sparse} and \eqref{eq:vectorial softmax gating sparse} as for \eqref{eq:FME density}, the iFME model density is now given by
\begin{equation}
f(y_i|u_i(\cdot);\bsvPsi) = \sum_{k=1}^{K} 
\pi_{k}(\bs_i;\bw) 
\phi(y_{i};\beta_{k,0} + {\bsgamma^{[d_1]}_{k}}^{\top}\hspace{-2pt}\bv_i,\sigma^2_k),
\label{eq:iFME density}
\end{equation}
where $\bsvPsi = (\bw^\top, \bspsi_1^\top, \ldots, \bspsi_K^\top)^\top$ is the parameter vector of the model to be estimated.
Thus, the iFME  model constructed upon the parameter vectors $\bsgamma_k$'s and $\bsomega_k$'s, for which the sparsity is assumed to obtain interpretable estimates.

\subsection{Regularized MLE via the EM-iFME algorithm}\label{ssec: iEM-FME}

In order to fit the iFME  model and to maintain the sparsity in $\bsgamma_k$ and $\bsomega_k$, the following EM-iFME algorithm is then developed to maximize the penalized log-likelihood function 
\begin{equation}
\cL(\bsvPsi) = \sum_{i=1}^{n}\log f(y_i|u_i(\cdot);\bsvPsi)
+
\text{Pen}_{\lambda,\chi}(\bsvPsi)
 \label{eq:loglik iFME}
\end{equation}
with the conditional iFME density $f(y_i|u_i(\cdot);\bsvPsi)$ is defined in \eqref{eq:iFME density} and
the new sparse and interpretable regularization term is given by 
\begin{equation}\label{eq:pen term FLiRTI}
\text{Pen}_{\lambda,\chi}(\bsvPsi) = \lambda \sum_{k=1}^{K} \left(\Vert \bsgamma_k^{[d_1]} \Vert_1 + \rho\Vert \bsgamma_k^{[d_2]} \Vert_1\right) + \chi \sum_{k=1}^{K-1} \left(\Vert \bsomega_k^{[d_1]} \Vert_1 + \varrho\Vert \bsomega_k^{[d_2]} \Vert_1\right),
\end{equation}
where $\rho$ and $\varrho$ are, respectively, the weights associated to the $d_2$th derivative of the expert and the gating parameter function. The appearance of the weighting parameters  $\rho$ and $\varrho$, besides the usual regularization parameters $\lambda$ and $\chi$, is motivated by the fact that  one may wish to place a greater emphasis on sparsity in the $d_2$th derivative than in the $d_1$th derivative of the parameter functions, or vice versa \textcolor{black}{(we will see the usage of them in the subsequent section of experimental study)}. In practice, the selection of $\rho$ and $\varrho$ is more about whether they are greater than or less than one (i.e., the emphasis on $d_2$th) rather than select an exact value.

Note that, firstly, unlike the previous FME-Lasso, in iFME model the regularization operates on the functional derivative $\bsgamma_k$'s rather than  the functional coefficients $\bseta_k$'s for the experts,
and on the functional derivatives  $\bsomega_k$'s rather than the functional coefficients $\bszeta_k$'s for the gating network. Secondly, maximizing the penalized log-likelihood function \eqref{eq:loglik iFME} with penalization in \eqref{eq:pen term FLiRTI} in iFME model must be coupled with the constrains \eqref{eq: FLiRTI constrains}. Follows are the two steps of the proposed EM-iFME algorithm.

\paragraph{E-Step.}The E-Step for the new iFME model calculates for each observation the conditional probability memberships of each expert $k$
\begin{equation}
\tau_{ik}^{(s)}= \Pro(Z_i=k|y_i,u_i(\cdot);\bsvPsi^{(s)}) = \frac{
\pi_{k}(\bs_i;\bw^{(s)})\, 
\phi(y_{i};\beta^{(s)}_{k,0} + \bv^{\top}_{i}{\bsgamma^{[d_1]}_k}^{(s)},{\sigma^2_k}^{(s)})
}{f(y_{i}|u_i(\cdot);\bsvPsi^{(s)})},
\label{eq:iFME post prob}
\end{equation}
where $f(y_{i}|u_i(\cdot);\bsvPsi^{(s)})$ is now calculated according to the iFME density given by \eqref{eq:iFME density}. 
 
\paragraph{M-Step.} 
The maximization is performed by separate maximizations w.r.t. the gating network parameters $\bw$ and the experts network parameters $\bspsi_k$'s.

\paragraph{\it Updating the gating network parameters.}
The maximization step for updating the gating network parameters 
$\bw=\left(\alpha_{1,0},{\bsomega^{[d_1]}_1}^\top,\ldots,\alpha_{K-1,0},{\bsomega^{[d_1]}_{K-1}}^{\hspace{-4pt}\top}\right)^{\hspace{-2pt}\top}$  consists of maximizing the function $\mathcal{Q}_{\chi}(\bw;\bsvPsi^{(s)})$ given by
\begin{eqnarray}
\label{eq:Q-function gating-network flirti}
\mathcal{Q}_{\chi}(\bw;\bsvPsi^{(s)}) =  Q(\bw;\bsvPsi^{(s)}) - \chi \sum_{k=1}^{K-1} \left(\Vert \bsomega_k^{[d_1]} \Vert_1 + \varrho\Vert \bsomega_k^{[d_2]} \Vert_1\right),
\end{eqnarray}
subject to
\begin{equation}\label{eq:gating-network constrain}
\bsomega_{k}^{[d_2]} = \bA_q^{[d_2]}{\bA_q^{[d_1]}}^{-1} \bsomega_{k}^{[d_1]},  \quad \forall k\in[K-1],
\end{equation}
where
{\small $Q(\bw;\bsvPsi^{(s)}) =   \sum_{i=1}^n\left[\sum_{k=1}^{K-1} \tau^{(s)}_{ik} \left(\alpha_{k,0} + {\bsomega_{k}^{[d_1]}}^\top \bs_i\right)  - \log\left(1 +   \sum_{k^\prime =1}^{K-1}  \exp\{\alpha_{k^\prime,0} + {\bsomega_{k^\prime}^{[d_1]}}^\top\bs_i\} \right) \right]\cdot$} This is a constrained version of the weighted regularized multinomial logistic regression problem, where the weights are the conditional probabilities $\tau^{(s)}_{ik}$. 

To solve it, in the same spirit as when updating the gating network in the previous EM-Lasso algorithm, we use an outer loop that cycles over the gating function parameters to update them one by one. 
However, to update a single gating function parameter $\bw_k=(\alpha_{k,0}, {\bsomega_{k}^{[d_1]}}^\top)^\top$, $k\in[K-1]$, since the maximization problem \eqref{eq:Q-function gating-network flirti} is now coupled with an additional constraint \eqref{eq:gating-network constrain}, rather than using a coordinate ascent algorithm as in EM-Lasso, we use the following alternative approach.
 For each single gating network $k$, using a Taylor series expansion, we partially approximate the smooth part of $\mathcal{Q}_{\chi}(\bw;\bsvPsi^{(s)})$ defined in \eqref{eq:Q-function gating-network flirti} w.r.t. $\bw_k$ at the current estimate $\bw^{(t)}$, then maximize the resulting objective function (w.r.t. $\bw_k$), subject to the corresponding constraint (w.r.t. $k$) in \eqref{eq:gating-network constrain}.
It corresponds to solving the following penalized weighted least squares problem with constraints,
\begin{equation}\label{gating penalized weighted LS problem}
\begin{aligned}
\underset{(\alpha_{k,0},\ \bsomega_k^{[d_1]}, \bsomega_{k}^{[d_2]})}{\max}
-&\frac{1}{2} \sum_{i=1}^n w_{ik} \left(c_{ik} - \alpha_{k,0} - \bs_i^{\top}\bsomega^{[d_1]}_k\right)^2
 - \chi \left(\Vert \bsomega_k^{[d_1]} \Vert_1 + \varrho\Vert \bsomega_k^{[d_2]} \Vert_1\right)\\
\text{subject to }\quad&\bsomega_{k}^{[d_2]} = \bA_q^{[d_2]}{\bA_q^{[d_1]}}^{-1} \bsomega_{k}^{[d_1]},
\end{aligned}
\end{equation}
where $w_{ik} = \pi_k(\bw^{(t)};\bs_i)\left(1-\pi_k(\bw^{(t)};\bs_i)\right)$ are the weights and
$c_{ik} = \alpha_{k,0}^{(t)} + \bs_i^{\top}{\bsomega^{[d_1]}_k}^{(t)} + \frac{\tau_{ik}^{(s)} - \pi_k(\bw^{(t)};\bs_i)}{w_{ik}}$ are the working responses computed given the current estimate $\bw^{(t)}$. 
This problem can be solved by quadratic programming (see, e.g., \cite{GainesKimZhou}) or by using the Dantzig selector \citep{candes2007dantzig}, which we opt to use in our experimental studies. 
The details of using Dantzig selector to solve problem \eqref{gating penalized weighted LS problem} are given in Appendix \ref{EM-iFME gating update}. 

Therefore, if $(\widetilde\alpha_{k,0}, \widetilde\bsomega_k^{[d_1]}, \widetilde\bsomega_{k}^{[d_2]})$ is an optimal solution to problem \eqref{gating penalized weighted LS problem}, then $\widetilde\bw_k = {(\widetilde\alpha_{k,0}, {{\widetilde\bsomega_k}^{[d_1]}}}^\top)^\top$ is taken as an update for the gating parameter vector $\bw_k$. We keep cycling over $k\in[K-1]$ until there is no significant increase in the regularized $Q-$function \eqref{eq:Q-function gating-network flirti}.

\noindent{\it Updating the experts network parameters.} The maximization step for updating the expert parameter vector $\bspsi_k=(\beta_{k,0}, {\bsgamma_k^{[d_1]}}^\top, \sigma_k^2)^\top$ consists  of solving the following  problem:
\begin{equation}\label{expert penalized weighted LS problem}
\begin{aligned}
\underset{(\beta_{k,0},\ \bsgamma_k^{[d_1]}, \bsgamma_k^{[d_2]}, \sigma_k^2)}{\max}\ \ 
&\sum_{i=1}^{n} \tau_{ik}^{(s)} \log \phi\left(y_{i};\beta_{k,0} + \bv_i^\top \bsgamma_k^{[d_1]},\sigma^2_k\right)
 - \lambda \left(\Vert \bsgamma_k^{[d_1]} \Vert_1 + \rho\Vert \bsgamma_k^{[d_2]} \Vert_1\right)\\
\text{subject to}\quad\ \  &\bsgamma_k^{[d_2]} = \bA_p^{[d_2]}{\bA_p^{[d_1]}}^{-1} \bsgamma_k^{[d_1]}.
\end{aligned}
\end{equation}
As in the previous EM-Lasso algorithm, we first fix $\sigma^2_k$ to its previous estimate and perform the update for $(\beta_{k,0}, \bsgamma_{k}^{[d_1]})$, which corresponds to solving a penalized weighted least squares problem with constraints. This is be performed by using the Dantzig selector, in the same manner as previously for solving problem \eqref{gating penalized weighted LS problem}. The corresponding technical details can be found in Appendix \ref{EM-iFME experts update}.

Once the $(\beta_{k,0}, \bsgamma_{k}^{[d_1]})$ are updated,  the straightforward update for the variance $\sigma^2_k$ is given by the standard estimate of a weighted Gaussian regression. More specifically, let $(\widetilde\beta_{k,0}, \widetilde\bsgamma_k^{[d_1]}, \widetilde\bsgamma_k^{[d_2]})$ be the solution to the problem \eqref{expert penalized weighted LS problem} (with $\sigma^2_k$ fixed to ${\sigma^2_k}^{(s)}$), the updates for expert parameter vector $\bspsi_k$ are given by
\begin{eqnarray*}
(\beta_{k,0}^{(s+1)}, {\bsgamma_k^{[d_1]}}^{(s+1)}) &=& (\widetilde\beta_{k,0}, \widetilde\bsgamma_k^{[d_1]}),\\
{\sigma^2_k}^{(s+1)} &=& \frac{\sum_{i=1}^n \tau_{ik}^{(s)} \left(y_i - \beta_{k,0}^{(s+1)} - \bv_i^{\top}{\bsgamma_k^{[d_1]}}^{(s+1)}\right)^2}{\sum_{i=1}^n \tau_{ik}^{(s)}}\cdot
\end{eqnarray*}

Thus, at the end of the M-Step, we obtain a parameter vector update  $\bsvPsi^{(s+1)} = (\bw^{(s+1)},\allowbreak \bspsi_1^{(s+1)}, \ldots, \bspsi_K^{(s+1)})$ for the next EM iteration, where $\bw^{(s+1)}$ and $\bspsi_k^{(s+1)}$, $k\in[K]$, are, respectively, the updates of the gating network parameters and the experts network parameters,  calculated by the two procedures described above. Alternating the E-Step and M-Step until convergence, i.e., when there is no longer a significant change in the values of the penalized log-likelihood (\ref{eq:loglik iFME}), leads to a penalized maximum likelihood estimate $\widehat{\bsvPsi}$ for $\bsvPsi$.

\noindent {\it Estimating the coefficient functions:} Finally, once the parameter vector of  iFME model has been estimated, the coefficient functions of the gating network $\alpha_k(t)$, $k\in[K-1]$  and the ones of the experts network $\beta_k(t)$,  $k\in[K]$, can be reconstructed by the following formulas
\begin{eqnarray}
\begin{aligned}
\widehat{\alpha}_{k}(t) &= \bsb_q(t)^\top \bA_q^{-1}\widehat{\bsomega}^{[d_1]}_{k},\\
\widehat{\beta}_{k}(t) &= \bsb_p(t)^\top \bA_p^{-1}\widehat{\bsgamma}^{[d_1]}_{k},
\end{aligned} \label{eq: iFME network reconstruction}
\end{eqnarray}
where $\widehat{\bsomega}^{[d_1]}_{k}$ and $\widehat{\bsgamma}^{[d_1]}_{k}$ are respectively the regularized maximum likelihood estimates for $\bsomega_{k}^{[d_1]}$ and $\bsgamma_{k}^{[d_1]}$.

\subsection{Non-linear regression and clustering with FME models}
Once the model parameters have been estimated, \textcolor{black}{one can then construct an estimate of the unknown functional non-linear regression function, as well as a clustering of the data into groups of similar pairs of observations.}

For the aim of functional non-linear regression, the unknown non-linear regression function with functional predictors is given by the following conditional expectation $\widehat y|u(\cdot) = \E[Y|U(\cdot);\widehat \bsvPsi]$, 
which is defined by 
$\widehat y_i|u_i(\cdot) =\sum_{k=1}^{K} 
\pi_{k}(\br_i;\widehat \bsxi)(\widehat \beta_{k,0} + \widehat \bseta_{k}^{\top} \bx_{i})$   
 for the FME model \eqref{eq:FME density},  
and by
$\widehat y_i|u_i(\cdot) =\sum_{k=1}^{K} 
\pi_{k}(\bs_i;\widehat \bw) 
 (\widehat \beta_{k,0} + {\bsgamma^{[d_1]}_{k}}^{\top}\bv_i)$  
for the iFME model \eqref{eq:iFME density}.

For clustering, a soft partition of the data into $K$ clusters, represented by the estimated posterior probabilities $\widehat \tau_{ik} = \Pro(Z_i=k|u_i,y_i;\widehat\bsvPsi)$, is obtained. A hard partition can also be computed according to the Bayes' optimal allocation rule. That is, by assigning each curve to the component having the 
highest estimated  posterior probability $\tau_{ik}$, defined by (\ref{eq:FME post prob}) for FME or by \eqref{eq:iFME post prob} for the iFME model, using the MLE $\widehat{\bsvPsi}$ of $\bsvPsi$:
\begin{equation}
\widehat{z}_i = \arg \max_{1\leq k \leq K} \tau_{ik}(\widehat{\bsvPsi}), \quad i\in[n],
\label{eq:MAP rule FME}\nonumber
\end{equation}where $\widehat{z}_i$ denotes the estimated cluster label for the $i$th curve.

\bigskip
 
\subsection{Tuning parameters and model selection}
In practice, appropriate values of the tuning parameters should be chosen. In using FME, this cover the selection of $K$, the number of experts, and $r$, $p$, and $q$, the dimensions of B-spline bases used to approximate, respectively, the predictors, the experts, and the gating network functions, although they can be chosen to be equal. 
For the FME-Lasso, additionally the $\ell_1$ penalty constants $\chi$ and $\lambda$ in \eqref{eq:Pen term} should be chosen. 
For the iFME model, 
the tuning parameters include also $d_1$, $d_2$, the two derivatives related to the sparsity constraints, and $\rho$, $\varrho$, the weights associated to the $d_2$th derivative of the expert and gating functions (see \eqref{eq:pen term FLiRTI}).

The selection of the tuning parameters can be performed by a cross--validation  procedure with a grid search scheme to select the best combination. 
An alternative is to use the well-known BIC criterion \citep{SchwarzBIC} or, in our paper, its extension, called modified BIC \citep{stadler2010} defined as
\begin{eqnarray}\label{BIC criterion}
\text{mBIC} = L(\widehat{\bsvPsi}) - \text{df}(\widehat{\bsvPsi})\frac{\log n}{2},
\end{eqnarray}
where $\widehat{\bsvPsi}$ is the obtained log-likelihood estimator (for the FME model) or penalized log-likelihood estimator (for the FME-Lasso and iFME models), and the number of degrees of freedom $\text{df}(\widehat{\bsvPsi})$ is the effective number of parameters of the model, given by
\begin{align*}
\text{df}(\widehat{\bsvPsi}) = 
\begin{cases}
    \text{df}(\bszeta) + (K-1) + \text{df}(\bseta) + K + K & \quad\text{for the FME and FME-Lasso models,}\\
    \text{df}(\bsomega) + (K-1) + \text{df}(\bsgamma) + K + K & \quad\text{for the iFME model},
\end{cases}
\end{align*}
in which the quantities $\text{df}(\bszeta)$, $\text{df}(\bseta)$, $\text{df}(\bsomega)$ and $\text{df}(\bsgamma)$ are, respectively, the counts for non-zero free coefficients in the vectors $\bszeta$, $\bseta$, $\bsomega$, and $\bsgamma$. Note that, because of the constraints \eqref{eq: FLiRTI constrains} for the iFME model, free coefficients in $\bsomega$ and $\bsgamma$ consist of only  the part related to the $d_1$ derivative. 
That is,
$\text{df}(\bszeta) = \sum_{k=1}^{K-1}\sum_{j=1}^{q}\bone_{\{\zeta_{kj}\neq 0\}}$, 
$\text{df}(\bseta) = \sum_{k=1}^{K}\sum_{j=1}^{p}\bone_{\{\eta_{kj}\neq 0\}}$, 
$\text{df}(\bsomega) = \sum_{k=1}^{K-1}\sum_{j=1}^{q}\allowbreak\bone_{\{\omega^{[d_1]}_{kj}\neq 0\}}$ and $\text{df}(\bsgamma) = \sum_{k=1}^{K}\sum_{j=1}^{p}\bone_{\{\gamma^{[d_1]}_{kj}\neq 0\}}$.
We apply both the BIC and the modified BIC in our experimental study.

\section{Experimental study}
\label{sec:Experiments}
We study the performances of the FME, FME-Lasso, and iFME models in regression and clustering problems by considering simulated scenarios and real data with functional predictors and scalar responses. 
The interests of this study consist of the prediction performance as well as the estimation of the functional parameters, i.e., the expert and gating functions in the ME model, along with the clustering partition of the considered heterogeneous data.

\subsection{Evaluation criteria}
We will use the following criteria, for where applicable, to assess and compare the performances of the models and the related algorithms. 
For regression evaluation, we use the \textit{relative prediction error} (RPE) and the \textit{correlation} (Corr) index 
\textcolor{black}{as standard criteria to evaluate the prediction performance, i.e.,
to quantify the relationship between the true and the predicted values of the scalar outputs}. 
The RPE is defined by $\text{RPE}=\sum_{i=1}^n (y_i-\widehat y_i)^2/\sum_{i=1}^ny_i^2$, where $y_i$ and $\widehat y_i$ are, respectively, the true and the predicted response of the $i$th observation in the testing set.
For clustering evaluation, we use the \textit{Rand index} (RI), the \textit{adjusted Rand index} (ARI), and the \textit{clustering error} (ClusErr), 
\textcolor{black}{as standard criteria to evaluate the clustering performance, i.e.,}
to quantify how similar the testing observations are presented in the true partition compared to the predicted partition. 
To evaluate the parameters estimation performance, we compute the \textit{mean squared error} (MSE) between the true and the estimated functional parameters.
The MSE between a true function $g(\cdot)$ and its estimate $\widehat g(\cdot)$ is defined by
\begin{eqnarray}
\text{MSE}\big(\widehat{g}(\cdot)\big) = \frac{1}{m} \sum_{j=1}^m \big(g(t_j) - \widehat{g}(t_j)\big)^2,
\label{eq:functional estimation error}
\end{eqnarray}with $m$ being the number of time points taken into account.
The function $g(\cdot)$ here corresponds to an expert function $\beta(\cdot)$, or a gating function $\alpha(\cdot)$. The parameter functions are reconstructed from the model parameters using  (\ref{eq: experts projection}) and (\ref{eq: gates projection}) for both FME and FME-Lasso models, and using  \eqref{eq: iFME network reconstruction}  for iFME model. 

The values of these criteria are averaged over $N$ Monte Carlo runs ($N=100$ for simulation, for the real data, see the corresponding section).
Note that, the average over $N$ sample replicates of $\text{MSE}\,\big(\widehat{g}(\cdot)\big)$ is equivalent to the usual Mean Integrated Squared Error (MISE) defined by  $\text{MISE}\,\big(\widehat{g}(\cdot)\big) = \E\left[ \int_{\cT} \big(\widehat g(t) - g(t)\big)^2dt \right]$, where the integral here is calculated numerically by a Riemann sum over the grid $t_1,\ldots,t_m$. 

\subsection{Simulation studies}
\subsubsection{Simulation parameters and experimental protocol}
\textcolor{black}{
We generate data from a $K$-component functional mixture of Gaussian experts model that relates  a scalar response $y\in\R$ to a univariate functional predictor $X(t), t\in \cT$ defined on a domain  $\cT\subset\R$.
The data generation protocol is detailed in Appendix \ref{sec: Data generating protocols}.
The parameters that were used in this data generating process \eqref{Data generating protocols} are described and provided in the
simulation parameters and experimental protocol section in Appendix \ref{subsection Similation 1}. 
We study different simulation scenarios  (sample sizes $n$, 
number of observations per time-series input $m$, 
noise levels $\sigma^2_\delta$,..) summarised in Section \ref{subsection Similation 1} and Table. 
\ref{Table: scenarios}. 
}

\begin{figure}[htp!]
    \centering
    \includegraphics[width=.8\linewidth]{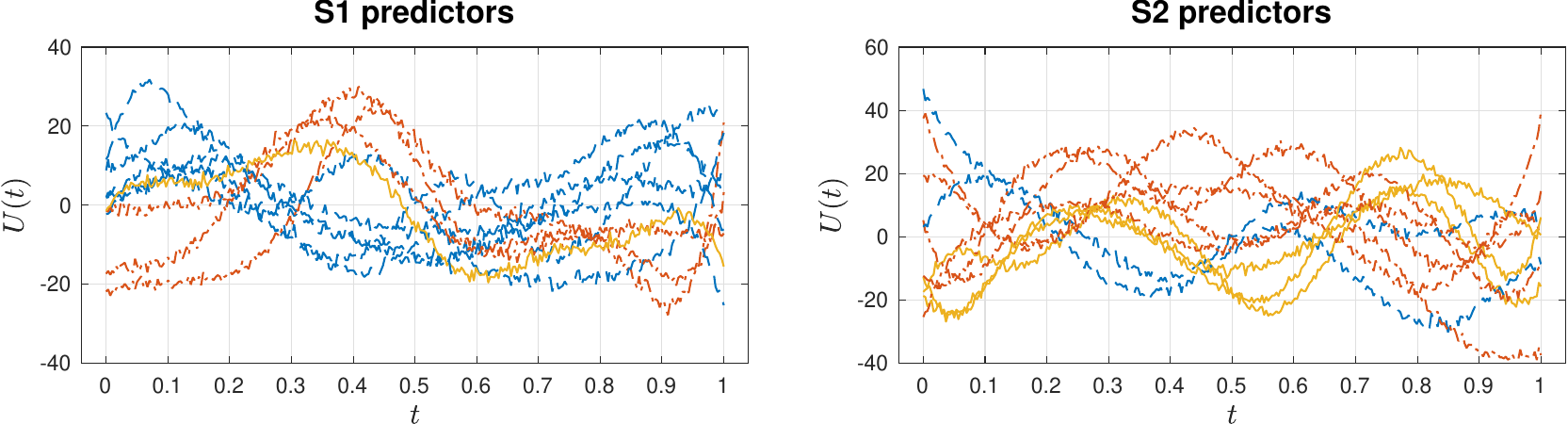}\\
    \vspace{0.3cm}
    \includegraphics[width=.8\linewidth]{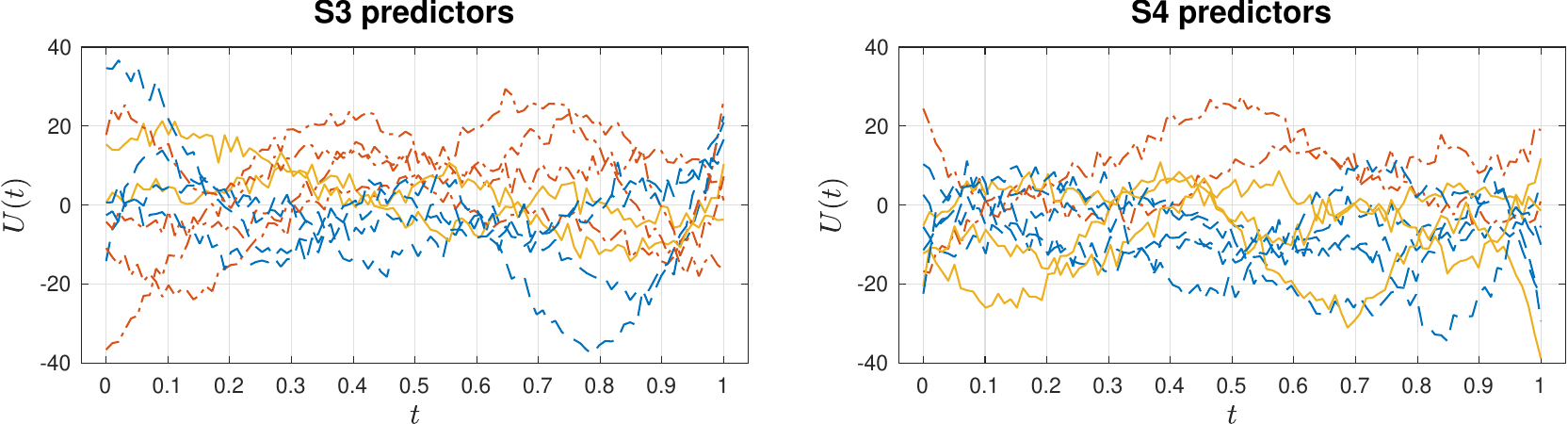}
    \caption{$10$ randomly taken predictors in scenarios $S1$ (large $m$ and small $\sigma_{\delta}$), $S2$ (small $m$ and small $\sigma_{\delta}$), $S3$ (large $m$ and large $\sigma_{\delta}$), and $S4$ (small $m$ and large $\sigma_{\delta}$). Here, the 
    noisy predictors $U_i(\cdot)$ are colored (blue, red, yellow) according to their true cluster labels $Z_i\in\{1,2,3\}$.}\label{Figure: random predictors S1S4}
    \end{figure}

Figure \ref{Figure: random predictors S1S4} displays, for each scenario, 10 randomly taken predictors colored according to their  corresponding clusters. 
For each run, the concerned dataset is randomly split into a training set and testing set of equal size, the model parameters are estimated using training set, with the tuning parameters selected by maximizing the modified BIC \eqref{BIC criterion}. The evaluation criteria are computed on testing set and reported for each model accordingly.
\color{black}    
    
\subsubsection{Some implementation details}
For all scenarios, for all datasets, we implemented the three proposed models with $10$ EM runs and with a tolerance of $10^{-6}$. For the iFME model, in principle, for each parameter function the two derivatives $d_1$ and $d_2$ to be penalized, and the weights for the latter (i.e., $\rho$ and $\varrho$) can be seen as tuning parameters. However, such an implementation could be computationally expensive in this simulation study with $400$ datasets in total and $10$ EM runs for each dataset. Therefore, we opted to fix $d_1$ and $d_2$ for all implementations ($d_1=0$ and $d_2=3$ for both expert and gating networks.) 
 and left $\rho$ and $\varrho$ to be selected in some sets of targeted values. The choices of the targeted values were made by the following straightforward arguments. 
Since $\beta_1(\cdot)$ and $\beta_2(\cdot)$ have zero-valued regions, the weight for their zeroth derivative in penalization term should be large, equivalently, the weight for the third derivative should be small, so $\rho$ is selected in a set of small values: $\rho\in\{10^{-2}, 10^{-3}, 10^{-4}\}$. 
On the other hand, for $\alpha_1(t)$ and $\alpha_2(t)$, we select $\varrho\in\{10, 10^2, 10^3\}$ as we should emphasize sparsity in their third derivative. 
%
%

\subsubsection{Simulation results}
\paragraph{Clustering and prediction performances.}
We report in Table \ref{Table: prediction performance S1S4} the results of regression and clustering tasks on simulated datasets in the four considered scenarios. 
\begin{table}[htp!]
\centering
\def\arraystretch{.8}
\begin{tabular}{r r *{4}{>{\arraybackslash}p{2.2cm}}}
\specialrule{1pt}{1pt}{1pt}
\multicolumn{2}{l}{}    & RPE & Corr & ARI & ClusErr \\
\hline
\multicolumn{2}{l}{\footnotesize $S1$ ($m=300,\sigma^2_{\delta}=1$)}    &  &  &  & \\
&FME        & $.1552_{(.1282)}$ & $.9188_{(.0602)}$ & $.7852_{(.0934)}$  & $.0899_{(.0434)}$ \\ 
&FME-Lasso  & $.1390_{(.1074)}$ & $.9224_{(.0603)}$ & $.7670_{(.1436)}$ & $.1112_{(.0891)}$  \\ 
&iFME       & $\Bs{.1334_{(.0977)}}$ & $\Bs{.9287_{(.0504)}}$ & $\Bs{.7997_{(.0796)}}$ & $\Bs{.0792_{(.0340)}}$   \\ 
\hline
\multicolumn{2}{l}{\footnotesize $S2$ ($m=100,\sigma^2_{\delta}=1$)}    &  &  &  & \\
&FME        & $.1600_{(.1698)}$ & $.9164_{(.0712)}$ & $\Bs{.7966_{(.0881)}}$  & $.0852_{(.0427)}$ \\ 
&FME-Lasso  & $.1656_{(.1316)}$ & $.9071_{(.0733)}$ & $.7455_{(.1520)}$ & $.1236_{(.0961)}$  \\
&iFME       & $\Bs{.1540_{(.0950)}}$ & $\Bs{.9192_{(.0476)}}$ & $.7955_{(.0822)}$ & $\Bs{.0820_{(.0341)}}$   \\ 
\hline
\multicolumn{2}{l}{\footnotesize $S3$ ($m=300,\sigma^2_{\delta}=4$)}    &  &  &  & \\
&FME        & $.1724_{(.1807)}$ & $.9110_{(.0778)}$ & $.7798_{(.0838)}$  & $.0918_{(.0415)}$ \\ 
&FME-Lasso  & $.1457_{(.1504)}$ & $.9197_{(.0785)}$ & $.7629_{(.1524)}$ & $.1115_{(.0928)}$  \\ 
&iFME       & $\Bs{.1432_{(.0860)}}$ & $\Bs{.9228_{(.0448)}}$ & $\Bs{.7987_{(.0899)}}$ & $\Bs{.0783_{(.0361)}}$   \\ 
\hline
\multicolumn{2}{l}{\footnotesize $S4$ ($m=100,\sigma^2_{\delta}=4$)}    &  &  &  & \\
&FME        & $.2251_{(.4462)}$ & $.9048_{(.0822)}$ & $\Bs{.7816_{(.1040)}}$  & $.0927_{(.0497)}$ \\ 
&FME-Lasso  & $\Bs{.1432_{(.1229)}}$ & $\Bs{.9188_{(.0683)}}$ & $.7526_{(.1654)}$ & $.1215_{(.1055)}$  \\ 
&iFME       & $.1639_{(.0978)}$ & $.9125_{(.0521)}$ & $.7798_{(.0848)}$ & $\Bs{.0864_{(.0325)}}$   \\ 
\specialrule{1pt}{2pt}{1pt}
\end{tabular}
\caption{\textcolor{black}{Evaluation criteria of FME, FME-Lasso and iFME models for test data in scenarios $S1,\ldots, S4$. The reported values are the averages of 100 Monte Carlo runs with standard errors in parentheses. The bold values correspond to the best solution.}}\label{Table: prediction performance S1S4}
\end{table}The mean and standard error of the relative predictions error (RPE) and correlation (Corr) summarize the regression performance, while the mean and standard error of the Rand index (RI), adjusted Rand index (ARI) and clustering error (ClusErr) summarize the clustering performance.

As we can see from Table \ref{Table: prediction performance S1S4}, all the models have very good performances on both regression and clustering tasks, with high correlation, RI, ARI, and small RPE and clustering error. The iFME appears to slightly have a better performance than the others in all scenarios. 
The low standard errors confirm the stability of the algorithms.

\begin{figure}[htp]
\centering
\subfloat{\includegraphics[width=.24\linewidth]{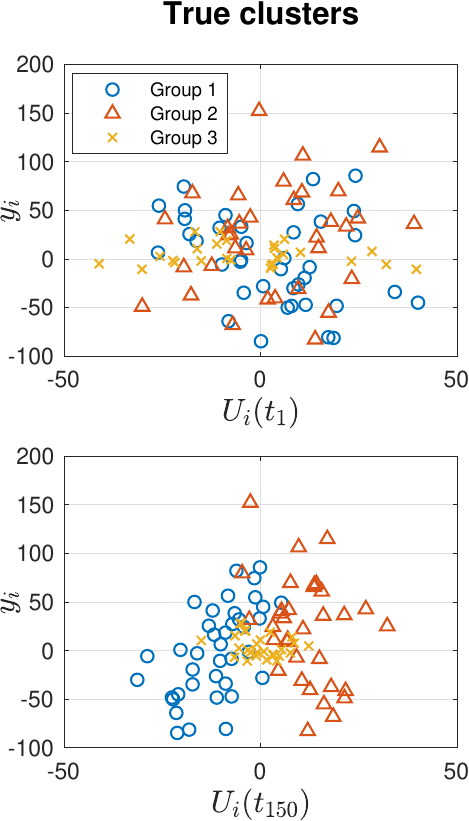}}
\subfloat{\includegraphics[width=.24\linewidth]{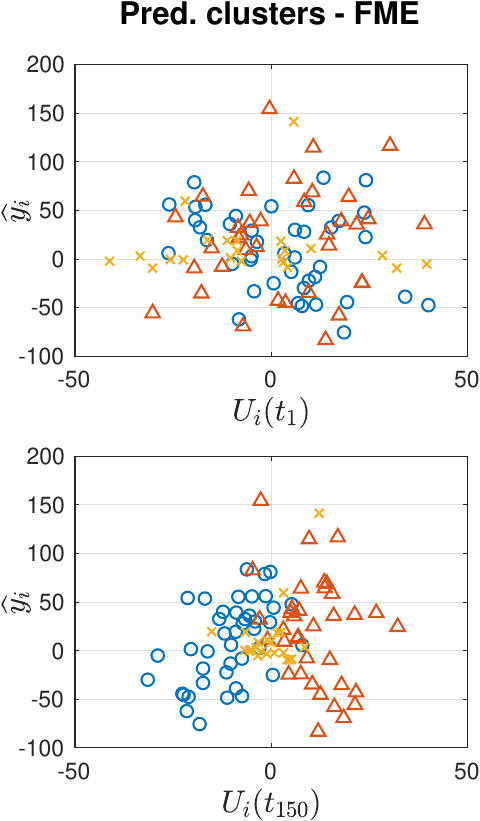}}
\subfloat{\includegraphics[width=.24\linewidth]{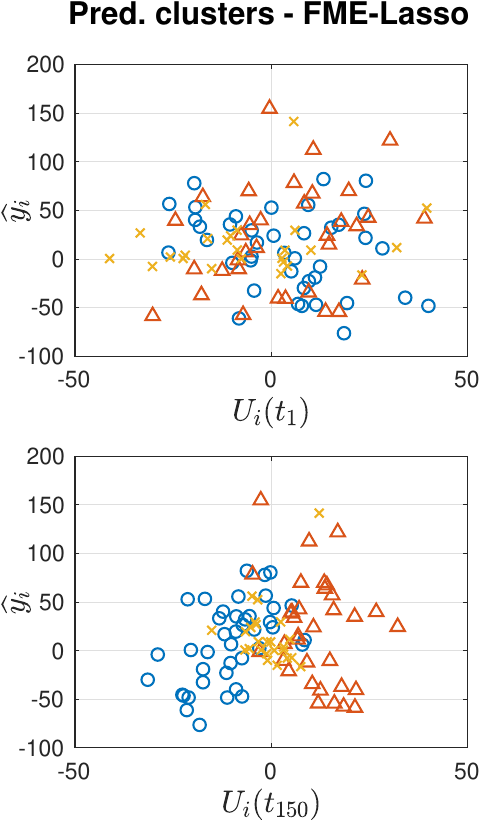}}
\subfloat{\includegraphics[width=.24\linewidth]{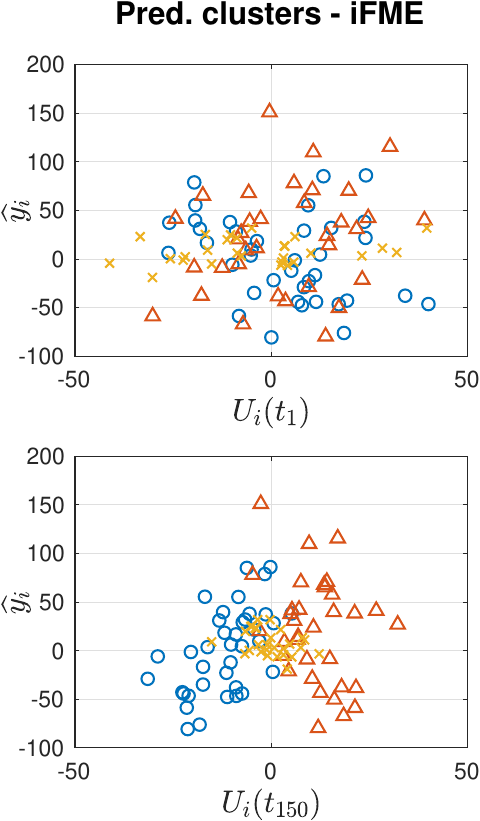}}
\caption{Scatter plots of $\widehat y_i$ against $U_i(t_1)$ (top panels) and $U_i(t_{50})$ (bottom panels) on a randomly selected dataset. Here, the clustering errors are $5.5 \%$, $4.75 \%$ and $5.0 \%$ for FME, FME-Lasso and iFME models, respectively.}\label{Figure: Cluster S1}
\end{figure}

Figure \ref{Figure: Cluster S1} shows the clustering results obtained by the models with highest  values of the modified BIC criterion, on a dataset selected in scenario $S1$. Here, we plotted the responses against the predictors at two specific time points: $t_1=0$ and $t_{50}=0.5$. The highly accurate predictions (in both regression and clustering) can be seen visually through Figure \ref{Figure: Cluster S1}. This figure also shows that it is difficult to cluster these data according to a few number of time observations, for example in $\R^2$, according to $\{(U_i(t_0),y_i)\}_{i=1}^n$ or $\{(U_i(t_{50}),y_i)\}_{i=1}^n$, which suggests using functional approaches.  

{\it Comparison with functional regression mixtures (FMR):} 
Finally, we compare our proposed models with the functional mixture regression (FMR) model proposed in \cite{FangYaoFMR2010}. In their approach, the functional predictors are first projected onto its eigenspace, then the obtained new coordinates are fed to the standard mixture regression model to estimate the weights and the coefficients of the $\widehat{\beta}(\cdot)$ in that eigenspace. They performed functional principal component analysis (FPCA) to obtain estimates for the eigenfunctions and the principal component scores (which serve as predictors). The number of relevant FPCA components are chosen automatically for each dataset by selecting the minimum number of components that explain $90\%$ of the total variation of the predictors. It is noticed that, in their simulation studies, the authors computed the scalar responses by using conditional prediction, i.e., the true $y_i$ were used to determine which cluster the observation belongs to. Then the predicted $\widehat y_i$ is calculated as the conditional mean of the density of the corresponding cluster. 
For comparison with that approach, we also used this strategy to make predictions in our models. 
We further  employed the FMR model with the B-spline functional representation, instead of the functional PCA, the number of B-spline functions is set to be the same as the number of basis functions used in our models. 
Table \ref{Table: S1 models comparison} shows the evaluation criteria corresponding to the considered models evaluated on $100$ datasets in scenario $S3$. Here, FMR-PC is the original model of \cite{FangYaoFMR2010} and FMR-B is the modified one with B-spline bases. 
As expected, FME, FME-Lasso, and iFME, 
which are more flexible compared to the the FMR model, allows to capture more complexity in the data, thanks to the predictor-depending mixture weights, and provides clearly better results than the FMR alternatives. 
\begin{table}[htp!]
\centering
\def\arraystretch{.9}
\begin{tabular}{r L{2.2cm} L{2.2cm} L{2.2cm} L{2.2cm}}
\specialrule{1pt}{30pt}{1pt}
& RPE & Corr & ARI & ClusErr\\
\hline
       FME & $.0614_{(.0896)}$ & $.9681_{(.0469)}$ & $.8523_{(.0934)}$ & $.0618_{(.0490)}$ \\ 
 FME-Lasso & $.0145_{(.0254)}$ & $.9924_{(.0133)}$ & $.9582_{(.0641)}$ & $.0174_{(.0325)}$ \\ 
      iFME & $\Bs{.0139_{(.0086)}}$ & $\Bs{.9928_{(.0046)}}$ & $\Bs{.9594_{(.0430)}}$ & $\Bs{.0151_{(.0166)}}$ \\
    FMR-B & $.0460_{(.0399)}$ & $.9768_{(.0201)}$ & $.7064_{(.1044)}$ & $.1124_{(.0434)}$ \\ 
   FMR-PC & $.0191_{(.0331)}$ & $.9902_{(.0164)}$ & $.8345_{(.0794)}$ & $.0612_{(.0361)}$ \\ 
\specialrule{1pt}{1pt}{1pt}
\end{tabular}
\caption{\textcolor{black}{Performance comparison of the models for datasets in scenarios $S3$. The reported values are the averages of 100 Monte Carlo samples with standard errors in parentheses.}}\label{Table: S1 models comparison}
\end{table}

\paragraph{Parameter estimation performance.}
To evaluate the parameter estimation performance, we consider the functional parameter functions estimation error
as defined by \eqref{eq:functional estimation error}.
This error between the true function and the estimated one, provides an evaluation of how well the proposed models reconstruct the hidden functional gating and expert networks.
\textcolor{black}{
In this evaluation, we considered scenario $S1$, i.e., $m=300$, $\sigma^2_{\delta}=1$. Moreover, in order to provide an idea of the impact of training size on parameter estimation, we run the models with different training sizes (share the same scenario $S1$) and report
}
the MSE for each function, for each model in Table \ref{Table: S1 MSE}. It shows that there are significant improvements, \textcolor{black}{even with small training size}, when using iFME model in estimating the gating network, compared to the FME and FME-Lasso model.

\begin{table}[htp!]
\centering
\def\arraystretch{.9}
\begin{tabular}{r r R{2.2cm} R{2.2cm} R{2.2cm} R{2.2cm} R{2.2cm}} 
\specialrule{1pt}{1pt}{1pt}
\multicolumn{2}{l}{} & $\widehat{\beta}_1(\cdot)$ & $\widehat{\beta}_2(\cdot)$ & $\widehat{\beta}_3(\cdot)$ & $\widehat{\alpha}_1(\cdot)$ & $\widehat{\alpha}_2(\cdot)$ \\
\hline
\multicolumn{2}{l}{\footnotesize Training size: $100$}    &  &  &  &  &  \\
&        FME & $11.68_{(56.63)}$ & $6.21_{(24.34)}$ & $8.57_{(27.60)}$ 
& $\nb[2]{7.23\mathrm{e}\splus04}{5.15\mathrm{e}\splus04}$ 
& $\nb[2]{5.90\mathrm{e}\splus06}{5.69\mathrm{e}\splus04}$  \\ 
&  FME-Lasso & $\Bs{0.53_{(0.19)}}$ & $\Bs{0.66_{(0.55)}}$ & $\Bs{0.62_{(0.79)}}$ & $17.06_{(28.68)}$ & $19.50_{(37.07)}$  \\ 
&       iFME & $1.09_{(0.78)}$ & $1.05_{(0.82)}$ & $2.44_{(3.06)}$ & $\Bs{5.76_{(7.78)}}$ & $\Bs{4.25_{(3.58)}}$  \\ 
\hline
\multicolumn{2}{l}{\footnotesize Training size: $200$}    &  &  &  &  &  \\
&        FME & $0.77_{(0.48)}$ & $0.61_{(0.30)}$ & $\Bs{0.19_{(0.20)}}$ & $8.55_{(11.99)}$ & $9.57_{(8.29)}$  \\ 
&  FME-Lasso & $\Bs{0.54_{(0.18)}}$ & $0.54_{(0.20)}$ & $0.31_{(0.81)}$ & $10.71_{(16.63)}$ & $12.64_{(16.33)}$  \\ 
&       iFME & $0.55_{(0.50)}$ & $\Bs{0.48_{(0.38)}}$ & $1.26_{(2.91)}$ & $\Bs{5.34_{(6.21)}}$ & $\Bs{2.81_{(2.82)}}$  \\  
\hline
\multicolumn{2}{l}{\footnotesize Training size: $300$}    &  &  &  &  &  \\
&        FME & $0.62_{(0.56)}$ & $0.59_{(0.50)}$ & $\Bs{0.16_{(0.18)}}$ & $12.58_{(19.83)}$ & $16.49_{(30.97)}$  \\ 
&  FME-Lasso & $0.60_{(0.52)}$ & $\Bs{0.57_{(0.44)}}$ & $0.17_{(0.19)}$ & $11.26_{(16.56)}$ & $13.15_{(24.80)}$  \\ 
&       iFME & $\Bs{0.56_{(0.43)}}$ & $0.59_{(0.50)}$ & $0.49_{(0.47)}$ & $\Bs{2.98_{(3.12)}}$ & $\Bs{3.16_{(3.02)}}$  \\ 
\hline
\multicolumn{2}{l}{\footnotesize Training size: $400$}    &  &  &  &  &  \\
&        FME & $0.54_{(0.23)}$ & $0.53_{(0.19)}$ & $\Bs{0.11_{(0.12)}}$ & $8.27_{(11.73)}$ & $7.08_{(8.53)}$  \\ 
&  FME-Lasso & $0.52_{(0.18)}$ & $0.51_{(0.16)}$ & $0.15_{(0.18)}$ & $6.09_{(8.98)}$ & $5.14_{(6.51)}$  \\ 
&       iFME & $\Bs{0.34_{(0.19)}}$ & $\Bs{0.38_{(0.25)}}$ & $0.74_{(0.62)}$ & $\Bs{3.06_{(4.21)}}$ & $\Bs{2.77_{(3.21)}}$  \\  
\hline
\multicolumn{2}{l}{\footnotesize Training size: $500$}    &  &  &  &  &  \\
&        FME & $0.49_{(0.22)}$ & $0.50_{(0.28)}$ & $\Bs{0.09_{(0.11)}}$ & $5.17_{(7.52)}$ & $3.90_{(5.69)}$  \\ 
&  FME-Lasso & $0.49_{(0.18)}$ & $0.49_{(0.21)}$ & $\Bs{0.09_{(0.11)}}$ & $4.62_{(7.28)}$ & $7.63_{(16.27)}$  \\ 
&       iFME & $\Bs{0.35_{(0.23)}}$ & $\Bs{0.40_{(0.25)}}$ & $0.55_{(0.63)}$ & $\Bs{2.11_{(2.82)}}$ & $\Bs{2.68_{(4.70)}}$  \\ 
\specialrule{1pt}{1pt}{1pt}
\end{tabular}
\caption{\textcolor{black}{Average of 100 Monte Carlo runs of MSE between the estimated functions resulted by FME, FME-Lasso and iFME models in $S1$ scenario.}}\label{Table: S1 MSE}
\end{table}


\color{black}
Now, in order to evaluate the designed sparsity of zeroth and third derivatives of the reconstructed functions, we compute the MSEs versus their true values, i.e., zeros, on each designed intervals. In particular, we divide the domain $\cT=[0,1]$ into three parts: $\cT_1=[0,0.3)$, $\cT_2=[0.3,0.7)$, and $\cT_3=[0.7,1]$, then for each model the MSEs on the corresponding intervals are reported in Table \ref{Table: S1 MSE sparsity}.
\color{black}


\begin{table}[htp!]
\centering
\def\arraystretch{.9}
\begin{tabular}{r R{1.3cm} R{1.4cm} R{1.3cm} R{1.3cm} R{1.8cm} R{1.8cm} R{1.8cm}}
\toprule
& \multicolumn{2}{c}{on $\cT_2$} & \multicolumn{2}{c}{on $\cT_1\cup \cT_3$} & \multicolumn{3}{c}{on $\cT$}\\
\cmidrule(lr){2-3}\cmidrule(lr){4-5}\cmidrule(lr){6-8}
& $\widehat\beta_1^{(0)}(\cdot)$  & $\widehat\beta_2^{(0)}(\cdot)$ 
& $\widehat\beta_1^{(3)}(\cdot)$ & $\widehat\beta_2^{(3)}(\cdot)$ 
& $\widehat\beta_3^{(3)}(\cdot)$ & $\widehat\alpha_1^{(3)}(\cdot)$  & $\widehat\alpha_2^{(3)}(\cdot)$ \\
\hline
\multicolumn{2}{l}{\footnotesize Training size: $100$}    &  &  &  &  & & \\
FME & $\nB{0.15}{0.26}$ & $\nb{0.91}{0.48}$ & $\nb{0.07}{0.15}$ & $\nB{0.08}{0.12}$ & $\nb[2]{1.37\mathrm{e}{\sminus}07}{3.96\mathrm{e}{\sminus}08}$ & $\nb[2]{3.32\mathrm{e}{\sminus}05}{9.24\mathrm{e}{\sminus}06}$ & $\nb[2]{2.62\mathrm{e}{\sminus}05}{1.08\mathrm{e}{\sminus}05}$  \\ 
 FME-Lasso & $\nb{0.19}{0.27}$ & $\nB{0.16}{0.26}$ & $\nB{0.07}{0.14}$ & $\nb{0.09}{0.15}$ & $\nB[3]{7.64\mathrm{e}{\sminus}09}{1.64\mathrm{e}{\sminus}09}$ & $\nb[2]{2.64\mathrm{e}{\sminus}08}{7.37\mathrm{e}{\sminus}09}$ & $\nb[2]{1.70\mathrm{e}{\sminus}08}{5.12\mathrm{e}{\sminus}09}$  \\ 
      iFME & $\nb{0.26}{0.40}$ & $\nb{0.30}{0.37}$ & $\nb{0.10}{0.20}$ & $\nb{0.11}{0.18}$ & $\nb[2]{2.44\mathrm{e}{\sminus}08}{6.48\mathrm{e}{\sminus}09}$ & $\nB[3]{1.33\mathrm{e}{\sminus}08}{4.05\mathrm{e}{\sminus}09}$ & $\nB[3]{4.51\mathrm{e}{\sminus}09}{2.46\mathrm{e}{\sminus}09}$  \\ 
\hline
\multicolumn{2}{l}{\footnotesize Training size: $200$}    &  &  &  &  & & \\
       FME & $\nb{0.56}{0.43}$ & $\nb{0.09}{0.22}$ & $\nb{0.10}{0.16}$ & $\nb{0.05}{0.14}$ & $\nB[3]{2.69\mathrm{e}{\sminus}11}{2.23\mathrm{e}{\sminus}11}$ & $\nB[3]{2.33\mathrm{e}{\sminus}10}{1.64\mathrm{e}{\sminus}10}$ & $\nb[2]{8.90\mathrm{e}{\sminus}10}{3.83\mathrm{e}{\sminus}10}$  \\ 
 FME-Lasso & $\nb{0.11}{0.21}$ & $\nb{0.09}{0.20}$ & $\nb{0.05}{0.15}$ & $\nb{0.05}{0.15}$ & $\nb[2]{6.04\mathrm{e}{\sminus}11}{3.62\mathrm{e}{\sminus}11}$ & $\nb[2]{3.38\mathrm{e}{\sminus}10}{2.36\mathrm{e}{\sminus}10}$ & $\nB[3]{6.41\mathrm{e}{\sminus}10}{2.96\mathrm{e}{\sminus}10}$  \\ 
      iFME & $\nB{0.08}{0.18}$ & $\nB{0.09}{0.17}$ & $\nB{0.05}{0.12}$ & $\nB{0.05}{0.12}$ & $\nb[2]{2.00\mathrm{e}{\sminus}09}{9.95\mathrm{e}{\sminus}10}$ & $\nb[2]{2.72\mathrm{e}{\sminus}09}{1.44\mathrm{e}{\sminus}09}$ & $\nb[2]{1.22\mathrm{e}{\sminus}09}{8.52\mathrm{e}{\sminus}10}$  \\ 
\hline
\multicolumn{2}{l}{\footnotesize Training size: $300$}    &  &  &  &  & & \\
       FME & $\nb{0.30}{0.25}$ & $\nb{0.25}{0.22}$ & $\nb{0.10}{0.16}$ & $\nb{0.09}{0.15}$ & $\nB[3]{2.30\mathrm{e}{\sminus}11}{1.72\mathrm{e}{\sminus}11}$ & $\nb[2]{4.61\mathrm{e}{\sminus}10}{2.55\mathrm{e}{\sminus}10}$ & $\nb[2]{1.42\mathrm{e}{\sminus}10}{1.01\mathrm{e}{\sminus}10}$  \\ 
 FME-Lasso & $\nb{0.28}{0.25}$ & $\nB{0.23}{0.21}$ & $\nb{0.09}{0.16}$ & $\nB{0.08}{0.15}$ & $\nb[2]{2.52\mathrm{e}{\sminus}11}{1.73\mathrm{e}{\sminus}11}$ & $\nB[3]{2.00\mathrm{e}{\sminus}10}{1.41\mathrm{e}{\sminus}10}$ & $\nB[3]{3.77\mathrm{e}{\sminus}10}{1.64\mathrm{e}{\sminus}10}$  \\ 
      iFME & $\nB{0.19}{0.25}$ & $\nb{0.19}{0.24}$ & $\nB{0.08}{0.16}$ & $\nB{0.08}{0.15}$ & $\nb[2]{1.26\mathrm{e}{\sminus}09}{5.94\mathrm{e}{\sminus}10}$ & $\nb[2]{7.26\mathrm{e}{\sminus}10}{5.41\mathrm{e}{\sminus}10}$ & $\nb[2]{1.05\mathrm{e}{\sminus}09}{7.15\mathrm{e}{\sminus}10}$  \\ 
\hline
\multicolumn{2}{l}{\footnotesize Training size: $400$}    &  &  &  &  & & \\
       FME & $\nb{0.12}{0.23}$ & $\nb{0.10}{0.23}$ & $\nb{0.05}{0.16}$ & $\nb{0.06}{0.16}$ & $\nB[3]{3.53\mathrm{e}{\sminus}11}{1.62\mathrm{e}{\sminus}11}$ & $\nb[2]{3.59\mathrm{e}{\sminus}10}{1.29\mathrm{e}{\sminus}10}$ & $\nB[3]{1.02\mathrm{e}{\sminus}10}{8.14\mathrm{e}{\sminus}11}$  \\ 
 FME-Lasso & $\nb{0.11}{0.22}$ & $\nb{0.10}{0.21}$ & $\nb{0.05}{0.16}$ & $\nb{0.05}{0.16}$ & $\nb[2]{3.55\mathrm{e}{\sminus}11}{1.90\mathrm{e}{\sminus}11}$ & $\nB[3]{1.63\mathrm{e}{\sminus}10}{9.71\mathrm{e}{\sminus}11}$ & $\nb[2]{1.54\mathrm{e}{\sminus}10}{1.03\mathrm{e}{\sminus}10}$  \\ 
      iFME & $\nB{0.08}{0.14}$ & $\nB{0.09}{0.16}$ & $\nB{0.03}{0.08}$ & $\nB{0.04}{0.08}$ & $\nb[2]{1.42\mathrm{e}{\sminus}08}{1.87\mathrm{e}{\sminus}08}$ & $\nb[2]{1.77\mathrm{e}{\sminus}08}{1.13\mathrm{e}{\sminus}08}$ & $\nb[2]{1.05\mathrm{e}{\sminus}08}{9.91\mathrm{e}{\sminus}09}$  \\ 
\hline
\multicolumn{2}{l}{\footnotesize Training size: $500$}    &  &  &  &  & & \\
       FME & $\nb{0.11}{0.22}$ & $\nb{0.10}{0.21}$ & $\nb{0.05}{0.17}$ & $\nb{0.05}{0.17}$ & $\nb[2]{4.66\mathrm{e}{\sminus}11}{2.12\mathrm{e}{\sminus}11}$ & $\nb[2]{2.70\mathrm{e}{\sminus}10}{8.98\mathrm{e}{\sminus}11}$ & $\nb[2]{2.83\mathrm{e}{\sminus}10}{1.03\mathrm{e}{\sminus}10}$  \\ 
 FME-Lasso & $\nb{0.09}{0.21}$ & $\nb{0.09}{0.20}$ & $\nb{0.05}{0.17}$ & $\nb{0.05}{0.17}$ & $\nB[3]{3.64\mathrm{e}{\sminus}11}{1.90\mathrm{e}{\sminus}11}$ & $\nB[3]{1.44\mathrm{e}{\sminus}10}{7.02\mathrm{e}{\sminus}11}$ & $\nB[3]{1.69\mathrm{e}{\sminus}10}{8.91\mathrm{e}{\sminus}11}$  \\ 
      iFME & $\nB{0.07}{0.14}$ & $\nB{0.08}{0.16}$ & $\nB{0.04}{0.08}$ & $\nB{0.04}{0.08}$ & $\nb[2]{1.34\mathrm{e}{\sminus}08}{1.55\mathrm{e}{\sminus}08}$ & $\nb[2]{1.58\mathrm{e}{\sminus}08}{1.03\mathrm{e}{\sminus}08}$ & $\nb[2]{1.20\mathrm{e}{\sminus}08}{8.34\mathrm{e}{\sminus}09}$  \\ 

\specialrule{1pt}{1pt}{1pt}
\end{tabular}
\caption{\textcolor{black}{MSE of the derivatives of reconstructed functions on the corresponding interested intervals, in which, $\cT_1=[0,0.3)$, $\cT_2=[0.3,0.7)$, $\cT_3=[0.7,1]$ and $\cT=[0,1]$. The reported values are the averaged of 100 Monte Carlo samples with standard errors in parentheses.}}\label{Table: S1 MSE sparsity}
\end{table}

The reported values show that, as expected, iFME model  is better than both FME and FME-Lasso in providing sparse solutions with respect to the derivatives of the parameter functions.

%

Table \ref{Table: intercepts variances} shows the means and the standard errors of the estimated intercepts and variances. Note that these coefficients are not considered in the penalization. For the intercepts $\beta_{k,0}$, all the models estimated them very well, while for the intercepts $\alpha_{k,0}$, iFME is slightly better than the others. For the variances, FME-Lasso gave the estimated values closest with the true values on average.
\begin{table}[htp!]
\centering
\def\arraystretch{.9}
\begin{tabular}{r R{1.4cm} R{1.3cm} R{1.3cm} R{1.5cm} R{1.5cm} R{1.2cm} R{1.2cm} R{1.2cm}}
\specialrule{1pt}{30pt}{1pt}
    & $\widehat{\beta}_{1,0}$ & $\widehat{\beta}_{2,0}$ & $\widehat{\beta}_{3,0}$ & $\widehat{\alpha}_{1,0}$ & $\widehat{\alpha}_{2,0}$ & $\widehat{\sigma}_1$ & $\widehat{\sigma}_2$ & $\widehat{\sigma}_3$ \\
\hline
      True value & $-5$ & $0$ & $5$ & $-10$ & $-10$ & $5$ & $5$ & $5$ \\
\hline
\multicolumn{2}{l}{\footnotesize Training size: $100$}    &  &  &  &  & & & \\
       FME & $\nb[3]{18.91}{-16.90}$ & $\nb[3]{3.25}{-3.18}$ & $\nb{8.87}{4.17}$ & $\nb[4]{30.74}{-31.37}$ & $\nb[4]{41.27}{-25.20}$ & $\nb{40.05}{32.38}$ & $\nb{25.15}{15.19}$ & $\nb{24.71}{21.47}$ \\ 
 FME-Lasso & $\nb[3]{9.37}{-6.48}$ & $\nB[4]{1.95}{-0.07}$ & $\nb{4.63}{5.51}$ & $\nb[5]{8.99}{-12.53}$ & $\nb[4]{10.99}{-12.69}$ & $\nb{10.98}{6.86}$ & $\nb{8.08}{6.73}$ & $\nb{20.60}{10.01}$ \\ 
      iFME & $\nB[4]{1.14}{-4.95}$ & $\nb{1.08}{0.21}$ & $\nB{0.90}{5.05}$ & $\nB[4]{3.85}{-9.44}$ & $\nB[4]{3.31}{-9.39}$ & $\nB{1.46}{5.70}$ & $\nB{1.49}{5.75}$ & $\nB{1.51}{4.70}$ \\ 
\hline
\multicolumn{2}{l}{\footnotesize Training size: $200$}    &  &  &  &  & & & \\
       FME & $\nb[3]{0.78}{-4.83}$ & $\nb{0.54}{0.14}$ & $\nb{0.38}{4.89}$ & $\nb[4]{26.36}{-23.77}$ & $\nb[5]{6.65}{-16.10}$ & $\nb{1.26}{5.77}$ & $\nB{0.95}{5.63}$ & $\nb{0.79}{5.24}$ \\ 
 FME-Lasso & $\nB[4]{0.66}{-4.97}$ & $\nB{0.60}{0.07}$ & $\nb{0.45}{4.99}$ & $\nb[5]{4.45}{-12.70}$ & $\nb[5]{4.38}{-13.72}$ & $\nb{1.23}{5.94}$ & $\nb{1.12}{5.99}$ & $\nB{1.30}{4.92}$ \\ 
      iFME & $\nb[3]{0.72}{-4.85}$ & $\nb{0.68}{0.19}$ & $\nB{0.42}{4.99}$ & $\nB[4]{3.15}{-9.30}$ & $\nB[4]{2.83}{-9.21}$ & $\nB{0.90}{5.77}$ & $\nb{1.02}{5.72}$ & $\nb{1.27}{5.28}$ \\ 
\hline
\multicolumn{2}{l}{\footnotesize Training size: $300$}    &  &  &  &  & & & \\
       FME & $\nB[4]{0.55}{-4.99}$ & $\nb[3]{0.54}{-0.06}$ & $\nb{0.26}{4.98}$ & $\nb[4]{13.09}{-16.82}$ & $\nb[4]{14.89}{-17.18}$ & $\nb{1.14}{5.92}$ & $\nb{0.94}{5.87}$ & $\nB{0.84}{5.34}$ \\ 
 FME-Lasso & $\nb[3]{0.54}{-5.03}$ & $\nB[4]{0.53}{-0.03}$ & $\nb{0.27}{4.98}$ & $\nb[5]{4.54}{-13.06}$ & $\nb[5]{4.96}{-12.96}$ & $\nb{1.10}{6.08}$ & $\nb{1.07}{5.99}$ & $\nb{0.89}{5.37}$ \\ 
      iFME & $\nb[3]{0.61}{-4.93}$ & $\nb{0.56}{0.04}$ & $\nB{0.29}{5.00}$ & $\nB[4]{2.48}{-8.96}$ & $\nB[4]{2.57}{-8.85}$ & $\nB{0.75}{5.54}$ & $\nB{0.72}{5.53}$ & $\nb{0.87}{5.53}$ \\ 
\hline
\multicolumn{2}{l}{\footnotesize Training size: $400$}    &  &  &  &  & & & \\
       FME & $\nB[4]{0.43}{-5.02}$ & $\nB{0.40}{0.01}$ & $\nb{0.28}{5.04}$ & $\nb[5]{9.93}{-16.37}$ & $\nb[5]{7.83}{-15.39}$ & $\nb{1.01}{5.99}$ & $\nb{1.09}{6.20}$ & $\nb{0.77}{5.47}$ \\ 
 FME-Lasso & $\nb[3]{0.43}{-5.03}$ & $\nb[3]{0.38}{-0.02}$ & $\nb{0.27}{5.02}$ & $\nb[5]{4.61}{-13.18}$ & $\nb[5]{4.57}{-12.65}$ & $\nb{0.94}{5.99}$ & $\nb{1.01}{6.13}$ & $\nB{0.84}{5.45}$ \\ 
      iFME & $\nb[3]{0.45}{-4.89}$ & $\nb{0.35}{0.05}$ & $\nB{0.27}{5.01}$ & $\nB[5]{2.79}{-10.17}$ & $\nB[4]{2.61}{-9.82}$ & $\nB{0.60}{5.08}$ & $\nB{0.58}{5.22}$ & $\nb{0.85}{5.57}$ \\ 
\hline
\multicolumn{2}{l}{\footnotesize Training size: $500$}    &  &  &  &  & & & \\
       FME & $\nB[4]{0.41}{-5.00}$ & $\nb[3]{0.38}{-0.01}$ & $\nB{0.25}{5.01}$ & $\nb[5]{7.26}{-14.09}$ & $\nb[5]{9.64}{-15.42}$ & $\nb{0.98}{6.09}$ & $\nb{0.93}{6.06}$ & $\nb{0.69}{5.44}$ \\ 
 FME-Lasso & $\nb[3]{0.40}{-4.99}$ & $\nB[4]{0.37}{-0.01}$ & $\nb{0.26}{5.01}$ & $\nb[5]{3.27}{-11.35}$ & $\nb[5]{4.15}{-12.38}$ & $\nb{0.98}{6.13}$ & $\nb{0.90}{6.06}$ & $\nB{0.68}{5.44}$ \\ 
      iFME & $\nb[3]{0.44}{-4.92}$ & $\nb{0.34}{0.06}$ & $\nb{0.27}{5.02}$ & $\nB[4]{1.93}{-9.23}$ & $\nB[4]{2.68}{-9.61}$ & $\nB{0.55}{5.23}$ & $\nB{0.55}{5.13}$ & $\nb{0.75}{5.60}$ \\ 
\specialrule{1pt}{1pt}{1pt}
\end{tabular}
\caption{\textcolor{black}{Intercepts and variances obtained by FME, FME-Lasso and iFME models in scenario $S1$. The reported values are the averages of 100 Monte Carlo samples with standard errors in parentheses.}}\label{Table: intercepts variances}
\end{table}

\begin{table}[htp!]
\centering
\def\arraystretch{.9}
\begin{tabular}{rrrrrrrrrr}
\toprule
& \multicolumn{3}{c}{log likelihood} & \multicolumn{3}{c}{RPE} & \multicolumn{3}{c}{Time (s)}\\
\cmidrule(lr){2-4}\cmidrule(lr){5-7}\cmidrule(lr){8-10}
& rand.  & zeros & LR    & rand.  & zeros & LR   & rand.  & zeros & LR\\
\midrule
FME 
& $\nBB[2]{24.77}{-215.52}$ & $\nbb[2]{18.51}{-221.03}$  & $\nbb[2]{69.30}{-285.23}$  
& $\nB{0.16}{0.19}$  & $\nb{0.13}{0.20}$  & $\nb{0.29}{0.38}$  
& $\nb{0.50}{0.21}$  & $\nB{0.12}{0.16}$  & $\nb{1.99}{0.52}$  \\ 
FME-Lasso 
& $\nBB[2]{21.98}{-235.35}$ & $\nbb[2]{18.54}{-239.47}$  & $\nbb[2]{63.09}{-244.99}$  
& $\nb{0.11}{0.14}$  & $\nB{0.10}{0.13}$  & $\nb{0.16}{0.20}$  
& $\nb{1.23}{1.13}$  & $\nB{2.48}{1.08}$  & $\nb{15.51}{4.63}$  \\ 
iFME 
& $\nBB[4]{7.83}{-254.65}$ & $\nbb[2]{13.76}{-256.26}$  & $\nbb[2]{80.38}{-336.87}$  
& $\nB{0.09}{0.13}$  & $\nB{0.09}{0.13}$  & $\nb{0.30}{0.35}$  
& $\nb{4.45}{6.58}$  & $\nb{3.65}{4.17}$  & $\nB{2.08}{3.91}$  \\ 
\bottomrule
\end{tabular}
\caption{\textcolor{black}{Comparison of different initialization strategies. The reported values are the mean and standard error (in parentheses) over 100 Monte Carlo runs.}}\label{table: initialization strategies}
\end{table}
{\color{black}
The initialization is crucial for the EM algorithm. In all of our experimental studies, the model parameter vector $\bsPsi$ of the FME model was initialized as follows (similar for the FME-Lasso and iFME models). Firstly, we perform a $K$-means algorithm on the predictors $\{\bx_i\}_{i=1}^n$, then for each estimated group $k$, $(\beta_{k,0}^{(0)}, \bseta_k^{(0)})$ is initialized as the solution to the linear regression problem $y_i=\beta_{k,0}+\bx_i^\top\bseta_k$, for $i$ belongs to group $k$th. The parameter $\sigma_k^{(0)}$ is then initialized as the estimated variance within group $k$th. For the gating parameter, we simply drawn each component of $\bsxi^{(0)}$ randomly from the uniform distribution $\mathcal U(0,1)$.

However, to see the impact of different initialization strategies, we performed a side experiment with data taken from $S1$ scenario. In particular, the initialization for expert parameters is fixed as described before, while that for gating parameter varies in ``rand.'', ``zeros'', and ``LR''. Here, ``rand.'' is the random strategy described above,
``zeros'' is the strategy where all coefficients of $\bsxi$ are initialized as zeros, i.e., we are putting equal weights for the experts; and ``LR'' is the strategy where we perform a logistic regression with the predictors are $\br_i$ (for FME, FME-Lasso) or $\bs_i$ (for iFME), and the responses are the labels resulted by $K$-means on them. Table~\ref{table: initialization strategies} shows the time to convergence, and the log-likelihood, RPE on testing sets corresponding with the three strategies. We can see that, in terms of log-likelihood and RPE, simple strategies as ``rand.'' and ``zeros'' perform quite well, while in terms of time to convergence, LR is a good choice for the iFME model.
}

\begin{figure}[ht]
\setcounter{subfigure}{-3}
\centering
\subfloat{\includegraphics[width=.3\linewidth]{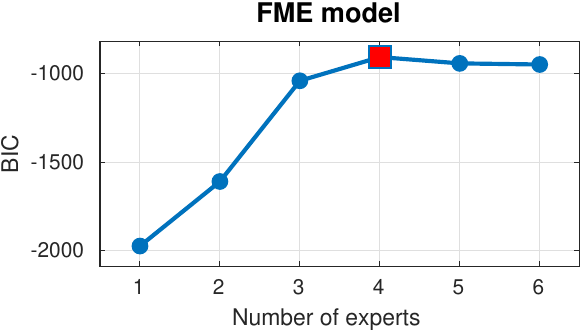}}\hfill
\subfloat{\includegraphics[width=.3\linewidth]{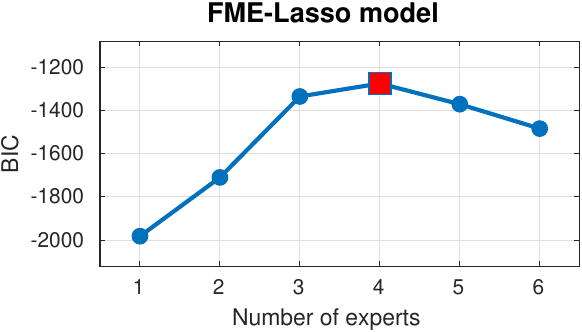}}\hfill
\subfloat{\includegraphics[width=.3\linewidth]{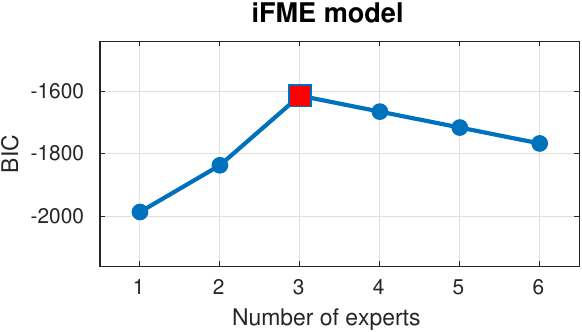}}\hfill\\
\subfloat[\label{fig:BIC S1 FME}]{\includegraphics[width=.3\linewidth]{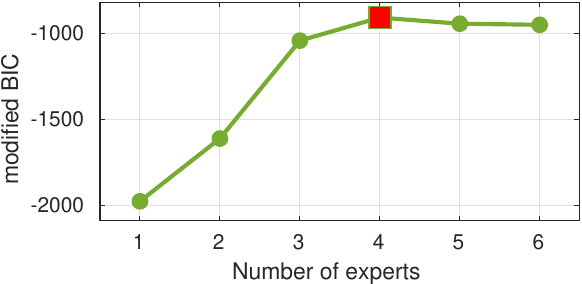}}\hfill
\subfloat[\label{fig:BIC S1 FME-Lasso}]{\includegraphics[width=.3\linewidth]{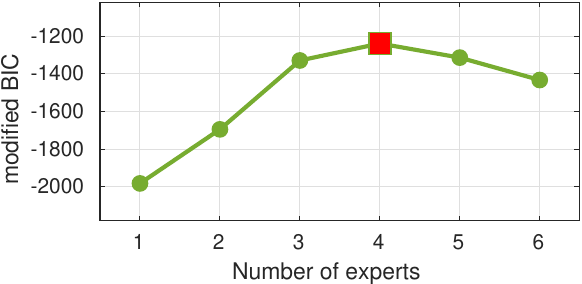}}\hfill
\subfloat[\label{fig:BIC S1 iFME}]{\includegraphics[width=.3\linewidth]{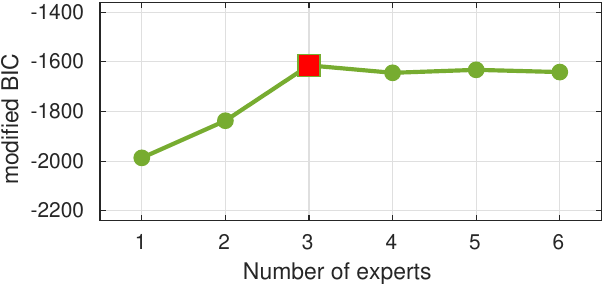}}\hfill
\caption{\textcolor{black}{Values of BIC (top) and modified BIC (bottom) for (a) FME, (b) FME-Lasso and (c) iFME model versus the number of experts $K$, fitted on a randomly taken dataset in scenario $S1$. Here, the square points correspond to the highest values.}}\label{fig: BIC curves of S1}
\end{figure}

Finally, to illustrate the selection of the number of expert components using BIC and/or modified BIC in this simulation study, we provide, in Figure \ref{fig: BIC curves of S1}, the plots of these criteria against the number of experts for each model. %
Here, we implemented the models with all fixed tuning parameters, except $K$ which varies in the set $\{1,\ldots,6\}$. We can observe that BIC selects the correct number $K=3$ for both FME-Lasso and iFME, while it selects $K=4$ for FME. 
However, the modified BIC selects $K=4$ for both FME and FME-Lasso, and selects the true number of components $K=3$ for iFME.

\subsection{Application to real data}
In this section, we apply the FME, FME-Lasso and iFME models to two well-known real datasets, Canadian weather and Diffusion Tensor Imaging (DTI). 
For each dataset, we perform clustering and
investigate the prediction performance, estimate the functional mixture of experts models with different number of experts $K$ and perform the selection of $K$ using modified BIC, and discuss  the obtained results. 

\subsubsection{Canadian weather data}
\begin{figure}[ht!]
\centering
\subfloat[\label{fig:CW}]{\includegraphics[width=.45\linewidth]{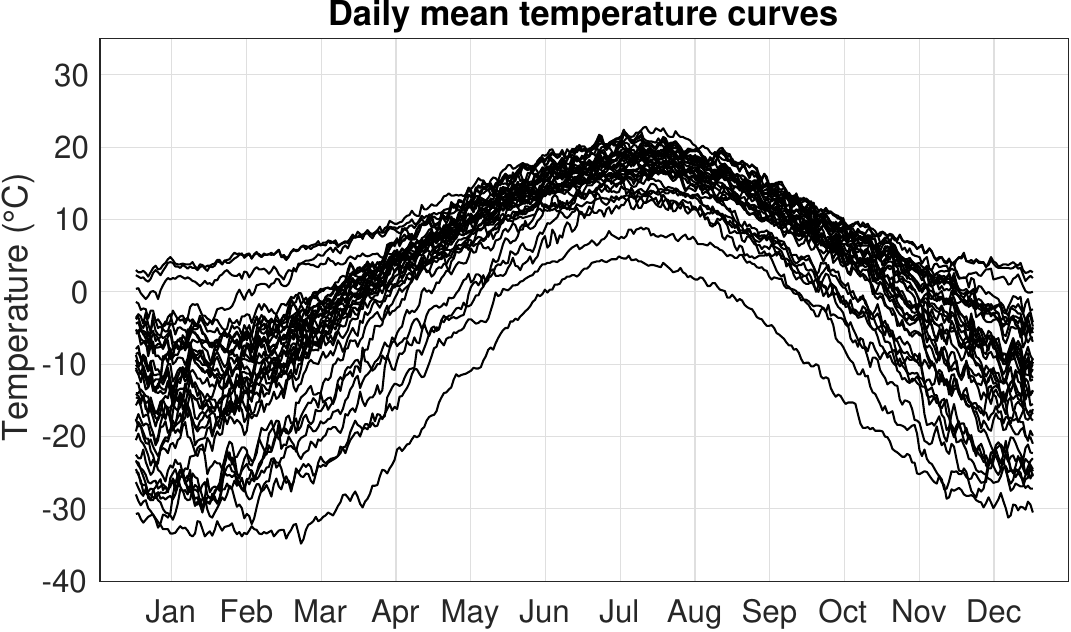}}\hspace*{.5cm}
\subfloat[\label{fig:CW station positions}]{\includegraphics[width=.48\linewidth]{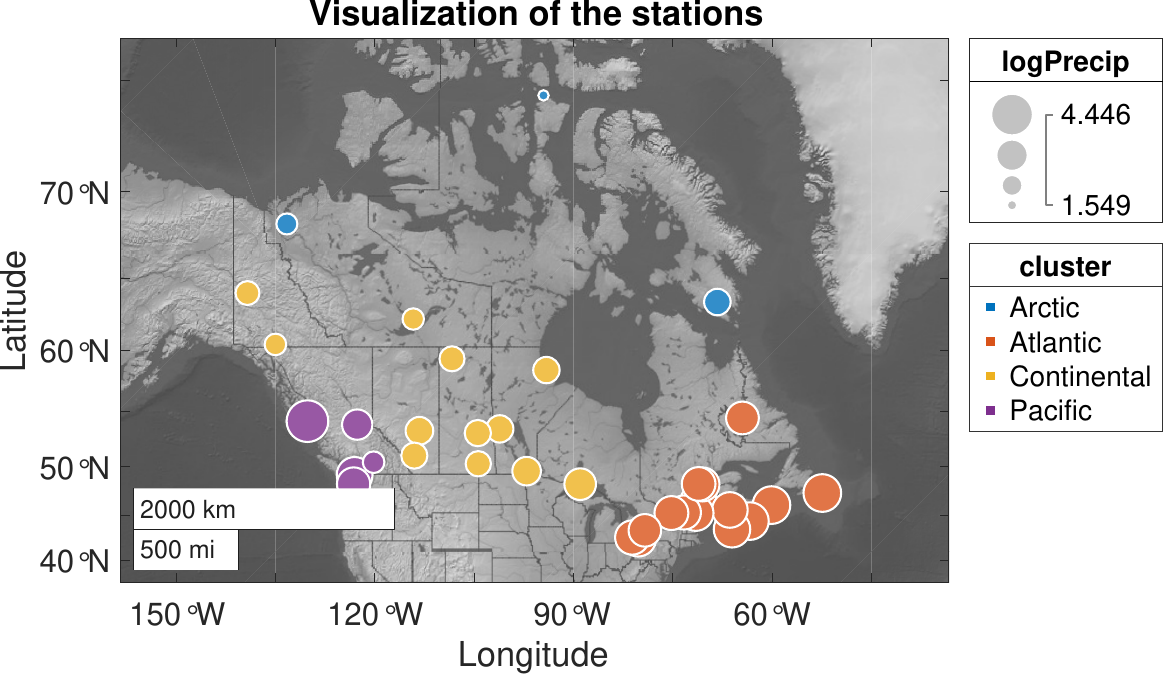}}\hfill
\caption{(a) $35$ daily mean temperature measurement curves; (b) Geographical visualization of the stations, in which the sizes of the bubbles corresponds to their $\log$ of precipitation values and the colors correspond to their climate regions.}
\end{figure}

Canadian weather data has been introduced in \cite{RamsayAndSilvermanFDA2005} and is also available in the R package fda.  This dataset consists of $m=365$ daily temperature measurements (averaged over the year 1961 to 1994) at $n=35$ weather stations in Canada, and their corresponding average annual precipitation (in $\log$ scale). The weather stations are located in \textcolor{black}{$4$ climate zones}: Atlantic, Pacific, Continental and Arctic (Figure \ref{fig:CW station positions}). 
In this dataset, presented in Figure \ref{fig:CW}, the noisy functional predictors $U_i(\cdot)$ are the curves of $365$ averaged daily temperature measurements, the scalar responses $Y_i$ are the corresponding total precipitation  at each station $i$, during the year, for 35 stations. Its station climate zone is taken as a cluster label (the cluster label $Z_i\in\{1,\ldots,4\}$.
The aim here is to use the daily temperature curves (functional predictors) to predict the precipitations (scalar responses) at each station. Moreover, in addition to predicting the precipitation values, we are interested in clustering the temperature curves (therefore the stations), as well as identifying the periods of time of the year that have effect on prediction for each group of curves. 

Firstly, in order to assess the prediction performance of the FME, FME-Lasso and iFME models on this dataset, we implement the models by selecting the tuning parameters, including the number of expert components $K$ in the set $\{1,2,3,4,5,6\}$, by maximizing the modified BIC criterion, given its performance as shown in the simulation study. We report in Table \ref{Table: Canadian corr sse} the results in terms of correlations, \textit{sum of squared errors} (SSE) and relative prediction errors. 
\begin{table}
\centering
\def\arraystretch{.9}
\begin{tabular}{r R{1.5cm} R{1.5cm} R{1.5cm}}
\specialrule{1pt}{1pt}{1pt}
 & Corr & SSE & RPE\\
\hline
FME & $0.690$& $14.410$ & $0.290$ \\
FME-Lasso & $0.632$ & $2.011$ & $0.045$ \\
iFME & $\Bs{0.944}$ & $\Bs{0.582}$ & $\Bs{0.012}$ \\
\specialrule{1pt}{1pt}{1pt}
\end{tabular}
\caption{7-fold cross-validated correlation (Corr), sum of squared errors (SSE) and relative prediction error (RPE) of predictions on Canadian weather data.}\label{Table: Canadian corr sse}
\end{table}According to the obtained results, iFME provides the best results w.r.t. all the criteria. The cross-validated RPE provided by iFME is only of $1.2\%$, the next is FME-Lasso with $4.5\%$, while the FME model has the worst RPE value. 
Note that in \cite{FLIRTI}, the authors applied their proposed model to Canadian weather data and obtained a 10-fold cross-validated SSE of $4.77$. 
Clearly, with the smaller cross-validated SSEs, the FME-Lasso and iFME models significantly improve the prediction. 

\begin{figure}[htbp]
\centering
\subfloat{\includegraphics[width=.32\linewidth]{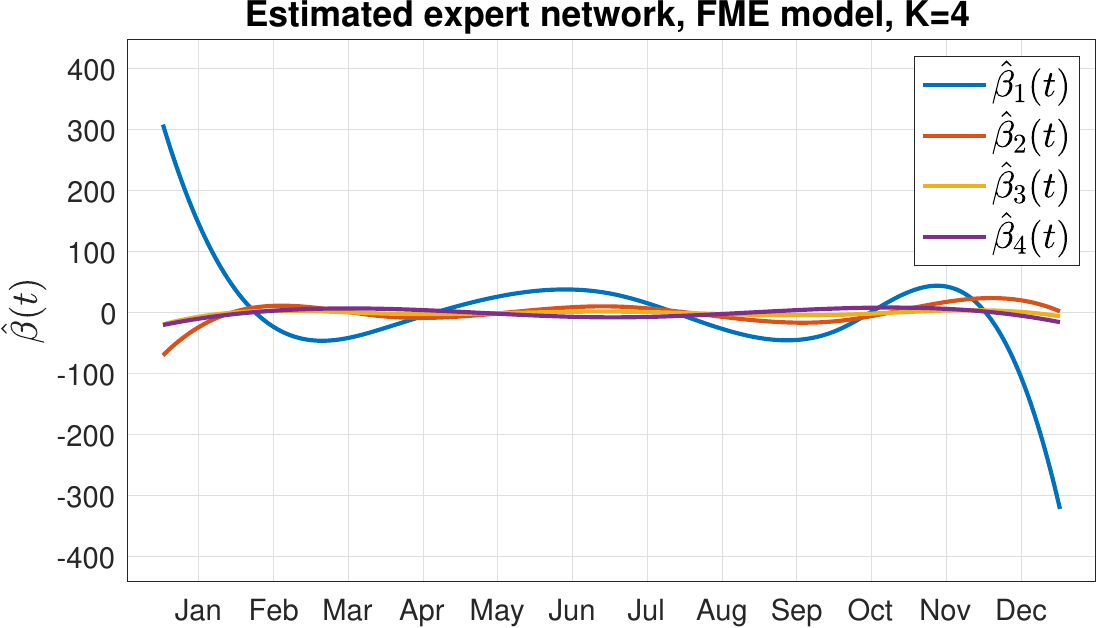}}\hspace*{.2cm}
\subfloat{\includegraphics[width=.32\linewidth]{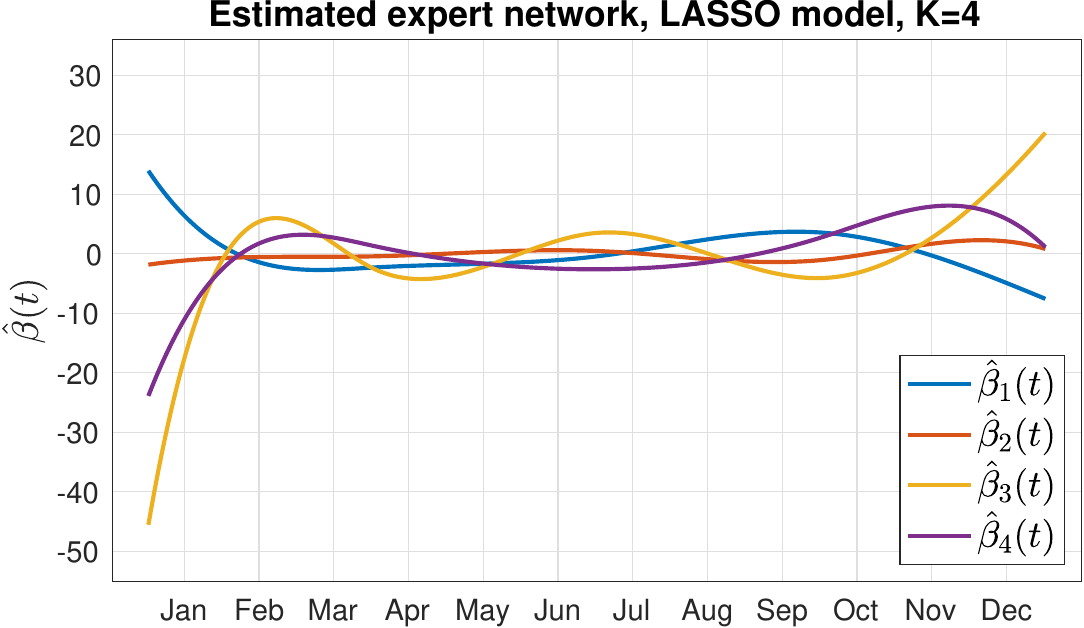}}\hspace*{.2cm}
\subfloat{\includegraphics[width=.32\linewidth]{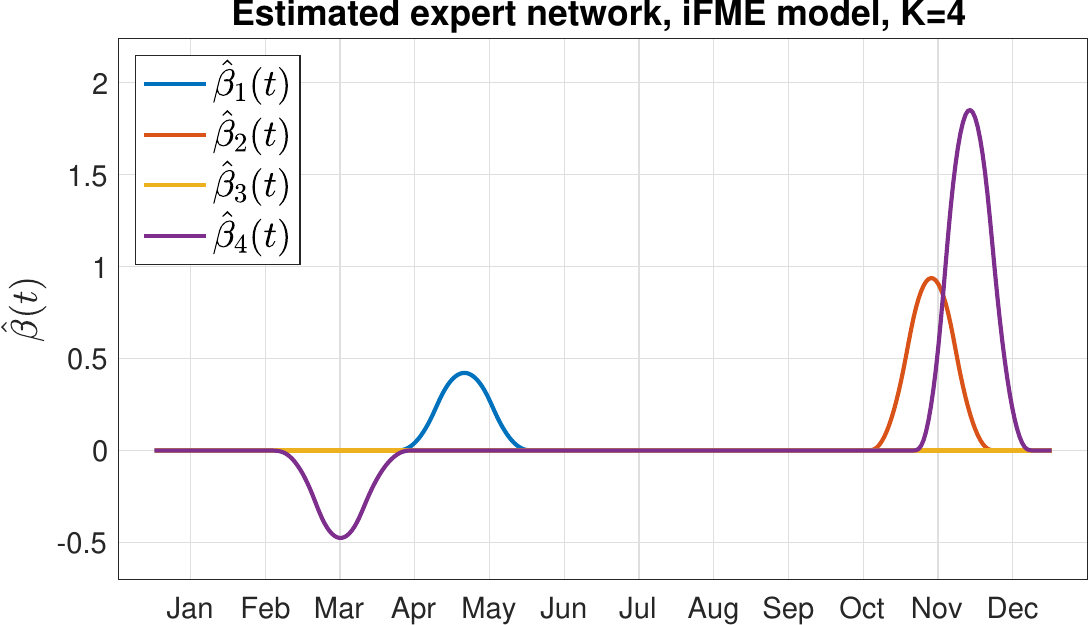}}\hfill
\subfloat{\includegraphics[width=.32\linewidth]{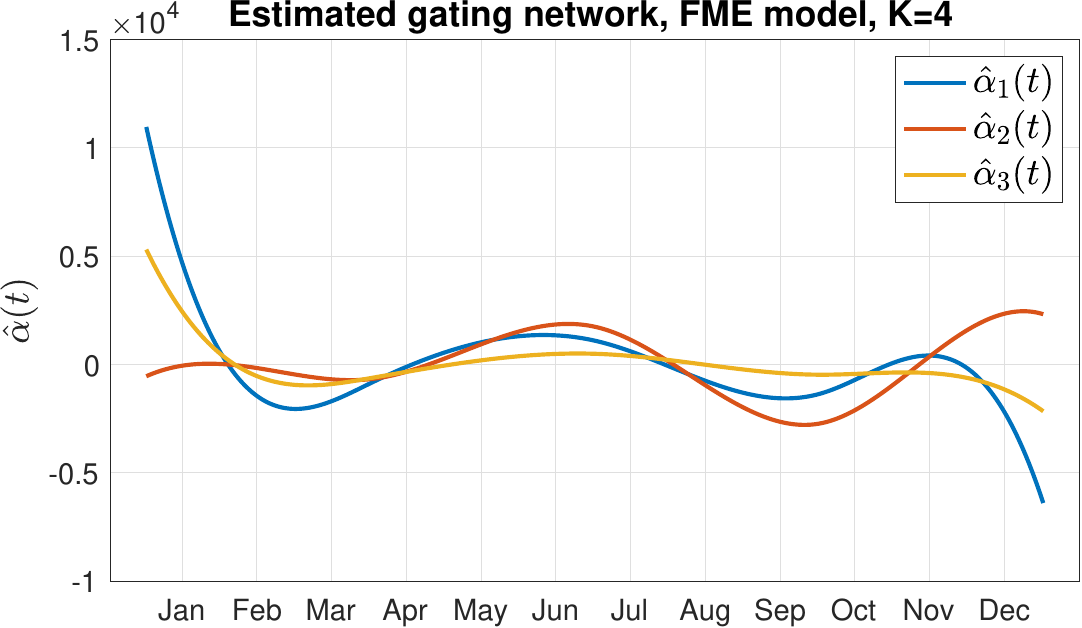}}\hspace*{.2cm}
\subfloat{\includegraphics[width=.32\linewidth]{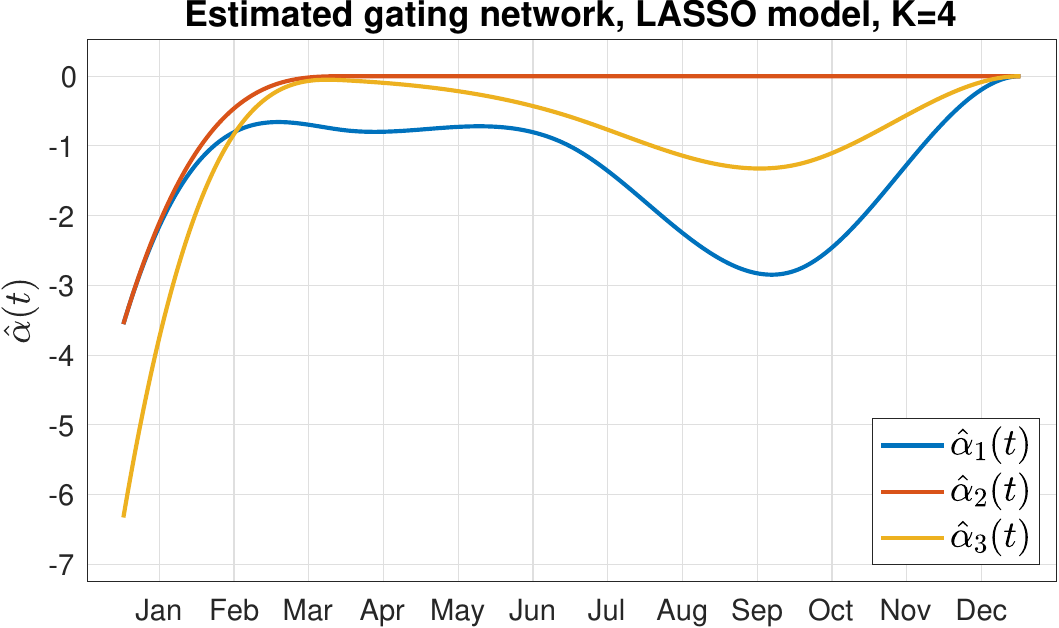}}\hspace*{.2cm}
\subfloat{\includegraphics[width=.32\linewidth]{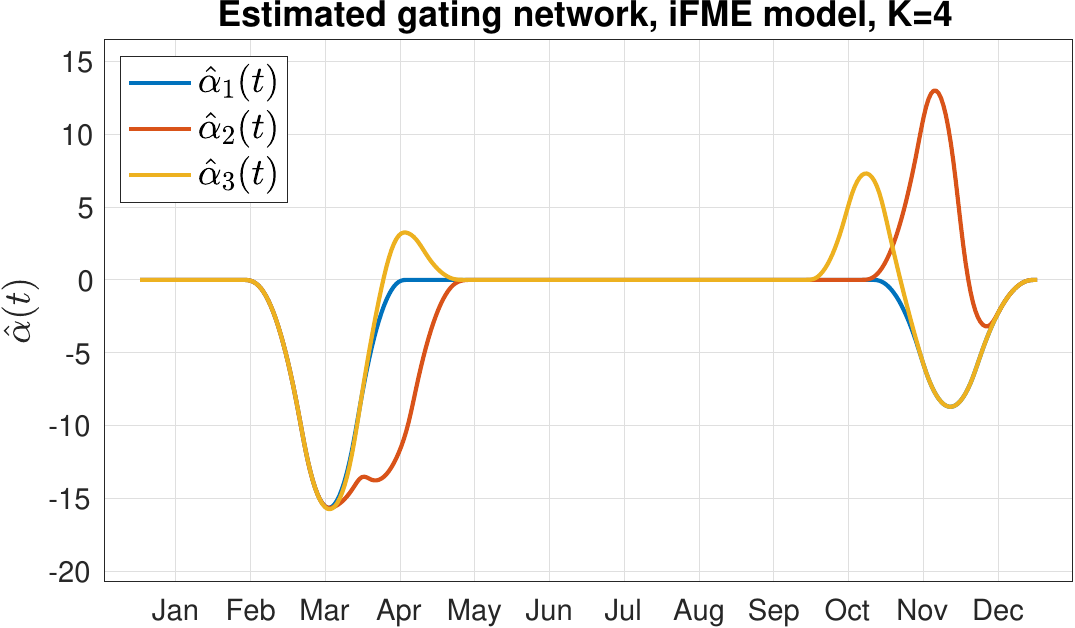}}\hfill
\subfloat{\includegraphics[width=.32\linewidth]{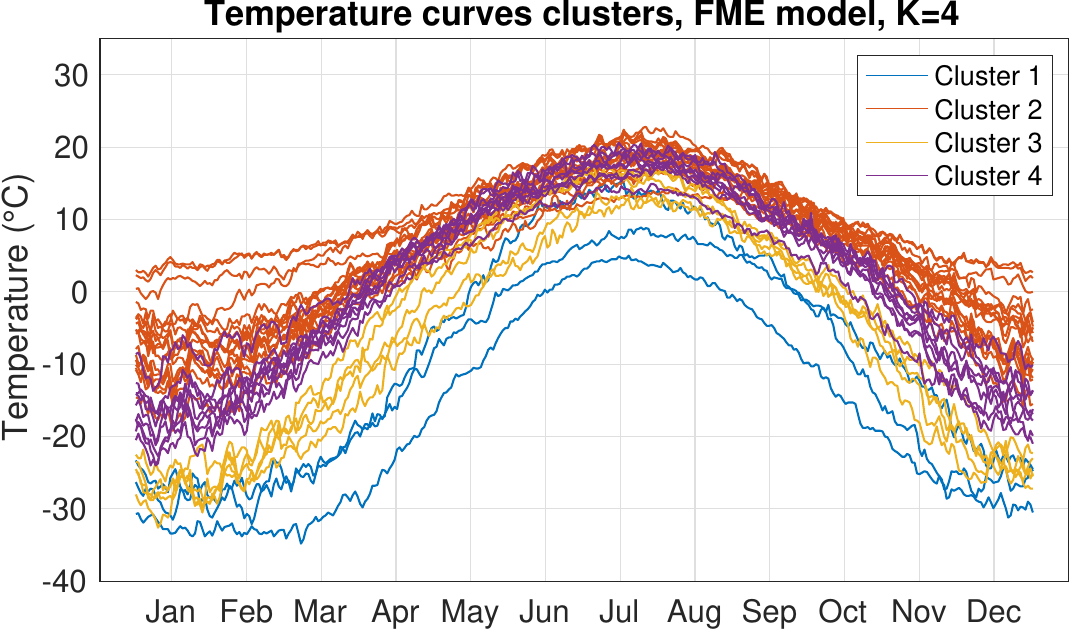}}\hspace*{.2cm}
\subfloat{\includegraphics[width=.32\linewidth]{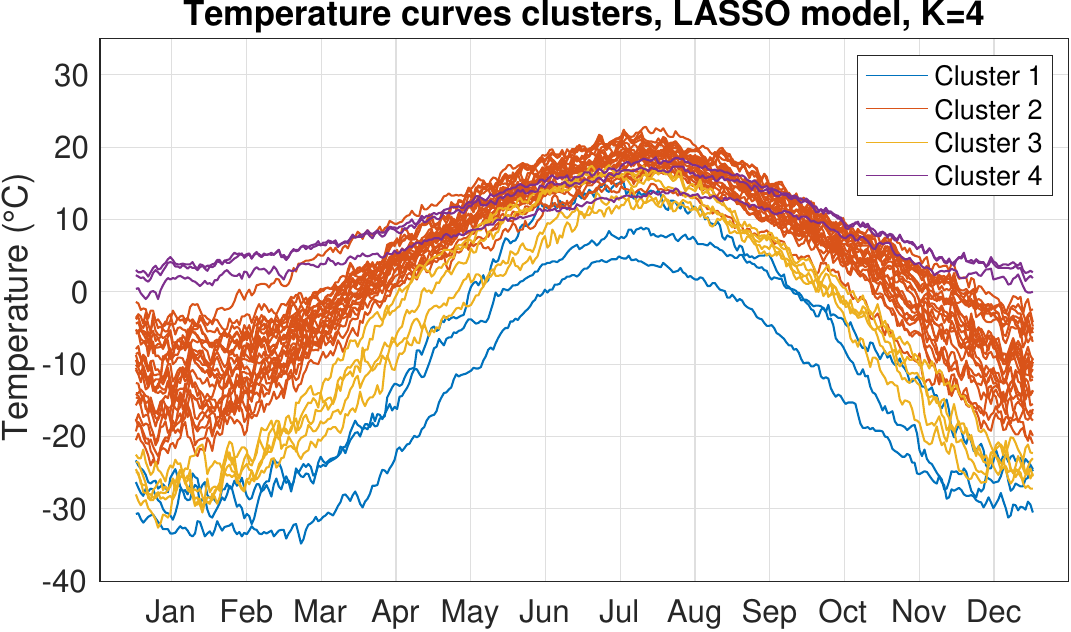}}\hspace*{.2cm}
\subfloat{\includegraphics[width=.32\linewidth]{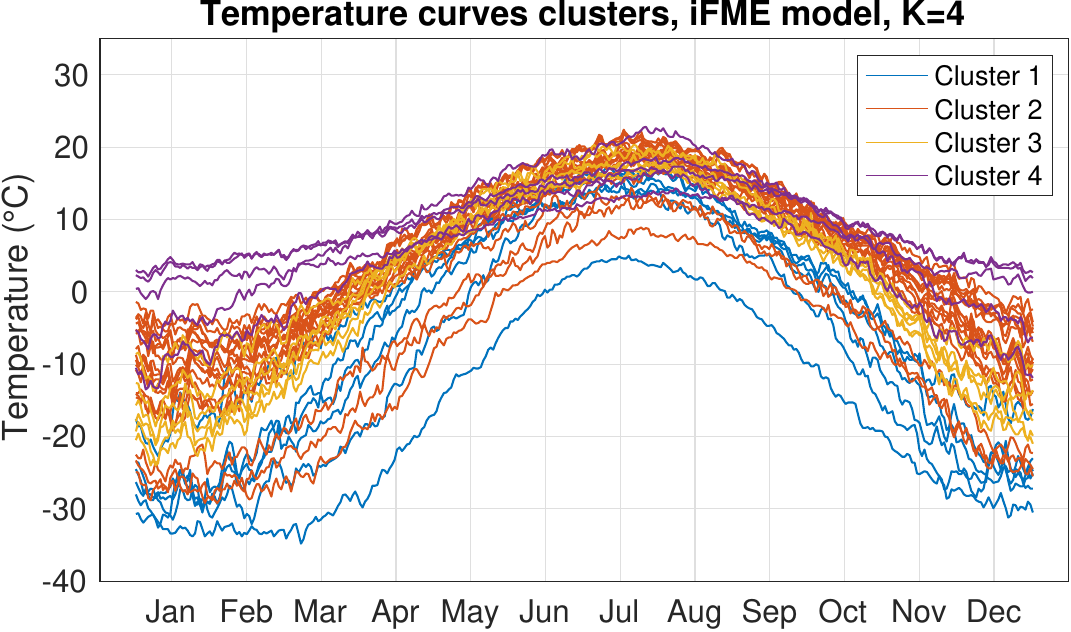}}\hfill
\subfloat{\includegraphics[width=.32\linewidth]{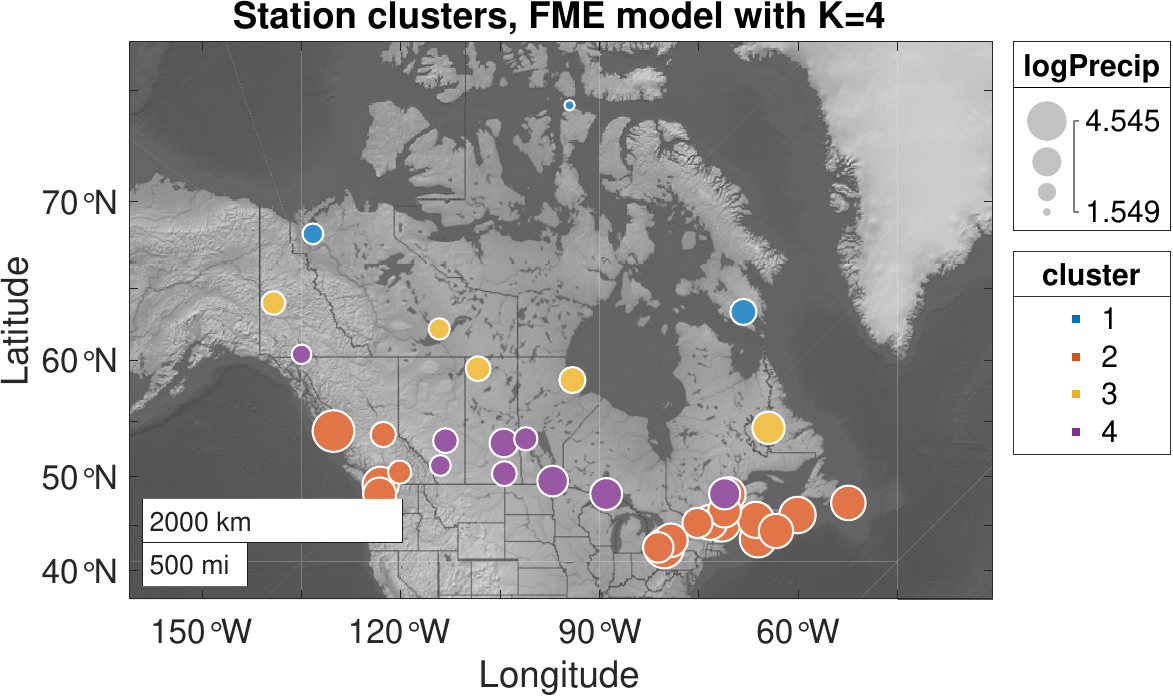}}\hspace*{.2cm}
\subfloat{\includegraphics[width=.32\linewidth]{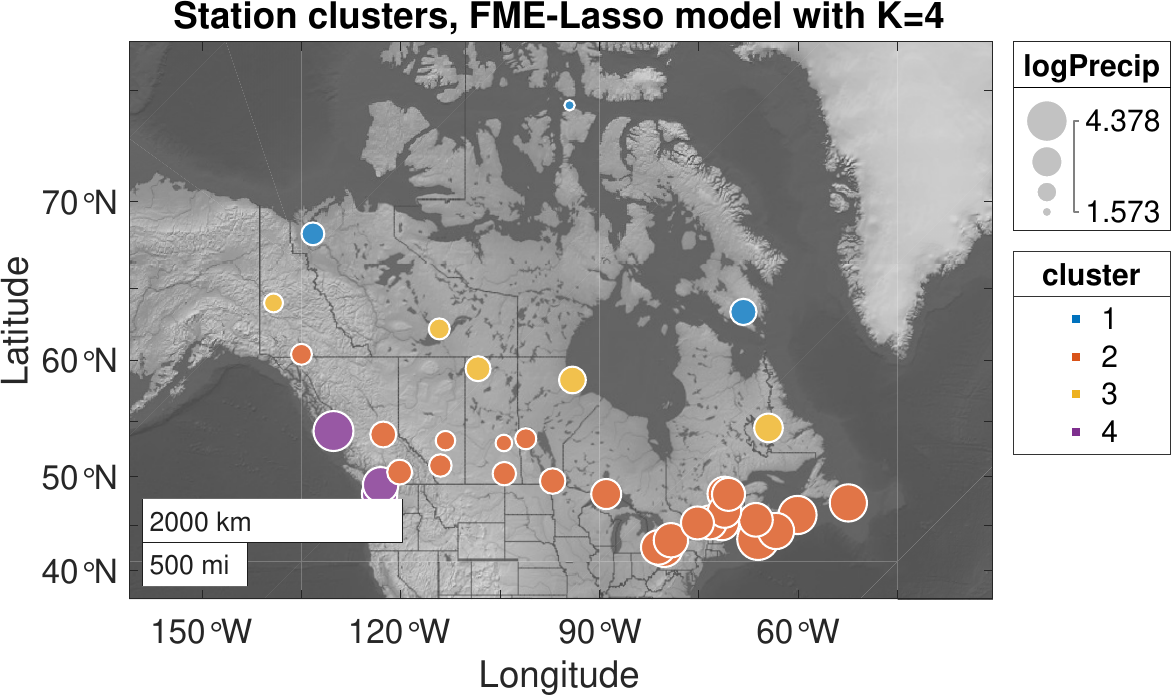}}\hspace*{.2cm}
\subfloat{\includegraphics[width=.32\linewidth]{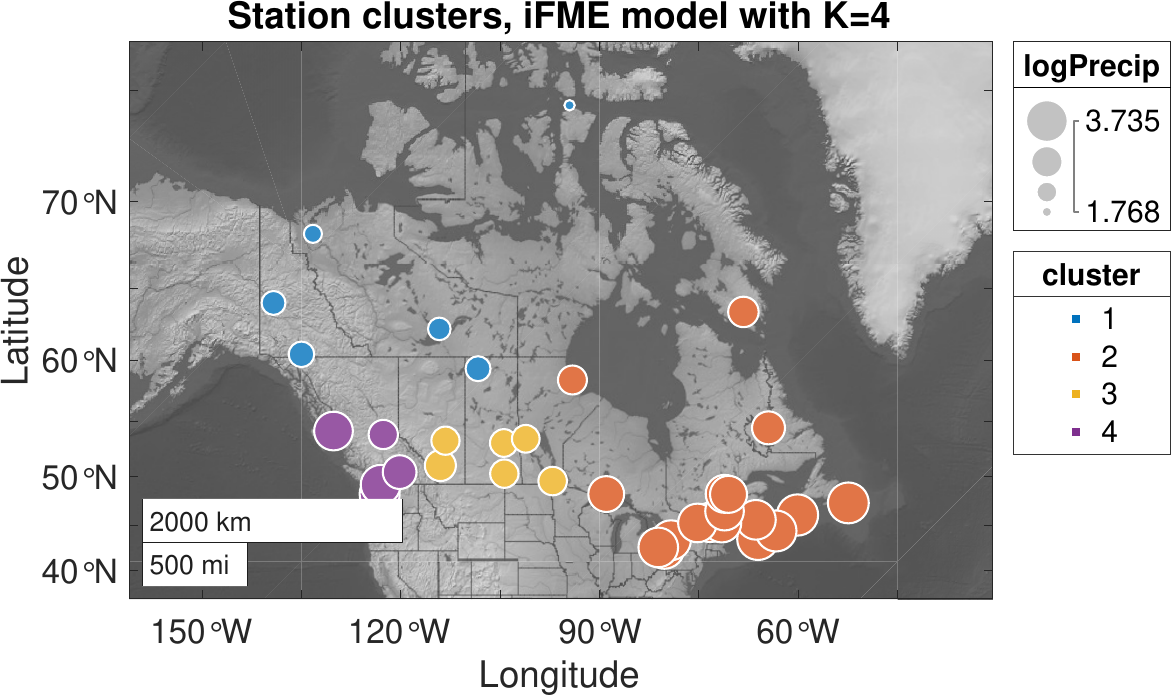}}\hfill
\caption{Results obtained by FME (left panels), FME-Lasso (middle panels) and iFME (right panels) on Canadian weather data with $K=4$. For each column, the panels are respectively the estimated functional experts network, estimated functional gating network, estimated clusters of the temperature curves and estimated clusters of the stations.}\label{Figure: CW with K4}
\end{figure}

Figure \ref{Figure: CW with K4} shows the obtained results with the FME, FME-Lasso and iFME models, 
with $K=4$, and with the derivatives $d_1=0$, $d_2=3$ for the iFME model.
The estimated experts functions and gating functions are presented in the two top panels of the curve,  
while the clustering for the temperature curves and the stations are shown in  the two bottom panels.
As we can see, all models provide reasonable clustering for the curves which may be corresponding to different complicated underlying meteorological forecasting mechanisms. 
Particularly, although not using any spatial information, merely temperature information, the obtained clustering for the stations is also comparable with the original labels of the stations. 
For example, the FME and FME-Lasso models identify exactly the Arctic stations, while iFME identifies exactly the Pacific stations, and all of the models provide reasonable spatially organized  clusters. 
However, what is interesting here is the shape of the expert and gating functions $\widehat\alpha(\cdot)$'s and $\widehat\beta(\cdot)$'s obtained by the models. 
While FME and FME-Lasso gave less interpretable estimations, 
iFME appears to  give, as it can be seen in the two top-right panels, piece-wise zero-valued and possibly quadratic estimated functions, which have a wide range of flat relationship from January to February and from June to September.

\begin{figure}[ht!]
\centering
\subfloat[]{\includegraphics[width=.4\linewidth]{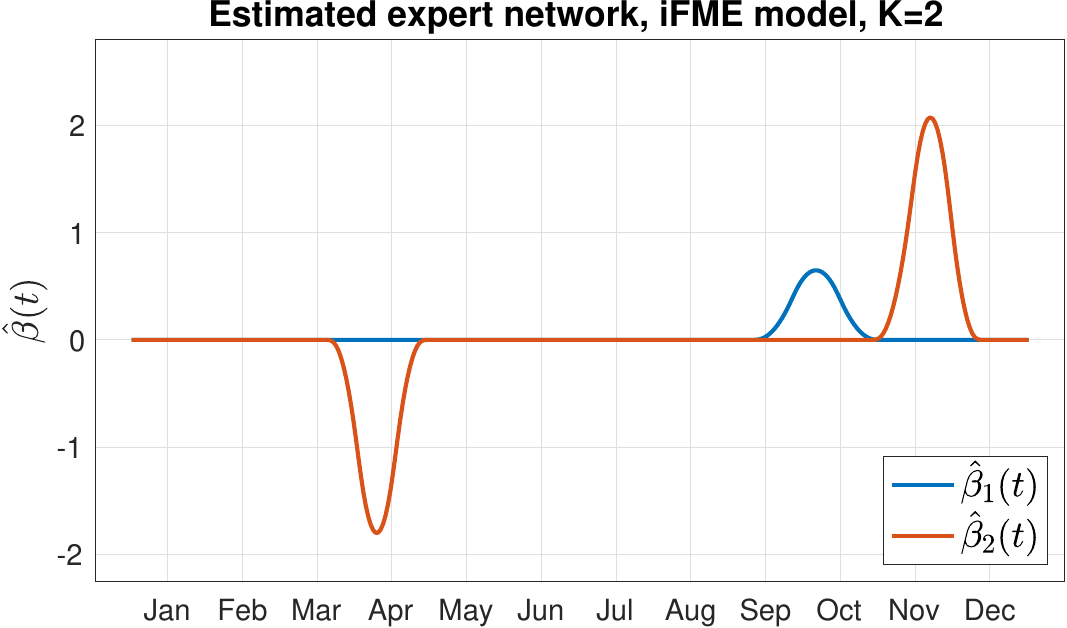}}\hspace*{.3cm}
\subfloat[]{\includegraphics[width=.4\linewidth]{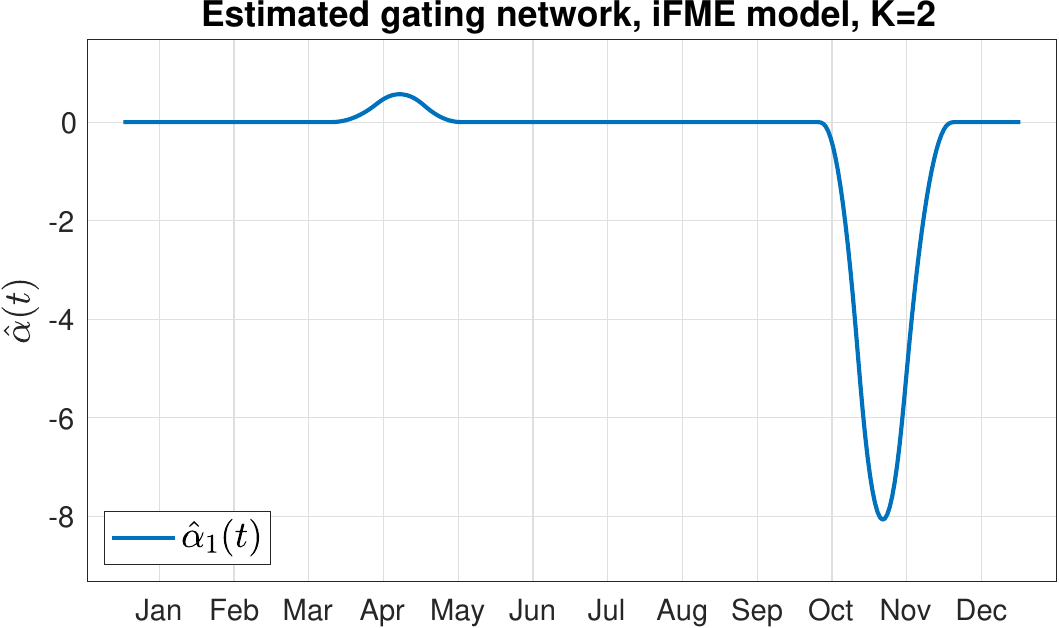}}\hfill
\subfloat[]{\includegraphics[width=.4\linewidth]{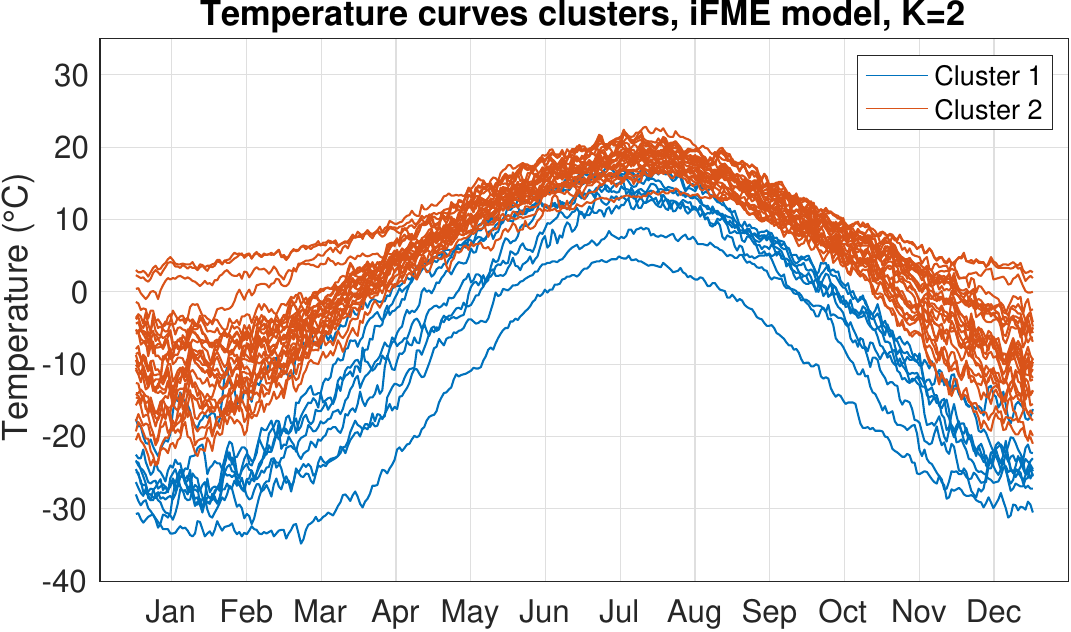}}\hspace*{.3cm}
\subfloat[]{\includegraphics[width=.4\linewidth]{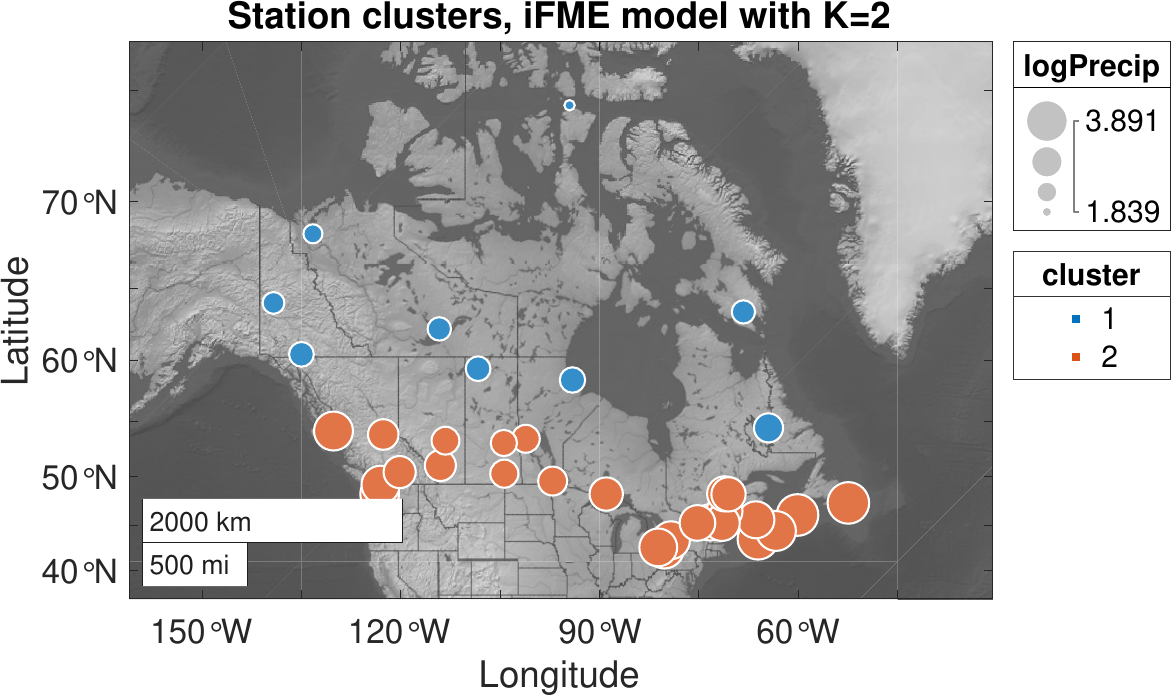}}\hfill
\caption{Results obtained by iFME model with $K=2$, $d_1=0$, $d_2=3$: (a) Estimated functional expert network, (b) Estimated functional gating network, (c) Estimated clusters of the temperature curves, and (d) Estimated clusters of the stations.}\label{Figure: iFME on CW K2}
\end{figure}

Motivated by the above results, on direction of identifying the periods of time of the year that truly have an effect on prediction, we implement the iFME model with $K=2$, and the derivative levels $d_1$ and $d_2$ are set to be the zeroth and the third derivatives. 
The reason for the choices of $d_1$ and $d_2$ is that the penalization on the zeroth derivative would take into account zero ranges in the expert and gating functions, while the penalization on the third derivative, would take into account the smoothness for the changes between the periods of times in the functions. 
The obtained results are shown in Figure \ref{Figure: iFME on CW K2}. 
As we can see, there are differences in the prediction mechanisms of the models between the northern stations and the southern stations. 
At southern stations, the obtained $\widehat{\beta}_2(t)$ shows that there is a negative relationship in the spring and a positive relationship in the late fall, but no relationship in the remaining period of the year. 
This phenomenon is concordant with the result obtained in \cite{FLIRTI}, where the authors obtained the same relationships in the same periods of time. 
However, our iFME model  additionally suggests that, at the northern stations, the relationship between temperature and precipitation may differ from that of southern stations. This may be explained by the differences in mean temperatures and climatic characteristics between the two regions.

\begin{figure}[ht!]
\centering
\subfloat[]{\includegraphics[width=.3\linewidth]{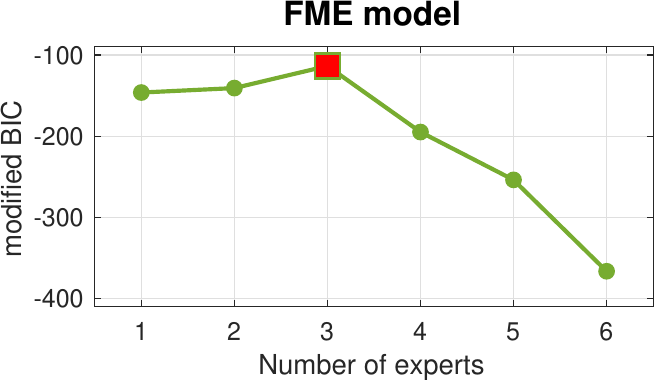}}\hfill
\subfloat[]{\includegraphics[width=.3\linewidth]{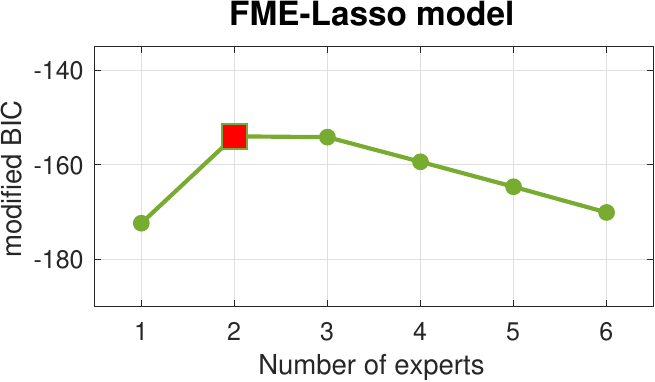}}\hfill
\subfloat[]{\includegraphics[width=.3\linewidth]{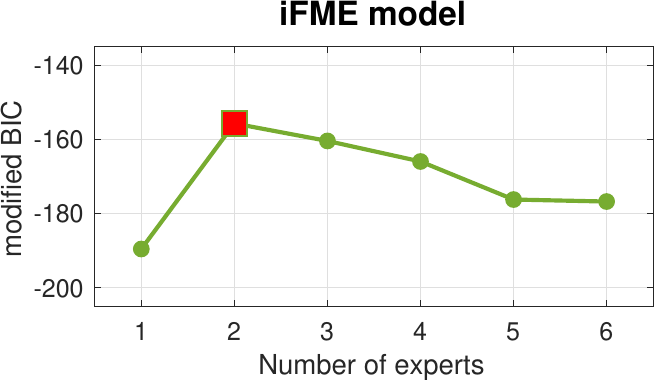}}\hfill
\caption{
Values of modified BIC for (a) FME, (b) FME-Lasso and (c) iFME model, versus the number of experts $K$, fitted on Canadian weather data. The square points correspond to highest values. Here, iFME is implemented with $d_1=0$, $d_2=3$ and $\rho=\varrho=100$.}\label{fig: BIC curves of CW}
\end{figure}

Finally, Figure \ref{fig: BIC curves of CW} displays the values of modified BIC for varying number of expert component for the proposed models on Canadian weather data. According to these values, FME-Lasso and iFME select $K=2$, while FME selects $K=3$.

\subsubsection{Diffusion tensor imaging data for multiple sclerosis subjects}

\begin{figure}[ht!]
\centering
\subfloat{\includegraphics[width=.9\linewidth]{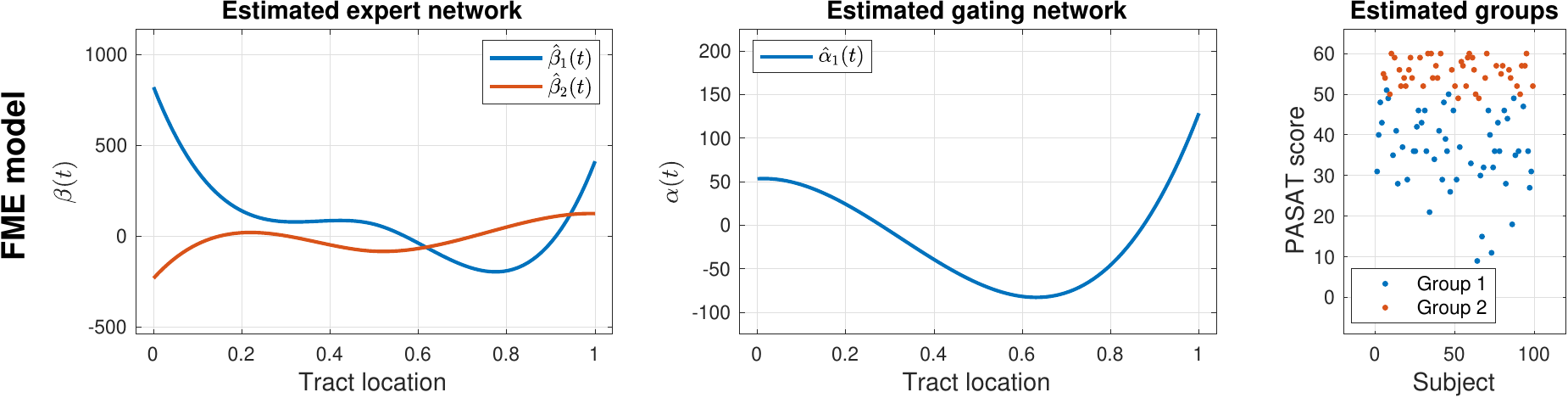}}\\
\subfloat{\includegraphics[width=.9\linewidth]{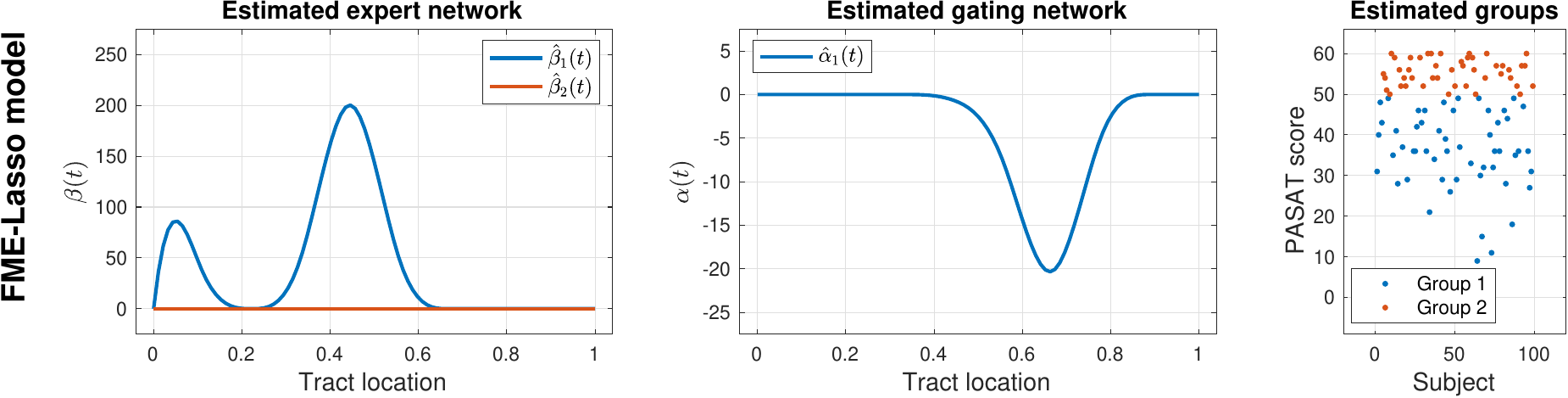}}\\
\subfloat{\includegraphics[width=.9\linewidth]{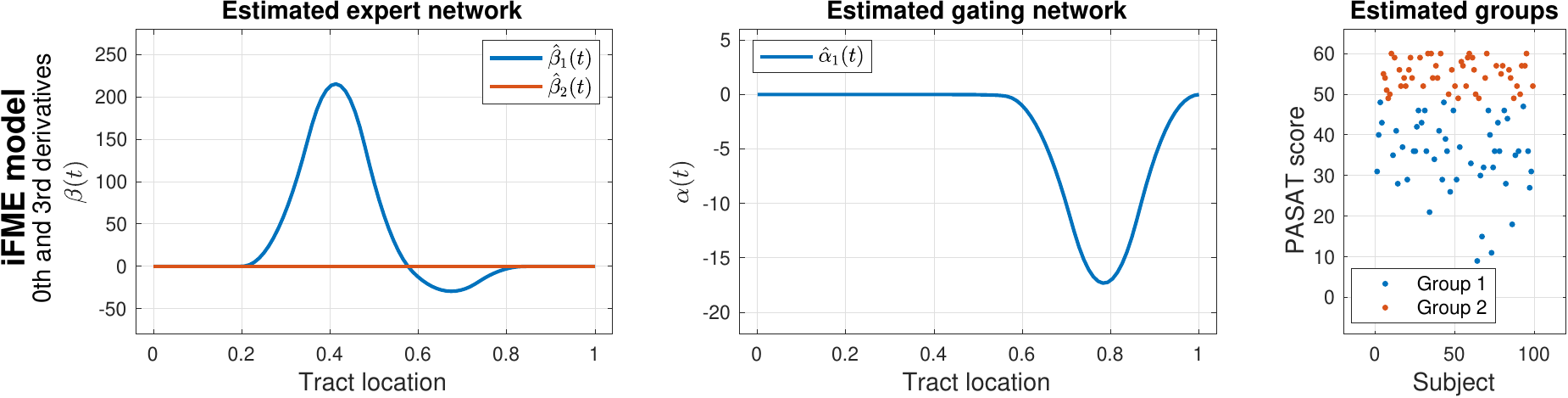}}\\
\subfloat{\includegraphics[width=.9\linewidth]{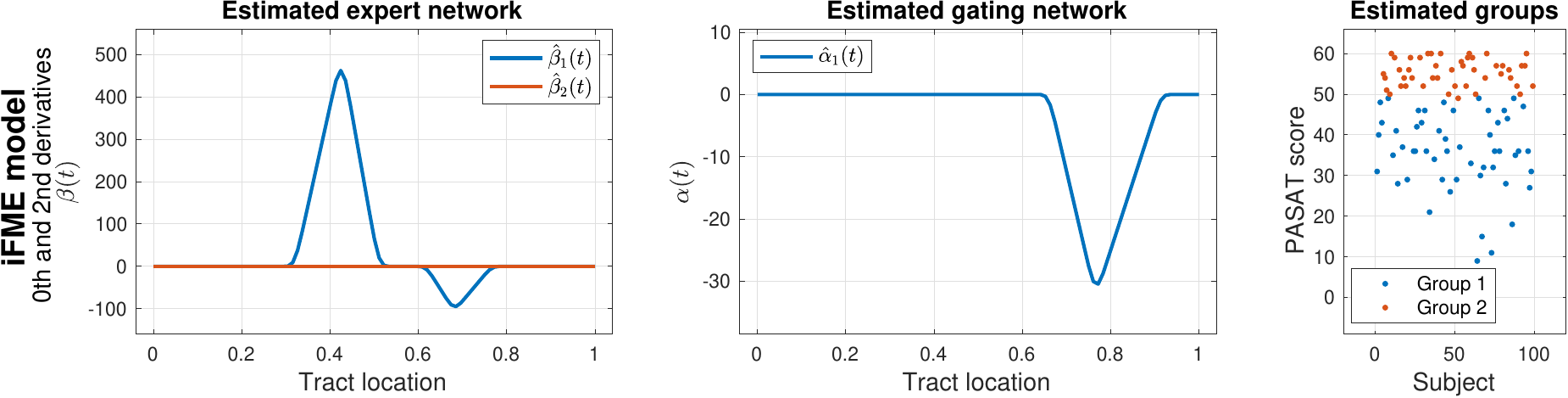}}
\caption{The estimated expert and gating coefficient functions, the estimated groups of the PASAT scores, resulted by FME, FME-Lasso and iFME models with $K=2$ for the DTI dataset. For iFME model, the upper is implemented with penalization on the zeroth and third derivatives, while the lower is with penalized zeroth and second derivatives.}\label{DTI: gating expert curves}
\end{figure}

\begin{figure}[ht!]
\centering
\subfloat[\label{figure: all FAPs}]{\includegraphics[width=.35\linewidth]{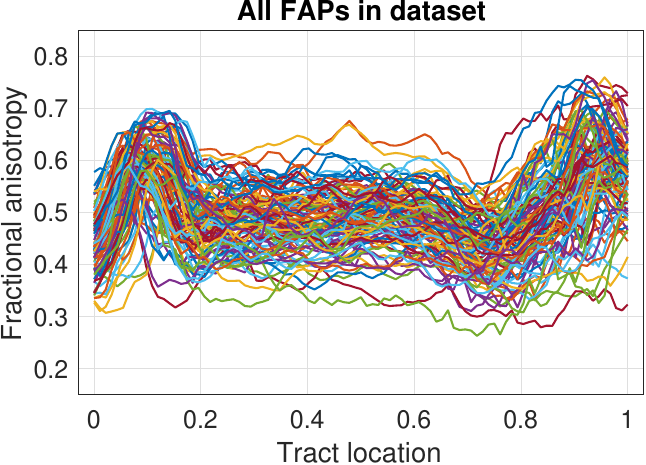}}\hspace{.5cm}
\subfloat[]{\includegraphics[width=.35\linewidth]{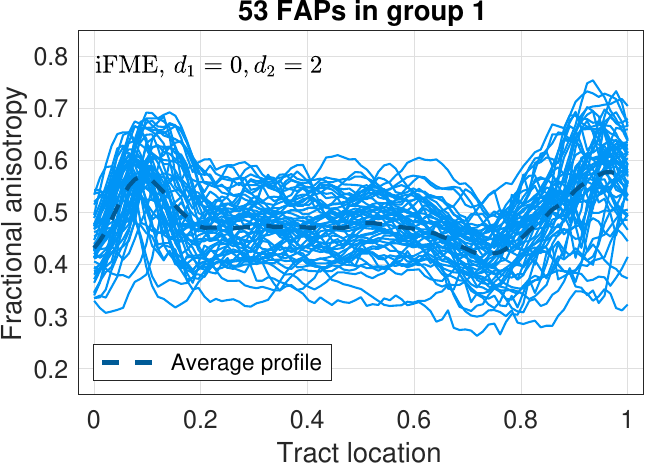}}\hfill
\subfloat[]{\includegraphics[width=.35\linewidth]{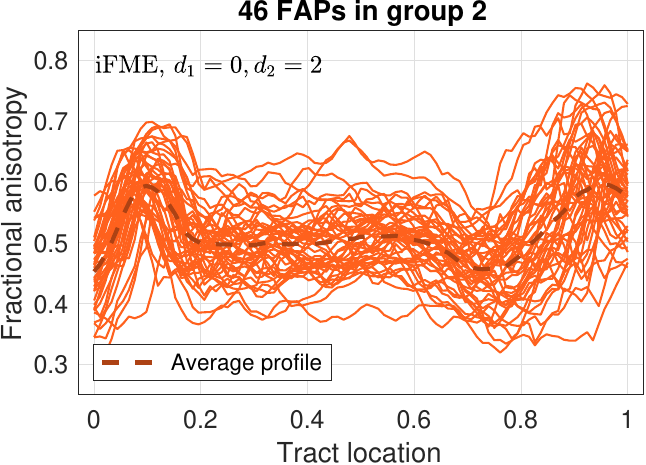}}\hspace{.5cm}
\subfloat[]{\includegraphics[width=.35\linewidth]{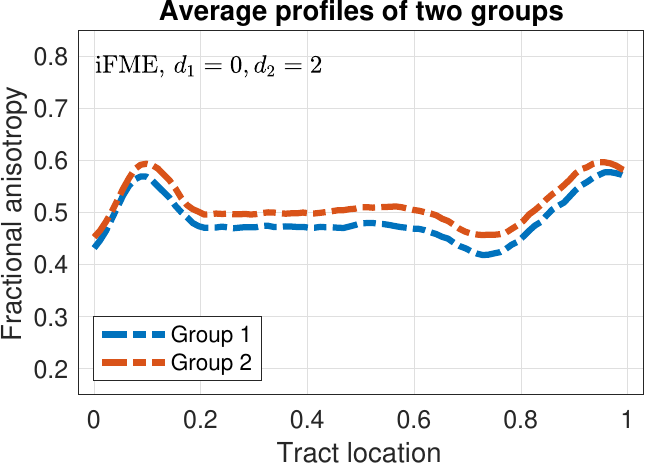}}
\caption{DTI data with (a) FAP curves of all subjects, (b) Cluster 1  and (c) Cluster 2, obtained by iFME model with $d_1=0$, $d_2=2$, and (d) the point-wise average of the curves in each of the two clusters.}
\label{DTI:curves}
\end{figure}

We now apply our proposed models to the diffusion tensor imaging (DTI) data for subjects with multiple sclerosis (MS), discussed in \cite{Goldsmith2012}. 
The data come from a longitudinal study investigating the cerebral white matter tracts of subjects with multiple sclerosis, recruited from an outpatient neurology clinic and healthy controls. 
We are interested in the underlying relationship between the fractional anisotropy profile (FAP) from the corpus callosum and the paced auditory serial addition test (PASAT) score, which is a commonly used examination of cognitive function affected by MS. 
The FAP curves are derived from DTI data, which are obtained by a Magnetic Resonance Imaging (MRI) scanner. Each curve is recorded at $93$ locations along the corpus callosum. The PASAT score is the number of correct answers out of $60$ questions, and thus ranges from $0$ to $60$. In our context, the FAP curves serve as the noisy functional predictors $U_i(\cdot)$   and the PASAT scores serve as the scalar responses $Y_i$. So this dataset consists of $n=99$ pairs $(U_i(\cdot),Y_i)$, $i\in[n]$, with each $U_i(\cdot)$ contains $m=93$ fractional anisotropy values. Figure \ref{figure: all FAPs}  shows all the predictors (FAP curves), and Figure \ref{DTI: gating expert curves} (most-right panels) shows the responses (PASAT Scores).

In \cite{CIARLEGLIO201686}, the authors applied their Wavelet-based functional mixture regression (WBFMR) model with two components to this dataset, and observed that there is one group in which there is no association between the FAP and the PASAT score for those subjects belonging to it. 
We accordingly fix $K=2$ in our models. 
Figure \ref{DTI: gating expert curves} displays the obtained results for each of the three models, and Figure \ref{figure: all FAPs} shows the functional predictors FAP curves clustered with the iFME model.
In this implementation, we tried iFME model with two different combinations of $d_1$ and $d_2$: $(d_1,d_2)=(0,2)$ and $(d_1,d_2)=(0,3)$. As expected, when $d_2$ is the second derivative, the reconstructed parameter functions are piecewise zero and linear, while when $d_2$ is the third derivative, the reconstructed functions have smooth changes along the tract location.

In Figure \ref{DTI: gating expert curves}, we have the three following observations. First, 
as it can be observed in Figure \ref{DTI: gating expert curves} right-panel,
 all models give a threshold of $50$ that clusters the PASAT scores, this is the same as the threshold observed in \cite{CIARLEGLIO201686}. 
Second, the absolute values of the coefficient functions $\widehat{\beta}_2(t)$'s are significantly smaller than those of $\widehat{\beta}_1(t)$'s, this is again the same with the result obtained by the WBFMR model. 
Third, when $\widehat{\beta}_2(t)$ is estimated as zero in the FME-Lasso and iFME models, the shape of $\widehat{\beta}_1(t)$ is almost the same as the shape obtained in \cite{CIARLEGLIO201686}, particularly, the peak at around the tract location of $0.42$. 
These confirm the underlying relationship between the fractional anisotropy and the cognitive function: higher fractional anisotropy values between the locations about $0.2$ to $0.7$ results in higher PASAT scores for subjects in Group $1$. 
The clustering of the FAP curves, resulted by the iFME model with $d_1=0$, $d_2=2$, is shown in Figure \ref{DTI:curves} (b)-(d).

Next, to compare with  \cite{CIARLEGLIO201686}, we investigate the prediction performance of the proposed models with respect to the leave-one-out cross validated relative prediction errors defined by
$\text{CVRPE}=\sum_{i=1}^n (y_i-\widehat y_i^{(-i)})^2/\sum_{i=1}^n y_i^2,$ where $y_i$ is the true score for subject $i$ and $\widehat y_i^{(-i)}$ is the score  predicted by the model fit on data without subject $i$. In this implementation, we keep fixing $K=2$ and select the other tuning parameters by maximizing the modified BIC.  The CVRPEs corresponding to the models are provided in Table \ref{Table: DTI CVRPE}. Note that, for comparison, in \cite{CIARLEGLIO201686}, the CVRPE of their WBFMR model is $0.0315$ and of the wavelet based functional linear model (FLM) is $0.0723$.
\begin{table}[htbp]
\centering
\def\arraystretch{.9}
\begin{tabular}{r R{2.7cm}}
\specialrule{1pt}{1pt}{1pt}
 & CVRPE\\
\hline
FME & $0.0273$ \\
FME-Lasso & $0.0280$ \\
iFME (with 0th and 3nd derivatives) & $0.0271$ \\
iFME (with 0th and 2nd derivatives) & $\Bs{0.0267}$ \\
\specialrule{1pt}{1pt}{1pt}
\end{tabular}
\caption{CVRPEs of the models on the DTI data.}\label{Table: DTI CVRPE}
\end{table}
 
\begin{figure}[!h]
\centering
\subfloat[]{\includegraphics[width=.3\linewidth]{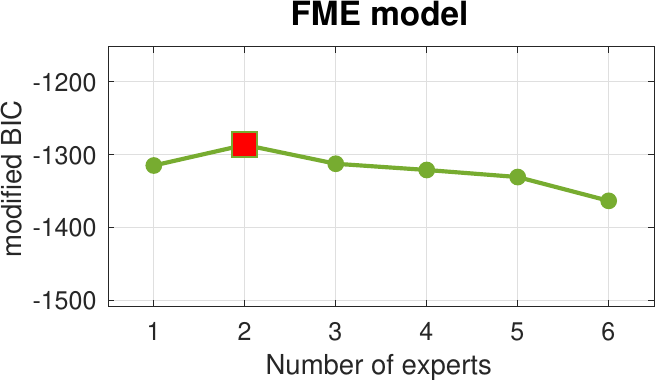}}\hfill
\subfloat[]{\includegraphics[width=.3\linewidth]{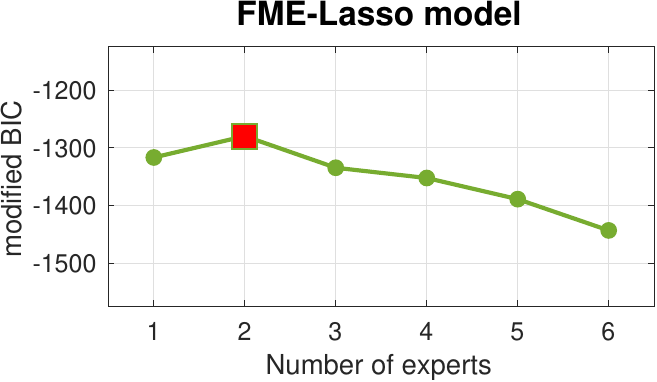}}\hfill
\subfloat[]{\includegraphics[width=.3\linewidth]{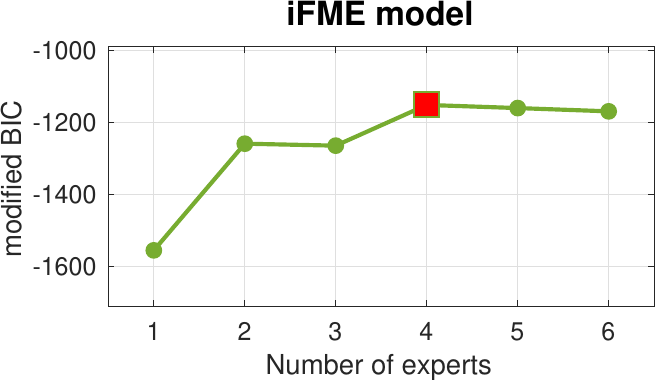}}\hfill
\caption{
Values of modified BIC  for (a) FME, (b) FME-Lasso and (c) iFME, versus the number of experts $K$, fitted on DTI data. The square points correspond to highest values.}\label{fig: BIC curves of DTI}
\end{figure}

Finally, we present in Figure \ref{fig: BIC curves of DTI} the selection of the number of experts $K$ with modified BIC. In this case, FME and FME-Lasso select $K=2$, and iFME selects $K=4$.

\section{Conclusion and discussion}
The first algorithm for mixtures-of-experts constructed upon functional predictors is presented in this paper. Beside the classic maximum likelihood parameter estimation, we proposed two other regularized versions that allow sparse and \textcolor{black}{notably} interpretable solutions, by regularizing in particular the derivatives of the underlying functional parameters of the model, after projecting onto a set of continuous basis functions. The performances of the proposed approaches are evaluated in data prediction and clustering via experiments involving simulated and two real-world functional data. 

The presented FME models can be extended in different ways. 
First direct extensions of the modeling framework presented here can be considered with categorical response, to perform supervised classification with  functional predictors, or with vector response, to perform multivariate functional regression. 
Then, it may be interesting to consider the extension of the FME model to  setting involving vector (or scalar) predictors and functional responses \citep{MullerFuncResponse2004}.

Another extension, which we intend also to investigate in the future, concerns the case when we observe pairs of functional data, i.e., a sample of $n$  functional data pairs \textcolor{black}{$\{X_i(t),Y_i(u)\}_{i=1}^n$, $t \in \cT \subset \R$, $u \in \mathcal U \subset \R$}, where $Y_i(\cdot)$ is a functional response, explained by a 
functional predictor $X_{i}(\cdot)$. 
The modeling with such FME extension then takes the form  
$Y_i(u) = \beta_{z_i,0}(u) + \int_t X_i(t) \beta_{z_i}(t,u) dt + \varepsilon_i(u)$, to explain the functional response $Y$ by the functional predictor $X$ via the unknown discrete variable $z$. The particularity with this model is that, for the clustering, as well as for the prediction, we model the relation between $Y$ at any time $u$ and the entire curve of $X$, or the entire curve of each variable $X_{ij}$ in the case of multivariate functional predictor $\bsX_i$.

\textcolor{black}{
Since we look for sparse estimates in the space of the derivatives of the functional parameters, and provided the resulting estimates with highly structured shapes, we believe that using the  fused LASSO \citep{TibshiraniFusedLasso2005} rather than the LASSO could be expected to take  accommodate more the targeted sparsity.  
 The choice of the regularization constants for the derivatives 
is more about the targeted penalization magnitude (i.e., very large or very small) rather than the exact values. 
Typically, in the case one has no idea about the shape of the functional experts, but still want them to be interpretable, then it is practical to consider the zeroth and third derivatives (since they produce smooth changes), with the grid for $\rho$ containing both very small and very large values. 
The same for the functional gating network.
 Furthermore, if we believe that the relation of the response on the predictor is very sparse, except over some small regions where they are linearly related, then we will penalize the zeroth and second derivatives of the functional experts, and we will put very small regularization on the second derivative. This could produce curves such as those in bottom panels of Figure \ref{DTI: gating expert curves}. For a more general tuning of the regularization constant values, we recommend performing a complete cross-validation study to select the best values of the regularization constants.
}

\color{black}

\section{Acknowledgements}
This research was partly supported by an Ethel Raybould Fellowship, University of Queensland (FC), 
the Australian Research Council grant number DP180101192 (GJM),
the French National Research Agency ANR grant SMILES ANR-18-CE40-0014 (FC and NTP), 
R\'egion Normandie grant RIN AStERiCs (FC and VHH),
Vietnam National University – Ho Chi Minh City (VNU- HCM)  grant number C2022-18-01 (VHH), 
and the exploratory research facility ``EXPLO'' at SystemX (FC).
 
\renewcommand{\baselinestretch}{1.3}
\normalsize
\setlength{\bibsep}{0.1pt}

\bibliographystyle{apalike}
{
\bibliography{REFERENCES}
}

\newpage
\thispagestyle{empty}
\appendix 
\renewcommand{\baselinestretch}{1}
\normalsize

\section{EM for the FME model\label{Appendix EM}}
 The FME model can be fitted by iteratively maximizing the observed-data log-likelihood \eqref{eq:loglik FME} iteratively via the EM algorithm.
 For FME, the EM takes the following form. 
The complete-data log-likelihood upon which the EM principle is constructed is defined by
\begin{equation}
\log L_c(\bsvPsi) = \sum_{i=1}^{n}\sum_{k=1}^{K} Z_{ik} \log \left[\pi_{k}(\br_i;\bsxi) 
\phi(y_{i};\beta_{k,0} + \bseta_{k}^{\top} \bx_{i},\sigma^2_k)
\right],
 \label{eq:complete log-lik FME}
\end{equation}$Z_{ik}$ being an indicator binary-valued variable such that $Z_{ik}=1$ if $Z_i=k$ (i.e., if the $i$th pair $(\bsx_i,\bsy_i)$ is generated from the $k$th expert component) and $Z_{ik}=0$, otherwise. 
 
\paragraph{E-step}
This step computes at each EM iteration $s$ the expectation of the complete-data log-likelihood (\ref{eq:complete log-lik FME}),  given the observed data $\cD$, and the current parameter vector $\bsvPsi^{(s)}$: 
\begin{eqnarray}
 Q(\bsvPsi;\bsvPsi^{(s)}) &=&  \E\left[\log L_c(\bsvPsi)|\cD;\bsvPsi^{(s)}\right]\nonumber\\
&=& \sum_{i=1}^{n}\sum_{k=1}^{K}\tau_{ik}^{(s)} 
\log 
\left[
\pi_{k}(\br_i;\bsxi)
\phi(y_{i};\beta_{k,0} + \bseta_{k}^{\top} \bx_{i},\sigma^2_k)
\right],
\label{eq:Q-function FME}
\end{eqnarray}where
$\tau_{ik}^{(s)}= \Pro(Z_i=k|y_i,u_i(\cdot);\bsvPsi^{(s)})$
is the conditional probability that the observed pair $(u_i(\cdot), y_i)$ is generated by  the $k$th expert. This step therefore only requires  the computation of the conditional probabilities $\tau^{(s)}_{ik}$ as defined by \eqref{eq:FME post prob}. 

\paragraph{M-step}
\label{sec: M-step EM-FME} 
This step updates the value of the parameter vector $\bsvPsi$ by maximizing the $Q$-function (\ref{eq:Q-function FME}) with respect to $\bsvPsi$, that is $\bsvPsi^{(s+1)} = \arg \max_{\bsvPsi} Q(\bsvPsi;\bsvPsi^{(s)}),$ via separate maximizations w.r.t. the gating network parameters, and the experts network parameters as follows.

\paragraph{Updating the the gating network parameters}
 Updating the the gating network's parameters $\bsxi$ consists of maximizing  w.r.t. $\bsxi$ the following function 
\begin{eqnarray}
\label{eq:Q-function gating-network}
Q(\bsxi;\bsvPsi^{(s)}) & = &  \sum_{i=1}^{n}\sum_{k=1}^{K}\tau_{ik}^{(s)} \log \pi_{k}(\br_i;\bsxi) \nonumber\\
&=  &\sum_{i=1}^n\left[\sum_{k=1}^{K-1} \tau^{(s)}_{ik} (\alpha_{k,0} + \bszeta^\top_{k}\br_i)  - \log\left(1 +   \sum_{k^\prime =1}^{K-1}  \exp\{\alpha_{k^\prime,0} + \bszeta^\top_{k^\prime}\br_i\} \right) \right] \cdot
\end{eqnarray}This consists of a weighted  multinomial logistic problem for which there is no closed-form solution. 
This can be performed by the Newton-Raphson (NR) algorithm which iteratively maximizes \eqref{eq:Q-function gating-network}
%
according to the procedure \eqref{eq. NR for Q gating}. 

Let us denote by $\bsxi_1,\ldots, \bsxi_{K-1}$ the parameter vectors $(\alpha_{1,0},\bszeta_1^{\top})^{\top},\ldots,(\alpha_{K-1,0},\bszeta_{K-1}^{\top})^{\top}$. Since there are $K-1$ parameter vectors to be estimated, the Hessian matrix $H(\bsxi;\bsvPsi^{(s)})$ is a block-matrix, consists of $(K-1)\times(K-1)$ blocks, in which each block $H_{k\ell}(\bsxi;\bsvPsi^{(s)})$, for $k,\ell\in[K-1]$, is given by:
\begin{eqnarray*}
H_{k\ell}(\bsxi;\bsvPsi^{(s)})
&=& \frac{\partial^2 Q(\bsxi; \bsvPsi^{(s)})}{\partial \bsxi_k \partial \bsxi_{\ell}^{\top}} 
=
- \sum_{i=1}^n \pi_{k}(\br_i;\bsxi^{(t)}) 
\left[\delta_{k\ell} - \pi_{l}(\br_i;\bsxi^{(t)})\right] \br_i \br_i^{\top}, 
\end{eqnarray*}
where $\delta_{k\ell}$ is the Kronecker symbol ($\delta_{k\ell}=1$ if $k=\ell$, 0 otherwise). The gradient vector consists of $K-1$ gradients  corresponding to the vectors $\bsxi_k$, for $k\in[K-1]$, and is given by
\begin{eqnarray*}
g (\bsxi;\bsvPsi^{(s)})
= \frac{\partial Q(\bsxi; \bsvPsi^{(s)})}{\partial \bsxi}
 = \left[g_1(\bsxi^{(t)}), \ldots, g_{K-1}(\bsxi^{(t)})\right]^{\top},
\end{eqnarray*}
where, for $k\in[K-1]$,
$g_k(\bsxi^{(t)})
= \frac{\partial Q(\bsxi; \bsvPsi^{(s)})}{\partial \bsxi_k} =
\sum_{i=1}^n \left[ \tau_{ik}^{(s)} - \pi_{k}(\bsxi^{(t)}; \br_i) \right] \br_i ^{\top}.$

\paragraph{Updating the experts network parameters}
Updating the experts network's parameters $\bstheta_k$ consists of maximizing the function $Q(\bstheta_k;\bsvPsi^{(s)})$ given by
\begin{eqnarray*}
\label{eq:Q-function expert-network}
 Q(\bstheta_k;\bsvPsi^{(s)}) &=& \sum_{i=1}^{n} \tau_{ik}^{(s)} \log \phi(y_{i};\beta_{k,0} + \bseta_{k}^{\top} \bx_{i},\sigma^2_k)
 \\ &=&
  - \frac{1}{2\sigma_k^2}\sum_{i=1}^{n} \tau_{ik}^{(s)} \left[y_{i} - (\beta_{k,0} + \bseta_{k}^{\top} \bx_{i})\right]^2 - \frac{n}{2}\log(2\pi \sigma_k^2)\cdot
\end{eqnarray*}Thus, updating  $\bstheta_{k} =  (\beta_{k,0},\bseta^\top_k, \sigma_{k}^{2})^\top$, consists of a weighted Gaussian regression problem where the weights are the  conditional memberships $\tau^{(s)}_{ik}$, and the updates are given by \eqref{eq:MLE FME beta and sigma2}.
%

\section{EM-Lasso for \texorpdfstring{$\ell_1$}{}-regularized MLE of the FME model\label{Appendix EM-Lasso}}
The EM-Lasso algorithm for the maximization of (\ref{eq:pen-loglik FME}) firstly requires the construction of the penalized complete-data log-likelihood
\begin{equation}
\cL_c(\bsvPsi) = \log L_c(\bsvPsi) - \text{Pen}_{\lambda,\chi}(\bsvPsi)
 \label{eq:pen complete log-lik FME}
\end{equation}where $\log L_c(\bsvPsi)$ is the non-regularized complete-data log-likelihood log-likelihood defined by
\eqref{eq:complete log-lik FME}.  
Thus, the EM algorithm for the FME model is implemented as follows. After starting with an initial solution $\bsvPsi^{(0)}$, it alternates between the two following steps, until convergence (when there is no longer a significant change in the values of the penalized log-likelihood (\ref{eq:pen-loglik FME})).
 
\paragraph{E-step.}
This step computes the expectation of the complete-data log-likelihood (\ref{eq:pen complete log-lik FME}),  given the observed data $\cD$, using the current parameter vector $\bsvPsi^{(s)}$: 
\begin{eqnarray}
 \mathcal{Q}_{\lambda,\chi}(\bsvPsi;\bsvPsi^{(s)}) &=&  \E\left[\cL_c(\bsvPsi)|\cD;\bsvPsi^{(s)}\right]
= Q(\bsvPsi;\bsvPsi^{(s)}) - \text{Pen}_{\lambda,\chi}(\bsvPsi),
\label{eq:Q-function FMElasso}
\end{eqnarray}
which only requires  the computation of the posterior probabilities of component membership $\tau^{(s)}_{ik}$ $(i\in[n])$, for each of the $K$ experts as defined by \eqref{eq:FME post prob}. 

\paragraph{M-step.}
\label{sec: M-step EM-FME-Lasso} 
This step updates the value of the parameter vector $\bsvPsi$ by maximizing the $Q$-function (\ref{eq:Q-function FMElasso}) with respect to $\bsvPsi$, that is, by computing the  parameter vector update 
\begin{equation}
\bsvPsi^{(s+1)} = \arg \max_{\bsvPsi} \mathcal{Q}_{\lambda,\chi}(\bsvPsi;\bsvPsi^{(s)}).
\label{eq:parameter update EM-FMElasso}
\end{equation} 
The maximization is performed by separate maximizations w.r.t. the gating network parameters and the experts network parameters.

\subsection{Updating the gating network parameters}
Updating the gating network parameters at the $s$th EM iteration consists of maximizing the following function 
\begin{eqnarray*}
\mathcal{Q}_{\chi}(\bsxi;\bsvPsi^{(s)}) & = &  Q(\bsxi;\bsvPsi^{(s)}) - \chi \sum_{k=1}^{K-1} \Vert\bszeta_{k}\Vert_1,
\end{eqnarray*}
with
\begin{eqnarray*}
Q(\bsxi;\bsvPsi^{(s)}) =  &\sum_{i=1}^n\left[\sum_{k=1}^{K-1} \tau^{(s)}_{ik} \left(\alpha_{k,0} + \bszeta^\top_{k}\br_i\right)  - \log\left(1 +   \sum_{k^\prime =1}^{K-1}  \exp\{\alpha_{k^\prime,0} + \bszeta^\top_{k^\prime}\br_i\} \right) \right]
\end{eqnarray*}
where $\bsxi = (\alpha_{1,0},\bszeta_1^{\top},\ldots,\alpha_{K-1},\bszeta_{K-1}^{\top})^{\top} \in \R^{(q+1)(K-1)}$ is the gating network parameter vector and $\bsvPsi^{(s)}$ is the current estimation of model's parameters. One can see this is equivalent to solving a weighted regularized multinomial logistic problem for which $\mathcal{Q}_{\chi}(\bsxi;\bsvPsi^{(s)})$ is its penalized log-likelihood. There is no closed-form solution for this kind of problem. We then use an iterative optimization algorithm  to seek for a maximizer of $\mathcal{Q}_{\chi}(\bsxi;\bsvPsi^{(s)})$, i.e., an update for the parameters of the gating network. The idea is to update only  a single gate at a time, while fixing other gate's parameters to their previous estimates. Again, to update that single gate, we only update one component at a time, while fixing the other components to their previous values. This procedure for updating the gating network parameters is supported by the methodology of coordinate ascent algorithm: if the objective function consists of a concave, differentiable function and a sum of concave functions then the maximizer can be achieved by iteratively maximizing with respect to each coordinate direction at a time.  

\paragraph{Coordinate ascent for updating the gating network.}
Suppose at the $s$th EM iteration, we wish to update the gates one by one such that it maximizes $\mathcal{Q}_{\chi}(\bsxi;\bsvPsi^{(s)})$. To do that, we create an outer loop, indexed by $t$, which cycles over the gates. For each single gate, say gate $k$, we partially approximate the smooth part of $\mathcal{Q}_{\chi}(\bsxi;\bsvPsi^{(s)})$ with respect to $(\alpha_{k,0}, \bszeta_{k})$ at $\bsxi^{(t)}$, then optimize the obtained objective function (with respect to $(\alpha_{k,0}, \bszeta_{k})$) by solving a penalized weighted least square problem using coordinate ascent algorithm. Note that $\bsxi^{(t)}$ denotes the current value of $\bsxi$ at the iteration $t$th of the outer loop, while $\bsxi^{(s)}$ is the value of $\bsxi$ before entering the outer loop.
 
 In particular, using Taylor expansion, one has a quadratic approximation for smooth part of $\mathcal{Q}_{\chi}(\bsxi;\bsvPsi^{(s)})$ with respect to $(\alpha_{k,0}, \bszeta_{k})$ at  $\bsxi^{(t)}$ given by
\begin{eqnarray*}
l_{k}(\alpha_{k,0}, \bszeta_k) = -\frac{1}{2} \sum_{i=1}^n w_{ik}(c_{ik} - \alpha_{k,0} - \br_i^{\top}\bszeta_k)^2 + C(\bsxi^{(t)}),
\end{eqnarray*}
where
\begin{eqnarray*}
w_{ik} &=& \pi_k(\bsxi^{(t)};\br_i)\left[1-\pi_k(\bsxi^{(t)};\br_i)\right], \quad \text{(weights)}\\
c_{ik} &=& \alpha_{k,0}^{(t)} + \br_i^{\top}\bszeta_k^{(t)} + \frac{\tau_{ik}^{(s)} - \pi_k(\bsxi^{(t)};\br_i)}{w_{ik}},\quad \text{(working response)}
\end{eqnarray*}
and $C(\bsxi^{(t)})$ is a function of $\bsxi^{(t)}$. After calculating the partial quadratic approximation $l_{k}(\alpha_{k,0}, \bszeta_k)$ about the current estimator $\bsxi^{(t)}$, we then solve the following penalized weighted least square problem
\begin{eqnarray}\label{gating optimization problem}
\underset{(\alpha_{k,0}, \bszeta_{k})}{\max}l_{k}(\alpha_{k,0}, \bszeta_k) - \chi\Vert \bszeta_k \Vert_1, \quad \chi>0,
\end{eqnarray}
to obtain an update for the parameters of gate $k$. 

As mentioned above, this problem could be solved by coordinate ascent algorithm. This means we will create an inner loop, indexed by $m$, cycles over the components of $(\alpha_{k,0}, \bszeta_k)$ and update them one by one until the objective function of \eqref{gating optimization problem} does not gain any significant increase. For each $j\in[q]$, using the soft-thresholding operator (see \cite{hastie2015statistical}, sec. 5.4), one can obtain the closed form update for $\bszeta_{kj}$ as follows
\begin{eqnarray*}
    \zeta_{kj}^{(m+1)} = \frac{\mathcal S_{\chi}
    \Big(
        \sum_{i=1}^n w_{ik}\rr_{ij}
        (c_{ik} - \tilde{c}_{ikj}^{(m)})
    \Big)}
    { \sum_{i=1}^n w_{ik} \rr_{ij}^2 },
\end{eqnarray*}
in which $\tilde{c}_{ikj}^{(m)} = \alpha_{k0}^{(m)} + \br_i^{\top}\bszeta_k^{(m)} - \zeta_{kj}^{(m)}\rr_{ij}$ is the fitted value excluding the contribution from $\rr_{ij}$, $\mathcal S_{\chi}(\cdot)$ is a soft-thresholding operator defined by $\mathcal S_{\chi}(u) = \sign(u)(\vert u \vert - \chi)_+$ and $(v)_+$ a shorthand for $\max\{v,0\}$. 
Note that at each iteration of the inner loop, only one component is updated while the others are kept to their previous values, that means $\zeta_{kh}^{(m+1)} = \zeta_{kh}^{(m)}$ for all $h\neq j$. For $\alpha_{k,0}$, the closed-form update is given by
\begin{eqnarray*}
    \alpha_{k,0}^{(m+1)} = \frac{ \sum_{i=1}^n w_{ik} 
    (c_{ik} - \br_i^{\top} \bszeta_k^{(m+1)}) 
    }{\sum_{i=1}^n w_{ik}}.
\end{eqnarray*}
Once the inner loop converges, the new values of $(\alpha_{k,0},\bszeta_{k})$ are used for the updating procedure of the next gate. When all the gates have their new values, i.e., after $K-1$ inner loops, we perform a backtracking line search before actually updating the gating network's parameters for the next $t$-indexed iteration. More precisely, the update is $\bsxi^{(t+1)} = (1-\nu)\bsxi^{(t)} + \nu\bar{\bsxi}^{(t)}$ where $\bar{\bsxi}^{(t)}$ is the output after $K-1$ inner loops and $\nu$ is backtrackingly determined to ensure $\mathcal{Q}_{\chi}(\bsxi^{(t+1)};\bsvPsi^{(s)}) > \mathcal{Q}_{\chi}(\bsxi^{(t)};\bsvPsi^{(s)})$.

We keep running the $t$-indexed loop until convergence, i.e., when there is no significant relative variation in $\mathcal{Q}_{\chi}(\bsxi;\bsvPsi^{(s)})$. Once $\alpha_{k,0}$ and $\zeta_{kj}$ reach their optimal values $\widetilde\alpha_{k,0}$ and $\widetilde\zeta_{kj}$ for all $k\in [K-1], j \in [q]$, the update for gating network's parameters is then $\bsxi^{(s+1)} = (\widetilde\alpha_{1,0},\widetilde\bszeta^{\top}_{1},\ldots, \widetilde\alpha_{K-1,0},\bszeta^{\top}_{K-1})^\top$.

\subsection{Updating the experts network parameters}
\label{Appendix: expert updates for EM-Lasso}

The maximization step for updating the expert parameters $\bstheta_k$ consists of maximizing the function 
$\mathcal{Q}_{\lambda}(\bstheta_k;\bsvPsi^{(s)})$ given by
\begin{eqnarray*}
 \mathcal{Q}_{\lambda}(\bstheta_k;\bsvPsi^{(s)}) &=& Q(\bstheta_k;\bsvPsi^{(s)}) - \lambda\Vert \bseta_k\Vert_1,
\end{eqnarray*}
with 
\begin{eqnarray*}
Q(\bstheta_k;\bsvPsi^{(s)}) = - \frac{1}{2\sigma_k^2}\sum_{i=1}^{n} \tau_{ik}^{(s)} \left[y_{i} - (\beta_{k,0} + \bseta_{k}^{\top} \bx_{i})\right]^2 - \frac{n}{2}\log(2\pi \sigma_k^2),
\end{eqnarray*}
where $\bstheta_k = (\beta_{k,0}, \bseta_k^{\top}, \sigma^2_k)^{\top} \in \R^{p+2}$ is the unknown vector and $\bsvPsi^{(s)}$ is the current estimation of model's parameters. There is no closed-form solution for this optimization problem, we then solve it by an iterative optimization algorithm similarly to updating the gating network parameters. We first perform the update for $(\beta_{k,0}, \bseta_{k})$ while fixing $\sigma^2_k$. This corresponds to solving a weighted LASSO problem  where the weights are the the posterior experts memberships $\tau_{ik}^{(s)}$. Once $(\beta_{k,0}, \bseta_{k})$ has new value, the variance $\sigma^2_k$ is updated straightforwardly by the standard update of a weighted Gaussian regression. 

More specifically, when ${\sigma_k^2}$, the variance of expert $k$, is fixed to ${\sigma_k^2}^{(s)}$, updating $(\beta_{k,0}, \bseta_k)$ consists of solving the following weighted LASSO problem:
\begin{eqnarray}\label{expert optimization problem}
\underset{(\beta_{k,0}, \bseta_k)}{\max} 
- \frac{1}{2{\sigma_k^2}^{(s)}}\sum_{i=1}^{n} \tau_{ik}^{(s)} \left[y_{i} - (\beta_{k,0} + \bseta_{k}^{\top} \bx_{i})\right]^2 - \frac{n}{2}\log(2\pi {\sigma_k^2}^{(s)}) - \lambda\sum_{j=1}^q |\bseta_{kj}|,\quad \lambda>0,
\end{eqnarray}
which can be solved by coordinate ascent algorithm. For each $j\in[p]$, the closed-form update for $\bseta_{kj}$ is given by
\begin{eqnarray*}
\eta_{kj}^{(m+1)} = \frac{\mathcal S_{\lambda{\sigma_k^2}^{(s)}}
    \Big(
        \sum_{i=1}^n \tau_{ik}^{(s)}\rx_{ij}
        (y_{i} - \tilde{y}_{ij}^{(m)})
    \Big)}
    { \sum_{i=1}^n \tau_{ik}^{(s)} \rx_{ij}^2 },
\end{eqnarray*}
in which $\tilde{y}_{ij}^{(m)} = \beta_{k,0}^{(m)} + \bx_i^{\top}\bseta_k^{(m)} - \eta_{kj}^{(m)}\rx_{ij}$ is the fitted value excluding the contribution from $\rx_{ij}$ and $\mathcal S_{\chi}(\cdot)$ is the soft-thresholding operator (see \cite{hastie2015statistical}, sec. 5.4). Here $m$ denotes the $m$th iteration of the coordinate ascent algorithm. The update for $\beta_{k,0}$ is
\begin{eqnarray*}
\beta_{k,0}^{(m+1)} = \frac{\sum_{i=1}^n\tau_{ik}^{(s)}(y_i - \bx_i^{\top}\bseta_{k}^{(m+1)})}{\sum_{i=1}^n\tau_{ik}^{(s)}}\cdot
\end{eqnarray*}
We keep updating the components of $(\beta_{k,0},\bseta_k)$ cyclically until the change in objective function of \eqref{expert optimization problem} is small enough. So, the update for $(\beta_{k,0}, \bseta_{k})$ in this EM iteration is then $(\beta_{k,0}^{(s+1)}, \bseta_{k}^{(s+1)}) = (\beta_{k,0}^{*}, \bseta_{k}^{*})$ where the latter is the optimal solution of \eqref{expert optimization problem}. Finally, the update for $\sigma^2_k$ is given by
\begin{eqnarray*}
{\sigma^2}^{(s+1)}_k = \frac{\sum_{i=1}^n \tau_{ik}^{(s)} (y_i - \beta_{k,0}^{(s+1)} - \bx_i^{\top}\bseta_k^{(s+1)})^2}{\sum_{i=1}^n \tau_{ik}^{(s)}}\cdot
\end{eqnarray*}


\color{black}

Hence, the update for the value of parameters vector $\bsvPsi$ at M-step, i.e., the solution to problem \eqref{eq:parameter update EM-FMElasso}, is $\bsvPsi^{(s+1)} = (\bsxi^{(s+1)}, \bstheta_1^{(s+1)}, \ldots, \bstheta_K^{(s+1)})$ where $\bsxi^{(s+1)}$ and $\bstheta_k^{(s+1)}, k\in[K]$, are solved by maximizing $\mathcal Q_{\chi}(\bsxi;\bsvPsi^{(s)})$ and $\mathcal Q_{\lambda}(\bstheta_k;\bsvPsi^{(s)})$, respectively, using the algorithms described above. The EM algorithm monotonically increases (\ref{eq:pen-loglik FME}). Furthermore, the sequence of parameter estimates generated by the EM algorithm converges toward a local maximum of the log-likelihood function \citep{dlr,McLachlanEM2008,Wu-convergence-EM}.

\color{black}
\section{EM-iFME for updating iFME model parameters}
\subsection{Updating the gating network parameters}\label{EM-iFME gating update}
This section presents the using of Dantzig selector to solve problem \eqref{gating penalized weighted LS problem}.
Let us simplify the subscript $k$ in the notations and rewrite the problem under matrix form as follows
\begin{equation}
\begin{aligned}\label{prob: gating iFME}
\underset{\WT{\bsomega}\in\R^{2q+1}}{\max\ }\quad
&-\frac{1}{2} \Vert \bc_{\Bs{w}} - \bX_{\Bs{w}} \WT{\bsomega} \Vert_2^2 
- \chi \Vert \Omega \WT{\bsomega} \Vert_1\\ 
\text{subject to }\quad&\bA \WT{\bsomega} = \Bs{0}_q,
\end{aligned}
\end{equation}
where $\WT{\bsomega}=(\alpha_0,\ {\bsomega^{[d_1]}}^\top, {\bsomega^{[d_2]}}^\top)^\top$ is the unknown coefficients vector, 
$\bc_{\Bs{w}}=(\sqrt{w_1}c_1, \ldots,\sqrt{w_n}c_n)^\top$ is the weighted working response vector, 
$\bX_{\Bs{w}} = [\sqrt{\Bs{w}} | \bS_{\Bs{w}} | \Bs{0}_{n\times q}]\in\R^{2q+1}$  is the weighted design matrix, 
$\Omega=\text{diag}(0,\Bs{1}_q^\top, \varrho\Bs{1}_q^\top)$ is the diagonal weighting matrix and 
$\bA=[\Bs{0}_{q} | \bA_q^{[d_2]}{\bA_q^{[d_1]}}^{-1} | - \Bs{I}_q]$ is the constraints matrix. 
Here, $\sqrt{\Bs{w}}=(\sqrt{w_1},\ldots,\sqrt{w_n})^\top$, 
$\bS_{\Bs{w}} = [\sqrt{w_1}\bs_1,\ldots,\sqrt{w_n}\bs_n]^\top$, with $\bs_i$ are the design vectors (see \eqref{eq: FLiRTI constrains}),
$\Bs{0}_{n\times q}\in\R^{n\times q}$ contains $0$'s, $\Bs{0}_{q}\in\R^{q}$ contains $0$'s, $\Bs{1}_q\in\R^q$ contains $1$'s
and $\Bs{I}_q$ is the identity matrix in $\R^{q\times q}$.
This problem can be viewed as the problem of finding a sparse solution via Lasso estimate for a linear regression model with constraints. Therefore, we can solve it alternatively by Dantzig selector estimate as the solution to the following problem
\begin{eqnarray*}
&\underset{\WT{\bsomega}\in\R^{2q+1}}{\max}&
-\Vert \Omega \WT{\bsomega} \Vert_1\\ 
&\text{subject to}&\left\{
\begin{array}{l}
\vert {\bX^\top_{\Bs{w}}}(\bc_{\Bs{w}} - \bX_{\Bs{w}} \WT{\bsomega}) \vert \leq \chi\Bs{1}_{2q+1},\\
\bA \WT{\bsomega} = \Bs{0}_q,
\end{array}
\right.
\end{eqnarray*}
where the absolute value operator is understood componentwise. By decomposing $\WT{\bsomega}$ into its positive and negative parts, $\WT{\bsomega} = \WT{\bsomega}_+ - \WT{\bsomega}_-$, 
the above problem becomes
\begin{equation}\label{eq:gating LP}
\begin{aligned}
\underset{(\WT{\bsomega}_+, \WT{\bsomega}_-)\in\R^{4q+2}}{\max}
&- [0,\Bs{1}_q^\top, \varrho\Bs{1}_q^\top, 0,\Bs{1}_q^\top, \varrho\Bs{1}_q^\top]
\begin{bmatrix}
 \WT{\bsomega}_+\\
 \WT{\bsomega}_-
\end{bmatrix}\\
\text{subject to\  }&\left\{
\begin{array}{l}
\begin{bmatrix}
\ \ {\bX^\top_{\Bs{w}}}{\bX_{\Bs{w}}} & -{\bX^\top_{\Bs{w}}}{\bX_{\Bs{w}}}\\
-{\bX^\top_{\Bs{w}}}{\bX_{\Bs{w}}} & \ \ {\bX^\top_{\Bs{w}}}{\bX_{\Bs{w}}}
\end{bmatrix}
\begin{bmatrix}
 \WT{\bsomega}_+\\
 \WT{\bsomega}_-
\end{bmatrix}
\leq
\begin{bmatrix}
\chi + {\bX^\top_{\Bs{w}}} \bc_{\Bs{w}}\\
\chi - {\bX^\top_{\Bs{w}}} \bc_{\Bs{w}}
\end{bmatrix},\\
\begin{bmatrix}
 \bA & -\bA
\end{bmatrix}
\begin{bmatrix}
 \WT{\bsomega}_+\\
 \WT{\bsomega}_-
\end{bmatrix}= \Bs{0}_{4q+2},\\
\WT{\bsomega}_+ \geq \Bs{0}_{2q+1},\quad \WT{\bsomega}_- \geq \Bs{0}_{2q+1},
\end{array}
\right.
\end{aligned}
\end{equation}
which is a standard linear program with $4q+2$ variables, therefore can be easily solved by available toolboxes for linear programming, e.g., Matlab's \textsf{linprog} function. Finally, the solution ${\WT{\bsomega}}^*=(\alpha_0^*,\ {\bsomega^{[d_1]*}}^\top, {\bsomega^{[d_2]*}}^\top)^\top$ to the original problem could be retrieved by the relation ${\WT{\bsomega}}^* = {\WT{\bsomega}_+^{*}} - \WT{\bsomega}_-^{*}$ where $({\WT{\bsomega}_+^{*\top}},\ {\WT{\bsomega}_-^{*\top}})^\top$ is the solution of \eqref{eq:gating LP}.

\subsection{Updating the expert network parameters}\label{EM-iFME experts update}
This section presents how to solve problem \eqref{expert penalized weighted LS problem}. Firstly, we fix $\sigma^2_k$ to its previous estimate and perform an update for $(\beta_{k,0}, \bsgamma_{k}^{[d_1]})$, it corresponds to solving a penalized weighted least square problem with constraints. From now on, let us simplify the subscript $k$ in the notations and rewrite the problem under matrix form as follows
\begin{equation}\label{prob: expert iFME}
\begin{aligned}
\underset{\WT{\bsgamma}\in\R^{2p+1}}{\max}\quad&
-\frac{1}{2} \Vert \by_{\bssigma\bstau} - \bX_{\bssigma\bstau} \WT{\bsgamma} \Vert_2^2 
- \lambda \Vert \Lambda \WT{\bsgamma} \Vert_1
- \frac{n_k}{2}\log(2\pi\sigma^2)\\
\text{subject to}&\quad  \bA \WT{\bsgamma} = \Bs{0}_p, \nonumber
\end{aligned}
\end{equation}
where $\WT{\bsgamma}=(\beta_0,\ {\bsgamma^{[d_1]}}^\top, {\bsgamma^{[d_2]}}^\top)^\top$ is the unknown coefficients vector, 
$\by_{\bssigma\bstau}=(\sigma\sqrt{\tau_1}y_1, \ldots,\sigma\sqrt{\tau_n}y_n)^\top\allowbreak\in\R^n$ is the weighted response vector, 
$\bX_{\bssigma\bstau} = \bssigma \odot [\sqrt{\bstau} | \bV_{\bstau} | \Bs{0}_{n\times p}]\in\R^{n\times (2p+1)}$  is the weighted design matrix, 
$\Lambda=\text{diag}(0,\Bs{1}_p^\top, \rho\Bs{1}_p^\top)$ is the diagonal weighting matrix,
$n_k=\sum_{i=1}^n \tau_{ik}$
and $\bA=[\Bs{0}_{p} | \bA_p^{[d_2]}{\bA_p^{[d_1]}}^{-1} | - \Bs{I}_p] \in\R^{p\times (2p+1)}$ is the constraints matrix. 
Here, $\sqrt{\bstau}=(\sqrt{\tau_1},\ldots,\sqrt{\tau_n})^\top\in\R^n$, 
$\bV_{\bstau} = [\sqrt{\tau_1}\bv_1,\ldots,\sqrt{\tau_n}\bv_n]^\top\in\R^{n\times p}$, with $\bv_i$ are the design vectors (see \eqref{eq: FLiRTI constrains}),
$\Bs{0}_{n\times p}\in\R^{n\times p}$ contains $0$'s, $\Bs{0}_{p}\in\R^{p}$ contains $0$'s, $\Bs{1}_p\in\R^p$ contains $1$'s
and $\Bs{I}_p$ is the identity matrix in $\R^{p\times p}$. As the last term in the objective function is independent of $\WT{\bsgamma}$, this problem is similar to the problem \eqref{prob: gating iFME} and then can be solved by Dantzig selector estimate as the solution to the following problem
\begin{equation}
\begin{aligned}
\underset{\WT{\bsgamma}\in\R^{2p+1}}{\max}\quad&
-\Vert \Lambda \WT{\bsgamma} \Vert_1\\ 
\text{subject to}\ &\left\{
\begin{array}{l}
\vert {\bX^\top_{\bssigma\bstau}}(\by_{\bssigma\bstau} - \bX_{\bssigma\bstau} \WT{\bsgamma}) \vert \leq \lambda\Bs{1}_{2p+1},\\
\bA \WT{\bsgamma} = \Bs{0}_p.
\end{array}
\right.
\end{aligned}\nonumber
\end{equation}
Similarly to the problem in the gating network update, by decomposing $\WT{\bsgamma}$ into its positive and negative parts, $\WT{\bsgamma} = \WT{\bsgamma}_+ - \WT{\bsgamma}_-$, 
the above problem becomes
\begin{equation}\label{eq:expert LP}
\begin{aligned}
\underset{(\WT{\bsgamma}_+, \WT{\bsgamma}_-)\in\R^{4p+2}}{\max}
&- [0,\Bs{1}_p^\top, \rho\Bs{1}_p^\top, 0,\Bs{1}_p^\top, \varrho\Bs{1}_p^\top]
\begin{bmatrix}
 \WT{\bsgamma}_+\\
 \WT{\bsgamma}_-
\end{bmatrix}\\
\text{subject to\  }&\left\{
\begin{array}{l}
\begin{bmatrix}
\ \ {\bX^\top_{\bssigma\bstau}}{\bX_{\bssigma\bstau}} & -{\bX^\top_{\bssigma\bstau}}{\bX_{\bssigma\bstau}}\\
-{\bX^\top_{\bssigma\bstau}}{\bX_{\bssigma\bstau}} & \ \ {\bX^\top_{\bssigma\bstau}}{\bX_{\bssigma\bstau}}
\end{bmatrix}
\begin{bmatrix}
 \WT{\bsgamma}_+\\
 \WT{\bsgamma}_-
\end{bmatrix}
\leq
\begin{bmatrix}
\lambda + {\bX^\top_{\bssigma\bstau}} \by_{\bssigma\bstau}\\
\lambda - {\bX^\top_{\bssigma\bstau}} \by_{\bssigma\bstau}
\end{bmatrix},\\
\begin{bmatrix}
 \bA & -\bA
\end{bmatrix}
\begin{bmatrix}
 \WT{\bsgamma}_+\\
 \WT{\bsgamma}_-
\end{bmatrix}= \Bs{0}_{4p+2},\\
\WT{\bsgamma}_+ \geq \Bs{0}_{2p+1},\quad \WT{\bsgamma}_- \geq \Bs{0}_{2p+1},
\end{array}
\right.
\end{aligned}
\end{equation}
which is a standard linear program with $4q+2$ variables and can be solved similarly to the gating network case. The solution ${\WT{\bsgamma}}^*=(\beta_0^*,\ {\bsgamma^{[d_1]*}}^\top, {\bsgamma^{[d_2]*}}^\top)^\top$ to the original problem is retrieved by the relation ${\WT{\bsgamma}}^* = {\WT{\bsgamma}_+^{*}} - \WT{\bsgamma}_-^{*}$ where $({\WT{\bsgamma}_+^{*\top}},\ {\WT{\bsgamma}_-^{*\top}})^\top$ is the solution of \eqref{eq:expert LP}.

Finally, the update for $\sigma^2_k$ is
\begin{eqnarray*}
{\sigma^2_k}^{(s+1)} &=& \frac{\sum_{i=1}^n \tau_{ik}^{(s)} (y_i - \beta_{k,0}^{(s+1)} - \bv_i^{\top}{\bsgamma_k^{[d_1]}}^{(s+1)})^2}{\sum_{i=1}^n \tau_{ik}^{(s)}},
\end{eqnarray*}
in which $\beta_{k,0}^{(s+1)},{\bsgamma_k^{[d_1]}}^{(s+1)} $ are the new updates for $\beta_{k,0}$ and $\bsgamma_k^{[d_1]}$.


\section{Detailed Data generating protocol, simulation parameters and experimental protocol}

\subsection{Data generating protocol} \label{sec: Data generating protocols}

In the simulated data, the data generation protocol is as follows. 
We consider a $K$-component functional mixture of Gaussian experts model that relates  a scalar response $y\in\R$ to a univariate functional predictor $X(t), t\in \cT$ defined on a domain  $\cT\subset\R$. 
Given the model parameters (defined in the next paragraph) $\bsbeta = \{\beta_{k,0},\beta_{k}(t), \sigma_k^2\}_{k=1}^K$ and $\bsalpha = \{\alpha_{k,0},\alpha_{k}(t)\}_{k=1}^{K}$, $t\in \cT$, we first construct the functional predictors $X_i(\cdot)$ for $i\in[n]$ using the representation defined in \eqref{eq: X projection}, with detailed parameterization in \eqref{eq:functiondal predictior simulation}.
Then, for each $i\in[n]$, conditional on the functional predictor $X_i(\cdot)$, a hidden categorical random variable $Z_i \in [K]$ is generated following the multinomial distribution $\mathcal{M}\Big(1, \big(\pi_1(X_i(t);\bsalpha),\ldots, \pi_K(X_i(t);\bsalpha)\big)\Big)$, where $\pi_k(X_i(t);\bsalpha)$ for $k\in[K]$ is given by \eqref{eq:gating Net FME}. Finally, conditional on $Z_i = z_i$ and $X_i(\cdot)$, the scalar response $Y_i$ is obtained by sampling from the Gaussian distribution with mean $\beta_{z_i,0}+\int_{\cT}X_i(t)\beta_{z_i}(t)dt$ and variance $\sigma^2_{z_i}$. The value $z_i$ is then the true cluster label of the predictor $X_i(\cdot)$.
This hierarchical generative process can be summarized as follows:
\begin{eqnarray}
\begin{aligned}
Y_i|Z_{i}=z_i,X_i(t) \ &\sim\  \mathcal{N}\Big( \beta_{z_i,0}+\int_{\cT}X_i(t)\beta_{z_i}(t)dt; \ \sigma^2_{z_i} \Big),\\
Z_i|X_i(t) \ &\sim\  \mathcal{M}\Big(1, \big(\pi_1(X_i(t);\bsalpha),\ldots, \pi_K(X_i(t);\bsalpha)\big)\Big).
\end{aligned}\label{Data generating protocols}
\end{eqnarray}
The true generated data, i.e., $\big(X_i(\cdot), Y_i, Z_i\big)_{i=1}^n$, are used for evaluating the prediction and clustering performance.

\color{black}
Finally, to mimic real data, in our simulations, the generated predictors $X_i(\cdot)$ are contaminated with   measurement errors, that means we will not use $X_i(\cdot)$ for analysis, but 
\begin{equation*}
U_i(t) = X_i(t) + \delta_i(t),
\end{equation*}
with $\delta_i(t)$ is an independent Gaussian noise with mean zero and constant variance $\sigma_{\delta}^2$ for all $t\in\cT$. We considered different noise levels $\sigma_{\delta}^2$ (see Table \ref{Table: scenarios}).

\subsection{Simulation parameters and experimental protocol}\label{subsection Similation 1}

The parameters that were used in the data generating process \eqref{Data generating protocols} are as follows. We consider $K=3$ components and a time domain $\cT=[0,1]$. 
\paragraph{Functional experts and gating parameters:}
The functional experts parameters are given by
\begin{eqnarray*}
\beta_1(t) &=&\left\{
\begin{array}{ll}
     -50(t-0.5)^2 + 4 & \quad\text{if } 0 \leq t < 0.3,\\
      0 &  \quad\text{if }0.3 \leq t < 0.7,\\
     50(t-0.5)^2 - 4 & \quad\text{if } 0.7 \leq t \leq 1,
\end{array}\right.\\
\beta_2(t) &=& -\beta_1(t),\\
\beta_3(t) &=& 100(t-0.5)^2 - 10, \quad 0 \leq t \leq 1,\\
(\beta_{1,0}, \beta_{2,0}, \beta_{3,0})^{\top}&=&(-5,0,5)^{\top},\\
(\sigma^2_1,\sigma^2_2,\sigma^2_3)^{\top} &=& (5,5,5)^{\top},
\end{eqnarray*}
and the functional gating network parameters are given by
\begin{eqnarray*}
\alpha_1(t) &=& 80(t-0.5)^2 - 8,\\
\alpha_2(t) &=& -\alpha_1(t),\quad \alpha_3(t) = \Bs{0}, \quad 0 \leq t \leq 1,\\
(\alpha_{1,0}, \alpha_{2,0}, \alpha_{3,0})^{\top}&=&(-10,-10,0)^{\top}.
\end{eqnarray*}
Note that to satisfy the identifiability condition (see \cite{JIANG19991253}), the experts are ordered, for instance $(\beta_{1,0}, \beta_1,\sigma^2_1)\prec \ldots \prec (\beta_{K,0},\beta_K,\sigma^2_K)$, and the last gating network parameters, $\alpha_{K,0},\alpha_K(t)$, are initialized, i.e. fixed to zeros. Here, ``$\prec$'' is the lexicographical order on $\R^{p+2}$.

\begin{figure}[ht!]
\centering
\subfloat{\includegraphics[width=.8\linewidth]{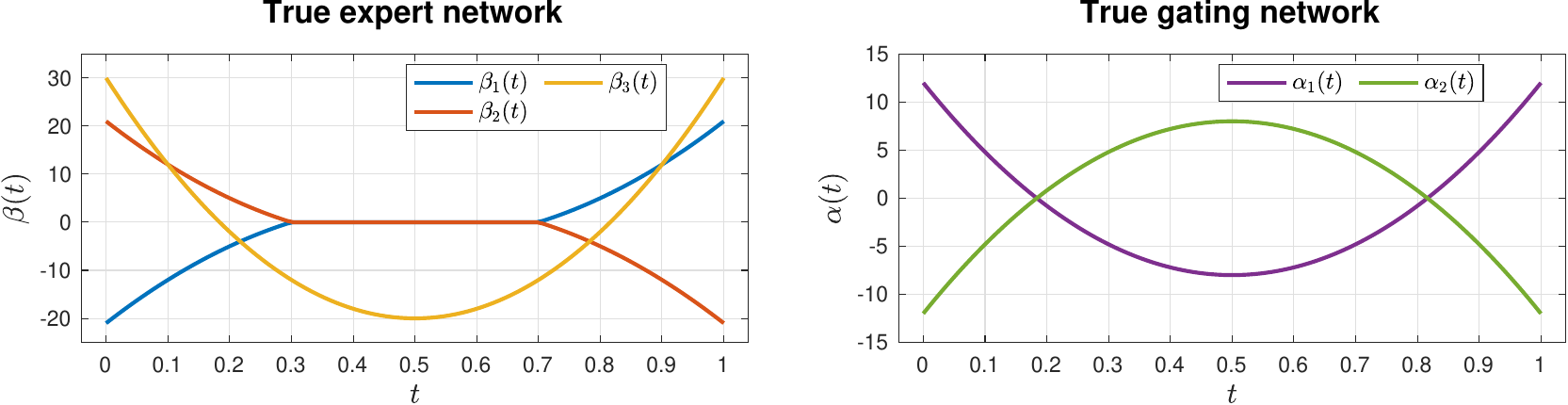}}
\caption{The true expert and gating functions used in simulations.}\label{Figure: True expert gating functions S1}
\end{figure}

As it can be seen in Figure \ref{Figure: True expert gating functions S1}, 
the expert parameter functions $\beta_1(t)$ and $\beta_2(t)$ have a flat region in the interval $0.3 \leq t < 0.7$, outside of which they are quadratic, while $\beta_3(t)$ and the gating parameter functions $\alpha_1(t)$, $\alpha_2(t)$ are all quadratic on the whole domain. 
By choosing \textcolor{black}{these functional experts and gating parameters}, we can later compare the sparsity in the zeroth and third derivatives of the solutions obtained by the proposed models.

\paragraph{Functional predictors parameters:} In this simulation, the functional predictors $X_i(\cdot)$ are constructed using the following formula,  
\begin{equation}
X_i(t) = \bsx_i^\top \bsb(t),\quad t\in \cT, 
\label{eq:functiondal predictior simulation}
\end{equation}
in which $\bsb(t) = \left[b_1(t), \ldots, b_{10}(t)\right]^\top$ is a $10$-dimensional B-spline basis $\bsx_{i}\in\R^{10}$ is a coefficient vector defined as $\bsx_{i} = \Bs{W}\Bs{v}_i$,
where $\Bs{W}\in\R^{10\times 10}$ is a  matrix of i.i.d. random values from the uniform distribution $\mathcal U(0,1)$ and $\Bs{v}_i\in\R^{10}$ is a vector of i.i.d. random values from the normal distribution $\mathcal N(0,10)$. 
Here, the matrix $\Bs{W}$ acts as a factor to facilitate the fluctuation of the generated $X_i(\cdot)$.

\paragraph{Noisy functional predictors:} 
Since in real practical situations we do not usually directly observe the true functional predictors $X(\cdot)$'s, but only a noisy and discretized  version of them, we thus consider several scenarios with different noise and sampling levels.
To this end, in the data generating protocol we consider {noisy versions} $U_i(t_j) = X_i(t_j) + \delta_i(t_j)$ of the functional predictors values $X_i(t_j)$, 
where $\delta_i(t_j)\sim\mathcal{N}(0,\sigma^2_{\delta})$ is a centred Gaussian noise with variance $\sigma^2_{\delta}$, for all $i\in[n]$, $j\in[m]$.
We investigate  simulated scenarios $S1,\ldots,S4$ with curve length $m$ and the noise level $\sigma^2_{\delta}$ of the functional predictors, including two levels of sampling $m\in\{\textcolor{black}{300,100}\}$, and two levels of noise $\sigma^2_{\delta}\in\{1,4\}$. The resulting four considered scenarios of simulated data are presented in Table \ref{Table: scenarios}.  
\begin{table}
\centering
\def\arraystretch{.9}
\begin{tabular}{r*{4}{>{\centering\arraybackslash}p{.9cm}}}
\specialrule{1pt}{1pt}{1pt}
Scenario & $S1$ & $S2$ & $S3$ & $S4$ \\
\hline
$\sigma^2_{\delta}$ & 1 & 1 & 4 & 4 \\
$m$ & \textcolor{black}{300} & \textcolor{black}{100} & \textcolor{black}{300} & \textcolor{black}{100} \\
\specialrule{1pt}{1pt}{1pt}
\end{tabular}
\caption{Simulation scenarios 
with different  noise level $\sigma^2_{\delta}$ and curve sampling level $m$.}\label{Table: scenarios}
\end{table}
For each considered scenario, we generate $N=100$ datasets, each dataset contains \textcolor{black}{$n$ pairs of $\big(U_i(t),Y_i\big)$, with $n\in\{200,400,600,800,1000\}$}.  
\end{document}